# Optical frequency combs on thin-film lithium niobate: from principles to applications

**Xiangzhi Xie,[a,*] Tianqi Lei,[b] Zeyu Luo,[a] Binbin Nie,[b] Yanyun Xue,[b] Yikun Chen,[a] Qi-Fan Yang,[b] Yaowen Hu,[b] and Cheng Wang[a,*]**

[a]Department of Electrical Engineering & State Key Laboratory of Terahertz and Millimeter Waves, City University of Hong Kong, Hong Kong, China
[b]State Key Laboratory for Mesoscopic Physics and Frontiers Science Center for Nano-optoelectronics, School of Physics, Peking University, Beijing, China

**Abstract** Optical frequency combs, consisting of mutually coherent and equally spaced spectral lines, provide a powerful means of synthesizing, distributing, and processing optical frequencies on chip. Among integrated photonic platforms, thin-film lithium niobate (TFLN) is particularly attractive for comb generation because it combines strong electro-optic (EO) functionality, accessible $\chi^{(2)}$ and $\chi^{(3)}$ nonlinearities, low-loss nanophotonic confinement, and versatile dispersion engineering. This review summarizes the physical foundations and representative realizations of TFLN combs, and discusses their applications in optical communication, microwave photonics, optical computing, and precision metrology. The integration of nonlinear functionality, EO control, and passive photonic components is expected to advance compact, energy-efficient comb sources and broaden the utility of TFLN in integrated photonics and quantum optics.

*Address all correspondence to Xiangzhi Xie, XZ.Xie@cityu.edu.hk; Cheng Wang, cwang257@cityu.edu.hk

## 1 Introduction

Optical frequency combs (OFCs) are spectra of phase-coherent and nearly equidistant lines that provide a coherent bridge between microwave and optical frequencies. Owing to this property, they have become essential tools for precision metrology[1], spectroscopy[2,3], coherent communications[4,5], microwave photonics[6], optical computing[7], and optical waveform synthesis[8]. Before the rise of integrated photonics, OFCs were mainly generated in bulk or fiber optic systems; prominent examples include mode-locked solid-state and fiber lasers[9,10], external modulator-based electro-optic (EO) combs[11], and phase-stabilized mid-infrared optical parametric oscillator combs[12]. These approaches formed the foundations of modern comb science and engineering, offering excellent coherence, spectral purity, pulse energy, and bandwidth. However, they typically require multiple discrete components, long optical paths, and sophisticated stabilization and packaging techniques, leading to unsatisfactory footprint, power consumption, and system cost[13].

Integrated photonics offers a route to address these bottlenecks by exploiting strong optical confinement, engineered dispersion and modal interactions, resonant field enhancement, and wafer-scale fabrication, thereby enabling comb generation in compact, energy-efficient, and scalable architectures[14]. This strategy has proven effective across various material platforms, with notable breakthroughs including beyond-octave-spanning soliton microcombs in silicon nitride[15,16], mid-infrared combs in silicon[17,18], ultra-efficient combs in AlGaAs[19,20], and soliton microcombs with self-referencing capability in aluminum nitride[21,22].

Despite this impressive progress, it remains a significant challenge for a single material system to simultaneously offer low optical loss, efficient and broadband EO control, accessible second- and third-order nonlinearities, and compatibility with dense photonic integration. Recently emerging thin-film lithium niobate (TFLN) provides a distinctive option, as it combines low-loss and sub-wavelength light confinement, dispersion engineering capability, EO addressability via a large Pockels effect, as well as strong $\chi^{(2)}$ and $\chi^{(3)}$ nonlinearities. As a result, TFLN supports multiple comb-generation mechanisms, along with a variety of linear and nonlinear optical functions, within a single low-cost manufacturable platform. This combination makes TFLN a unique and versatile platform for frequency-comb generation and applications, which we detail in this review.

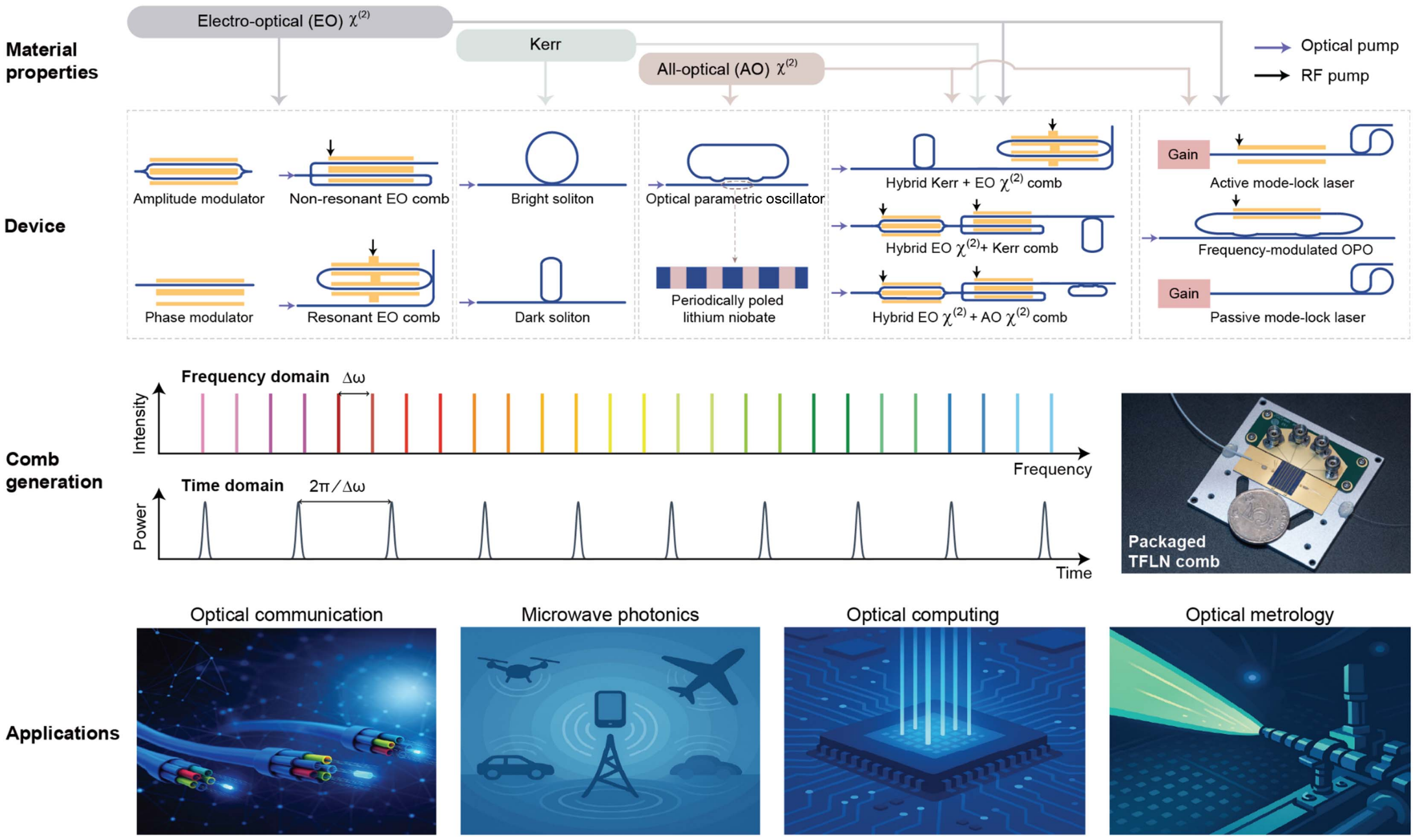

**Fig. 1** Thin-film lithium niobate (TFLN) comb technology and its applications. TFLN comb devices driven by electro-optic (EO), Kerr, and all-optical (AO) $\chi^{(2)}$ effects. TFLN combs can be generated via several mechanisms: pure EO using an external radio frequency (RF) drive, pure Kerr with an optical pump, and AO $\chi^{(2)}$ processes. Hybrid nonlinearities have also been employed, and mode-locking mechanisms are compatible with all of these approaches. The application landscape enabled by TFLN combs includes optical communication, microwave photonics, optical computing, and optical metrology.

Existing reviews have mainly focused on integrated comb generation across material platforms and application scenarios[14,23–26], on the theory, characteristics, and applications of EO comb generation[27–29], or on the material foundations and broader EO and nonlinear device capabilities of the TFLN platform[30–33]. This review provides a unified account of the generation mechanisms, underlying physics, device strategies, and emerging applications specifically for TFLN-based OFCs. Its structure follows the logic illustrated in Fig. 1, progressing from material properties to device-level comb generation and, finally, to system-level applications. We begin in Sec. 2 by outlining the fundamental material properties of TFLN, including the origins of its strong EO and Kerr nonlinearities. The basic operating principles of EO and Kerr comb generation are also introduced. In Sec. 3, we discuss EO-mediated comb-generation mechanisms and advanced EO interaction schemes that enable them. In Sec. 4, we survey recent progress in Kerr comb generation, which leverages third-order optical nonlinearity and dispersion engineering designs. In Sec. 5, we review hybrid nonlinearities used for comb generation, together with the associated laser dynamics. In Sec. 6, we focus on emerging applications enabled by OFC sources, including high-capacity optical communications, microwave photonics, and optical computing. In the final section, we also discuss future perspectives and possible development routes for TFLN-comb-based technologies. We note that, while this review mainly focuses on TFLN, lithium tantalate ($LiTaO_3$) has recently emerged as a compelling alternative because it offers comparable EO properties, substantially reduced crystal birefringence, and potentially better power-handling capability. We therefore also discuss recent advances in $LiTaO_3$ photonics and their potential benefits in relevant sections.

## 2 TFLN Material Properties for Comb Generation

Lithium niobate ($LiNbO_3$, LN) is a uniaxial birefringent dielectric crystal that has long been used in optics and photonics. It crystallizes in the non-centrosymmetric trigonal point group 3m and is a robust room-temperature ferroelectric with spontaneous polarization along the crystallographic *c*-axis, often labeled as the Z-axis[34], as shown in Fig. 2(a). What distinguishes LN from other popular photonic materials is its non-centrosymmetric crystalline structure. In this structure, Li and Nb cations are displaced relative to the oxygen framework along the crystallographic *c*-axis, forming a polar axis and two opposite ferroelectric domain orientations that break inversion symmetry.

As a result, LN exhibits non-zero second-order nonlinear optical coefficients that otherwise vanish in centrosymmetric materials. In a generalized description, the electric polarization inside the LN crystal can be expressed as[35]

$$P = \varepsilon_0 \chi^{(1)} E + \varepsilon_0 \chi^{(2)} E^2 + \varepsilon_0 \chi^{(3)} E^3 + \ldots, \tag{1}$$

where $P$ is the induced electric polarization, $\varepsilon_0$ is the vacuum permittivity, and $E$ is the applied electric field, which may include optical, RF, or DC field components. The tensors $\chi^{(1)}$, $\chi^{(2)}$, and $\chi^{(3)}$ denote the first-, second-, and third-order electric susceptibilities, respectively. Specifically, the first $E$ term describes the ordinary linear optical response and determines the ordinary and extraordinary refractive indices of birefringent LN. The second $E^2$ term describes a nonlinear optical response produced by two input fields, leading to three-wave mixing processes. Importantly, non-zero second-order nonlinearity is only allowed in non-centrosymmetric crystals like LN due to broken inversion symmetry. Depending on the participating field frequencies, this term can manifest either as an EO response, when an optical field interacts with a DC or RF electric field, or as all-optical $\chi^{(2)}$ nonlinear optical processes. These effects are distinct in operation but share the same physical origin. The third-order term represents the material response produced by three input fields, which gives rise to Kerr and related nonlinear optical effects.

In the following subsections, we provide more detailed descriptions of these nonlinear processes and their roles in various comb-formation mechanisms.

## 2.1 EO Effect and EO Comb Generation

### 2.1.1 EO effect in lithium niobate

In the case where one electric field of the second-order nonlinear interaction is provided by an external DC or RF electric field, the nonlinear polarization $P^{(2)}$ induces a shift in the dielectric permittivity tensor that is linearly proportional to the external electric field. In the principal-axis coordinate system, this linear EO response is typically expressed in terms of the optical impermeability tensor ($\eta_{ij}$), i.e.,

$$\Delta\eta_{ij} = \sum_{k=1}^{3} r_{ijk} E_k, \tag{2}$$

where $r_{ijk}$ is the electro-optic tensor, and $E_k$ is the applied electric field. In the principal-axis coordinate system, the unperturbed impermeability tensor is diagonal, with $\eta_{11} = \frac{1}{n_x^2}$, $\eta_{22} = \frac{1}{n_y^2}$, $\eta_{33} = \frac{1}{n_z^2}$, and $\eta_{ij}(i \neq j) = 0$, where $r_{ijk}$ is the EO tensor, and $E_k$ is the applied electric field. The numerical indices $i$, $j$, and $k \in (1, 2, 3)$, which correspond to the crystal axes X, Y, and Z, respectively. In other words, the EO tensor ($r_{ijk}$) measures how efficiently an electric field along direction $k$ changes the $ij$-component of the optical impermeability tensor.

For nearly lossless and non-magnetic materials such as LN, the optical impermeability tensor is symmetric, i.e., $\eta_{ij} = \eta_{ji}$. The index pair $ij$ therefore only contains six independent components and is conventionally contracted into a single Voigt

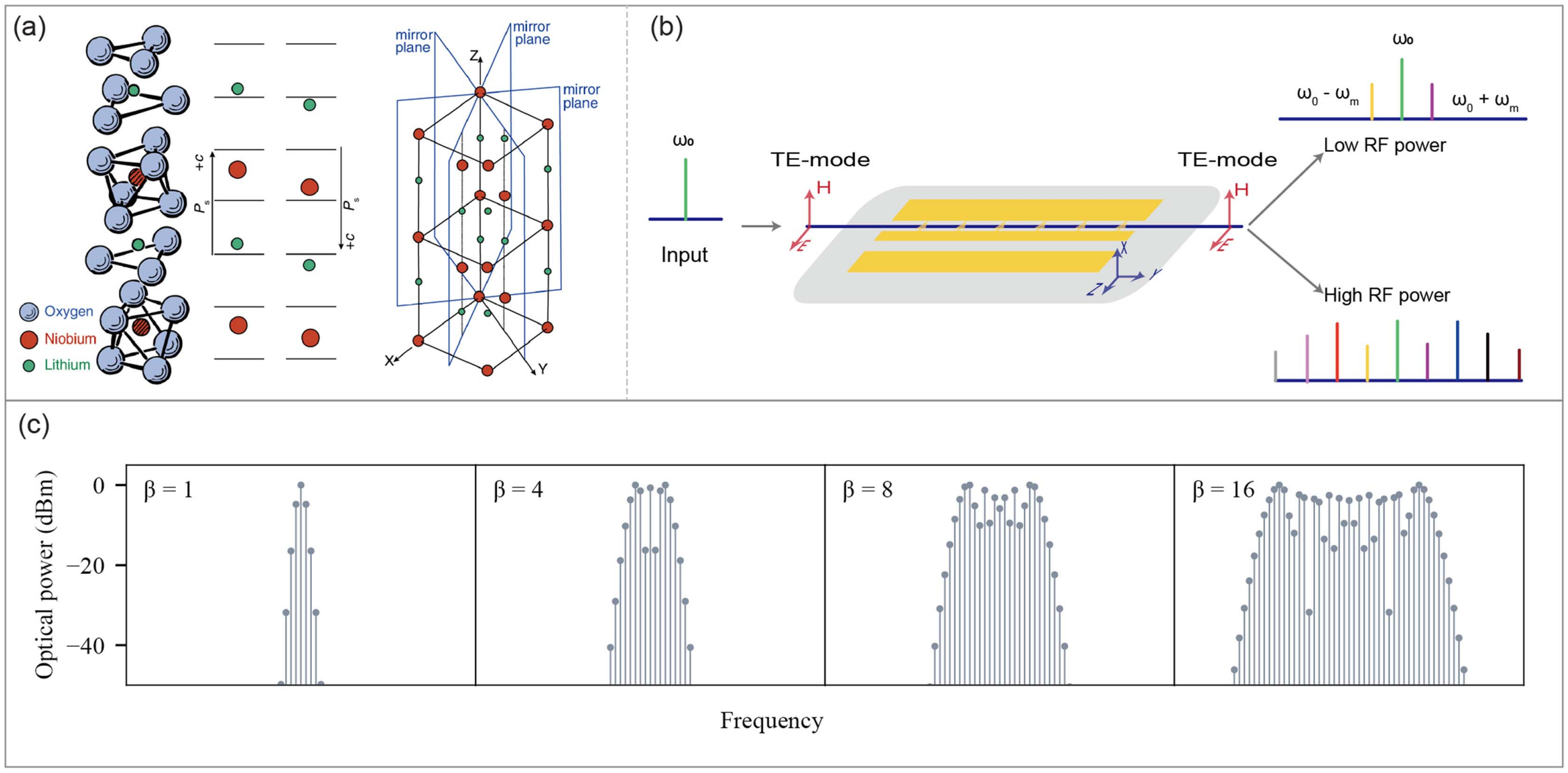

**Fig. 2** (a) Schematic illustration of lithium niobate (LN) crystal, showing positions of the lithium and niobium atoms with respect to the oxygen octahedra in the ferroelectric phase ($T < T_c$). The crystal exhibits three mirror planes and belongs to the 3m class. (b) Schematic illustration of the EO modulation process in an X-cut Y-propagating LN crystal, where the TE mode with a major field component along the Z direction interacts with the in-plane RF electric field through the Pockels effect, generating optical sidebands. Increasing RF power boosts optical–microwave energy exchange, which further generates higher-order sidebands and forms an EO frequency comb. (c) Simulated EO comb spectra for different modulation depths, illustrating the broadening of the comb as $\beta$ increases. Adapted with permission from Ref. [30], © 2021 Optica Publishing Group.

index ($h = 1, 2, \ldots, 6$), which corresponds to $ij$ indices of 11, 22, 33, 23/32, 13/31, and 12/21, respectively. This contraction is a notation choice rather than a new physical degree of freedom, but conveniently reduces the third-order tensor expression in Eq. (2) into a more manageable matrix description[35]:

$$\Delta\eta_{ij} = \sum_{h=1}^{3} r_{hk} E_k. \tag{3}$$

For crystals with the trigonal (3m) point-group symmetry like LN, several components in the EO tensor are forced to zero, leading to the material-specific Pockels relation

$$\begin{pmatrix} \Delta\eta_1 \\ \Delta\eta_2 \\ \Delta\eta_3 \\ \Delta\eta_4 \\ \Delta\eta_5 \\ \Delta\eta_6 \end{pmatrix} = \begin{pmatrix} 0 & -r_{22} & r_{13} \\ 0 & r_{22} & r_{13} \\ 0 & 0 & r_{33} \\ 0 & r_{42} & 0 \\ r_{42} & 0 & 0 \\ -r_{22} & 0 & 0 \end{pmatrix} \cdot \begin{pmatrix} E_1 \\ E_2 \\ E_3 \end{pmatrix}. \tag{4}$$

Among the remaining non-zero elements, $r_{33}$ is the largest and most widely used EO coefficient in LN, with a typical value of approximately 31 pm/V[36]. Since $r_{33}$ is a diagonal term that links electric fields only along the crystal Z-axis, this leads to significant design convenience. The simple rule of sum is to align all major electric fields, whether RF or optical, along the Z-axis, e.g., the case shown in Fig. 2(b) for an X-cut LN crystal. Since $\eta_3$ is associated with the extraordinary refractive index $n_e$, i.e., $\eta_3 = \frac{1}{n_e^2}$, in the small-signal limit, the Pockels effect in most TFLN devices can be simplified to

$$\Delta n_e \approx -\frac{1}{2} n_e^3 r_{33} E_z. \tag{5}$$

### *2.1.2 EO comb generation mechanism*

According to Eq. (5), the EO interaction strength is mainly determined by material properties and the external electric field. As a result, the TFLN platform enables a significantly stronger EO response than its bulk counterpart, since the sub-wavelength optical confinement allows close electrode placement that leads to an increased electric field for a certain applied voltage. This strengthened EO interaction reduces the required drive voltage or RF power, making TFLN particularly attractive for EO modulation and comb generation purposes.

While in principle a TM-like mode in a Z-cut TFLN waveguide can also effectively make use of the $r_{33}$ coefficient through top–bottom electrodes, the X-cut, Y-propagating architecture as shown in Fig. 2(b) is more commonly deployed due to its natural compatibility with RF coplanar waveguides. Here, a TE-like optical mode with its major electric field polarized along the crystal Z-axis propagates through the gap of a pair of planar electrodes that create a modulation electric field also along the Z-direction. This induces an effective-index change that accumulates over the interaction length $L$, producing an optical phase shift $\Delta\phi = \Delta n_{\text{eff}} \cdot k_0 L$, where $k_0$ is the free-space wavevector and $\Delta n_{\text{eff}}$ is the effective-index change. Under an applied voltage $V$, this phase shift can be written as

$$\Delta\phi = \pi \frac{V}{V_\pi}, \tag{6}$$

where $V_\pi$ is the half-wave voltage, defined as the voltage required to induce a $\pi$-phase shift. It is a key EO figure of merit that depends on electrode geometry, optical-microwave overlap, interaction length, and the effective EO coefficient. A lower $V_\pi$ generally leads to stronger EO modulation and more efficient EO comb generation.

When the applied voltage is an RF signal, the induced optical phase becomes time-dependent, $\Delta\phi(t)$. For a sinusoidal drive $V_m \cos(\omega_m t + \phi_0)$, where $V_m$ is the peak drive voltage, $\omega_m$ is the RF angular frequency, and $\phi_0$ is the initial phase[37,38], the modulation-induced additional phase can be expressed as

$$\Delta\phi(t) = \pi \frac{V_m}{V_\pi} \cos(\omega_m t + \phi_0). \tag{7}$$

For simplicity, we define the phase modulation depth as $\beta = \pi \frac{V_m}{V_\pi}$, so that $\Delta\phi(t) = \beta \cos(\omega_m t + \phi_0)$. For a monochromatic optical input field $E_{\text{in}} = \xi e^{i\omega_0 t}$, where $\xi$ and $\omega_0$ are the complex amplitude and frequency of the optical input field, respectively. The output optical field can be written as

$$\begin{aligned} E_{\text{out}} &= e^{i\Delta\phi(t)} E_{\text{in}}(t) = \xi e^{i\omega_0 t} e^{i\Delta\phi(t)} \\ &= \xi e^{i\omega_0 t} \exp[i\beta \cos(\omega_m t + \phi_0)]. \end{aligned} \tag{8}$$

In the small-signal limit, $\beta \ll 1$, the phase-modulated field can be approximated to the first order as $E_{\text{out}}(t) = \xi e^{i\omega_0 t}(1 + i\beta \cos(\omega_m t + \phi_0))$. The expression can be further written as

$$E_{\text{out}}(t) = \xi e^{i\omega_0 t} + i\frac{\beta}{2}\xi e^{i(\omega_0+\omega_m)t+i\phi_0} + i\frac{\beta}{2}\xi e^{i(\omega_0-\omega_m)t-i\phi_0}. \tag{9}$$

This result shows that a weak EO phase modulation transfers a small fraction of the carrier power into two first-order optical sidebands at $\omega_0 \pm \omega_m$ [as illustrated in the top right inset of Fig. 2(b)]. Physically, this process can also be interpreted as coherent energy transfer between the optical carrier and the RF drive. An optical photon at $\omega_0$ may absorb or emit an RF photon through sum-frequency or difference-frequency generation processes, producing frequency-converted photons at $\omega_0 \pm \omega_m$.

For larger modulation depths, however, the full expression in Eq. (9) must be retained and can be described by the Jacobi–Anger expansion

$$E_{\text{out}}(t) = \xi e^{i\omega_0 t} \sum_{n=-\infty}^{\infty} i^n J_n(\beta) e^{in(\omega_m t + \phi_0)}, \tag{10}$$

where $J_n(\cdot)$ denotes the $n$-th order Bessel function of the first kind, and $|J_n(\beta)|^2|\xi|^2$ gives the optical power in the $n$-th sideband. As $\beta$ grows, optical power is redistributed from the carrier into progressively higher-order sidebands, further broadening the spectrum [Fig. 2(c)]. These newly generated frequency components are equally spaced by the RF driving frequency, forming an EO comb centered around the input optical carrier. From the energy-quanta perspective, this corresponds to cascaded sum- and difference-frequency generation processes that generate new photons with energies further away from the pump photon.

In summary, the center frequency of an EO comb is determined by the input optical frequency, whereas its comb spacing (or FSR) is set by the RF driving frequency and the achievable comb span is mainly governed by the modulation depth ($\beta$), as shown in Fig. 2(c).

### 2.2 Kerr Nonlinearity and Kerr Comb Generation

If all electric fields in Eq. (1) are provided by optical input signals, the resulting all-optical nonlinear interactions can also serve as the foundation for frequency comb generation. Among these nonlinear terms, the third-order or Kerr effect is the most commonly used mechanism for all-optical microcomb generation in TFLN, for which we provide a more detailed description here. Frequency combs can also be generated based on all-optical second-order nonlinearity, which is, however, relatively less exploited and will be discussed in Sec. 5.2. In this section, we first review the material Kerr response and cavity-enhanced four-wave mixing (FWM) process in LN[39]. We then discuss the effect of Raman scattering in LN, which could be significantly enhanced in high-$Q$ resonators and compete with the Kerr process, leading to additional design constraints for Kerr comb generators[40–44].

#### 2.2.1 Kerr nonlinearity in lithium niobate

The third-order nonlinear polarization can be specifically written as

$$P_i^{(3)} = \varepsilon_0 \sum_{jkl} \chi_{ijkl}^{(3)} E_j E_k E_l, \tag{11}$$

where $\chi_{ijkl}^{(3)}$ is a rank-4 third-order susceptibility tensor that connects four optical fields, leading to so-called FWM processes [Fig. 3(a)]. A Kerr comb generation process is typically initiated from a degenerate FWM process, where two degenerate pump photons are converted into signal and idler photons.

For a quasi-monochromatic optical field, the instantaneous electronic contribution, i.e., the Kerr effect in the narrow sense, is commonly expressed as an intensity-dependent refractive index:

$$n = n_0 + n_2 I, \tag{12}$$

where $n_0$ and $n_2$ are the linear and nonlinear refractive indices and $I$ is the optical intensity[35,39]. In LN, reported effective values of $n_2$ in the near-infrared are typically on the order of $10^{-19}$ $\mathrm{m^2/W}$, with a representative value around $1.8 \times 10^{-19}$ $\mathrm{m^2/W}$, comparable to that of $Si_3N_4$, one of the most widely used integrated Kerr-comb platforms[30,40,45,46]. In TFLN, however, this quantity should, rigorously speaking, not be a single scalar. Because LN is a uniaxial crystal with point-group symmetry 3m [Fig. 2(a)], its third-order nonlinearity tensor depends on the orientation of the optical field relative to the crystal axes. In Z-cut TFLN devices, a quasi-TE mode with an in-plane electric field mainly samples the nonlinear responses perpendicular to the Z-axis, whereas X-cut devices experience an azimuthally varying projection of the optical field onto the Z-axis. In the latter case, both Kerr nonlinearity and Raman effect (to be discussed in Sec. 4.3) vary with the azimuthal angle, necessitating more careful design considerations [Fig. 3(b)]. In addition, experimentally extracted "effective $n_2$" values may include not only the purely electronic Kerr contribution, but also delayed Raman and phase-mismatched cascaded $\chi^{(2)}$ contributions if these channels are not explicitly separated[40,45,47,48]. Nevertheless, the reported effective $n_2$ values could still provide a good approximation for back-of-the-envelope calculation of Kerr comb generation in LN.

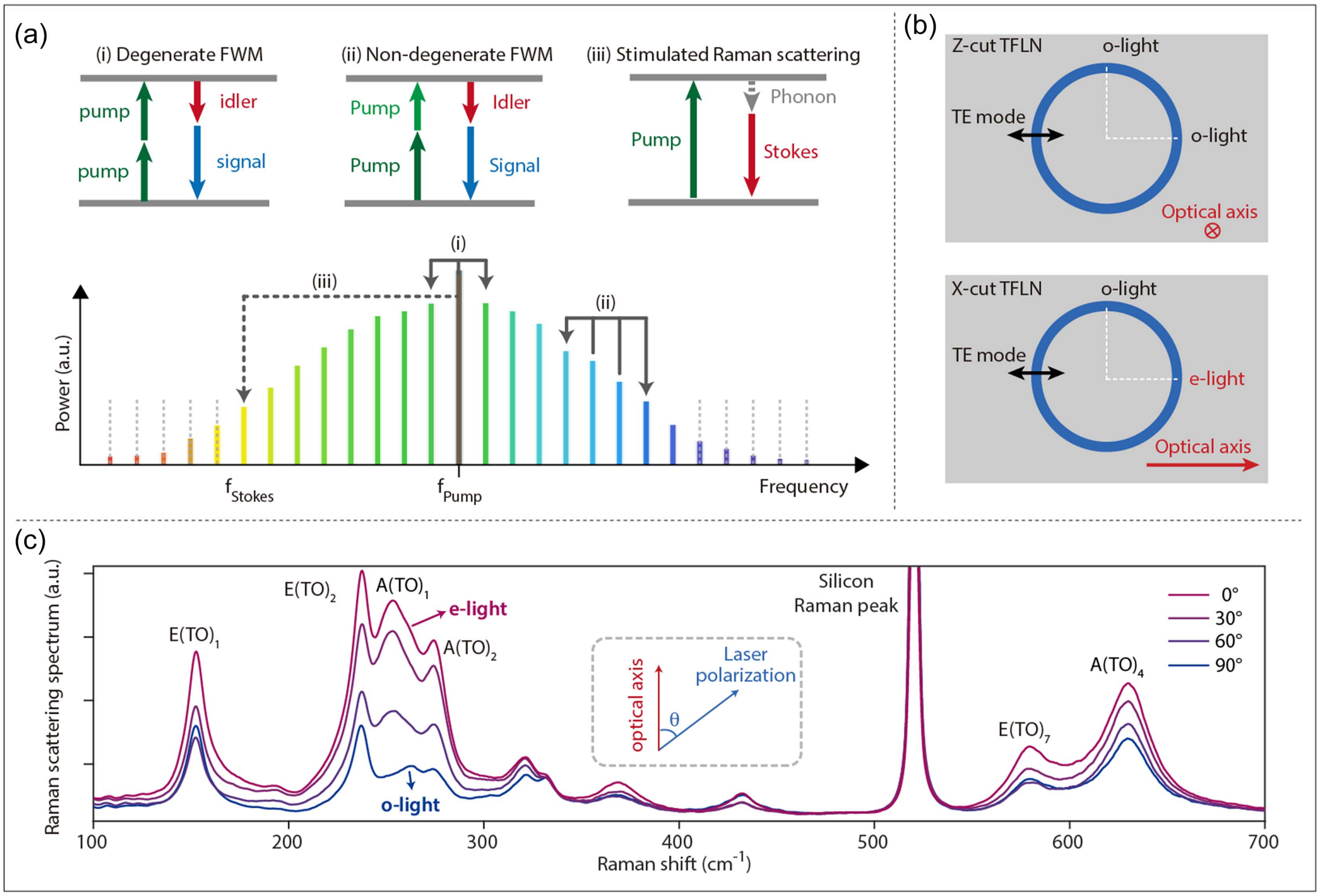

**Fig. 3** Kerr and Raman responses in TFLN for Kerr microcomb generation. (a) Schematics of Kerr four-wave mixing (FWM) and Raman scattering. (b) Schematic illustration of microresonators on Z-cut and X-cut TFLN. (c) Polarization-dependent Raman spectra of TFLN. Adapted from Ref. [41] under CC BY 4.0.

Similar to other integrated photonic platforms, TFLN waveguides provide tight optical confinement and thus significantly enhanced Kerr nonlinear processes compared to free-space or weakly confined waveguide scenarios.

#### 2.2.2 Kerr comb generation mechanism in TFLN microresonators

When such a waveguide is folded into a high-$Q$ microresonator, resonant recirculation further boosts the intracavity field. At sufficiently high pump power, the Kerr nonlinearity provides parametric gain for signal and idler sidebands through degenerate FWM. Once this parametric gain exceeds the total cavity loss, these sidebands start to oscillate, leading to Kerr parametric oscillation and subsequent cascaded FWM for comb formation[49,50]. Dispersion determines the phase matching and mode alignment of this process, and will be discussed in Sec. 4.2. A commonly used estimate of the Kerr parametric oscillation threshold is[51]

$$P_{\mathrm{th,Kerr}} = \frac{\pi n_0 \omega_0}{4\eta n_2} \frac{A_{\mathrm{eff}}}{D_1 Q^2}, \tag{13}$$

where $\omega_0$ is the angular frequency of the pump resonance, $A_{\mathrm{eff}}$ is the effective optical mode area, $Q = \omega_0/\kappa$ is the loaded quality factor, $D_1/2\pi$ is the FSR, and $\eta = \kappa_{\mathrm{ex}}/\kappa$ is the coupling efficiency, with $\kappa = \kappa_{\mathrm{int}} + \kappa_{\mathrm{ex}}$ the total loss rate, $\kappa_{\mathrm{int}}$ the intrinsic loss rate, and $\kappa_{\mathrm{ex}}$ the external coupling rate to the bus waveguide. This scaling makes the design logic clear: tighter confinement, higher $Q$, and more efficient pump coupling all reduce the required pump power. With the maturation of low-loss nanofabrication, Zhu *et al.* (2024) reported monolithic TFLN microresonators with intrinsic quality factors as high as $2.9 \times 10^7$[52], making low-threshold Kerr-FWM operation practical on this platform.

#### 2.2.3 Raman effect in TFLN resonators

Raman scattering is a well-known effect that could meaningfully compete with and impair the Kerr comb generation process in TFLN resonators. If the Raman threshold is comparable to or below the Kerr threshold, pump energy could be effectively diverted to Stokes waves, suppressing broadband FWM, destabilizing soliton formation, or redirecting comb dynamics. Since high-$Q$ cavities enhance both Kerr and Raman processes[40,43,44], mitigation strategies to suppress stimulated Raman scattering should usually focus on dispersion engineering, crystal-cut and polarization selection, and cavity-design choices that avoid strong overlap between Raman-gain peaks and cavity modes.

For a single Raman-active mode, stimulated Raman scattering (SRS), or Raman lasing, occurs when the Raman gain exceeds the cavity loss, and the corresponding intracavity pump-energy threshold can be expressed as

$$E_{\mathrm{th},R} = \frac{\kappa_R}{2g_R}, \tag{14}$$

where $\kappa_R$ is the dissipation rate of the cavity mode closest to the Raman-gain maximum and $g_R$ is the effective Raman gain experienced by that mode[40,53]. Thus, the practically important consideration in TFLN Kerr comb design is not $n_2$ alone, but the relative ordering of the Kerr and Raman oscillation thresholds under the actual cavity, crystal-cut, and polarization conditions.

As summarized in Fig. 3(c), the Raman responses in LN are strongly anisotropic and originate from multiple phonon modes. Raman gain is generally stronger when the optical field is polarized along the Z-axis, and weaker when it is perpendicular to the Z-axis[40–44]. In Z-cut TFLN microresonators, the Z-axis is perpendicular to the chip plane, so an in-plane quasi-TE mode typically samples a relatively uniform Raman response along the resonator circumference. In X-cut TFLN, by contrast, the angle between the circulating optical field and the Z-axis varies continuously along the azimuth, leading to oscillating Raman strength over one round trip. This anisotropy complicates nonlinear dynamics, but it also provides a route to mitigate Raman scattering through appropriate choices of crystal cut, propagation direction, and guided-mode polarization, which will be discussed in more detail in Sec. 4.3[41,42,54].

## 3 EO Comb Generation in TFLN

EO combs are attractive for integrated frequency-comb generation because they are driven by external microwave fields through a deterministic modulation process. Unlike combs that rely on complex nonlinear dynamical processes, EO combs can provide stable, coherent, and microwave-synchronized comb formation with well-defined drive conditions. This externally driven nature also facilitates precise phase control, straightforward RF synchronization, and flexible comb engineering at the system level.

We next review various EO comb generation designs and implementations on TFLN, including non-resonant schemes and resonantly enhanced architectures. Non-resonant EO combs generally offer better operational flexibility, broad frequency-spacing accessibility, and relatively flat comb spectra, but their modulation efficiency is limited, often requiring high RF drive power or long interaction lengths to generate broad bandwidths. Resonantly enhanced EO combs can substantially improve modulation efficiency by using optical and/or microwave resonant enhancement, but this improvement usually comes at the cost of reduced tunability, because both optical cavities and microwave resonators impose additional resonance-matching constraints. Moreover, optical-cavity-enhanced EO combs often require further cavity, dispersion, or modulation engineering to improve spectral flatness. We therefore discuss how stronger light–microwave interaction, improved optical cavity design, and resonator–electrode co-optimization can improve comb bandwidth, flatness, efficiency, and integration density.

### 3.1 Flat-Top Non-Resonant EO Combs

#### 3.1.1 Principle of space–time duality for flat-top EO comb generation

It therefore retains the broad repetition-rate tunability, high pump-to-comb conversion efficiency, and compatibility with continuous-wave (CW) pump wavelengths that make EO combs attractive[55–57]. Here, optical pump-to-comb conversion efficiency denotes the ratio of the power in the newly generated comb sidebands, excluding the residual pump, to the input optical pump power. This metric should be distinguished from the overall system-level power efficiency, which additionally accounts for the electrical power. As discussed in Sec. 2.1.2, a simple sinusoidal phase modulation of a CW optical carrier [Fig. 2(b)] can already produce coherent optical sidebands spaced by the RF drive frequency. However, the spectral envelope of such a phase-modulated comb follows a non-flat Bessel function, whereas its temporal waveform is also not compressed. Flat-top EO combs are desirable for multi-line applications such as wavelength-division-multiplexed optical communications, dual-comb spectroscopy, waveform generation, and parallel ranging, where uniform line powers improve usable channel count and reduce spectral equalization. In the time domain, such spectra can also be phase-aligned and compressed into high-quality pulse trains for high-peak-power and ultrafast applications.

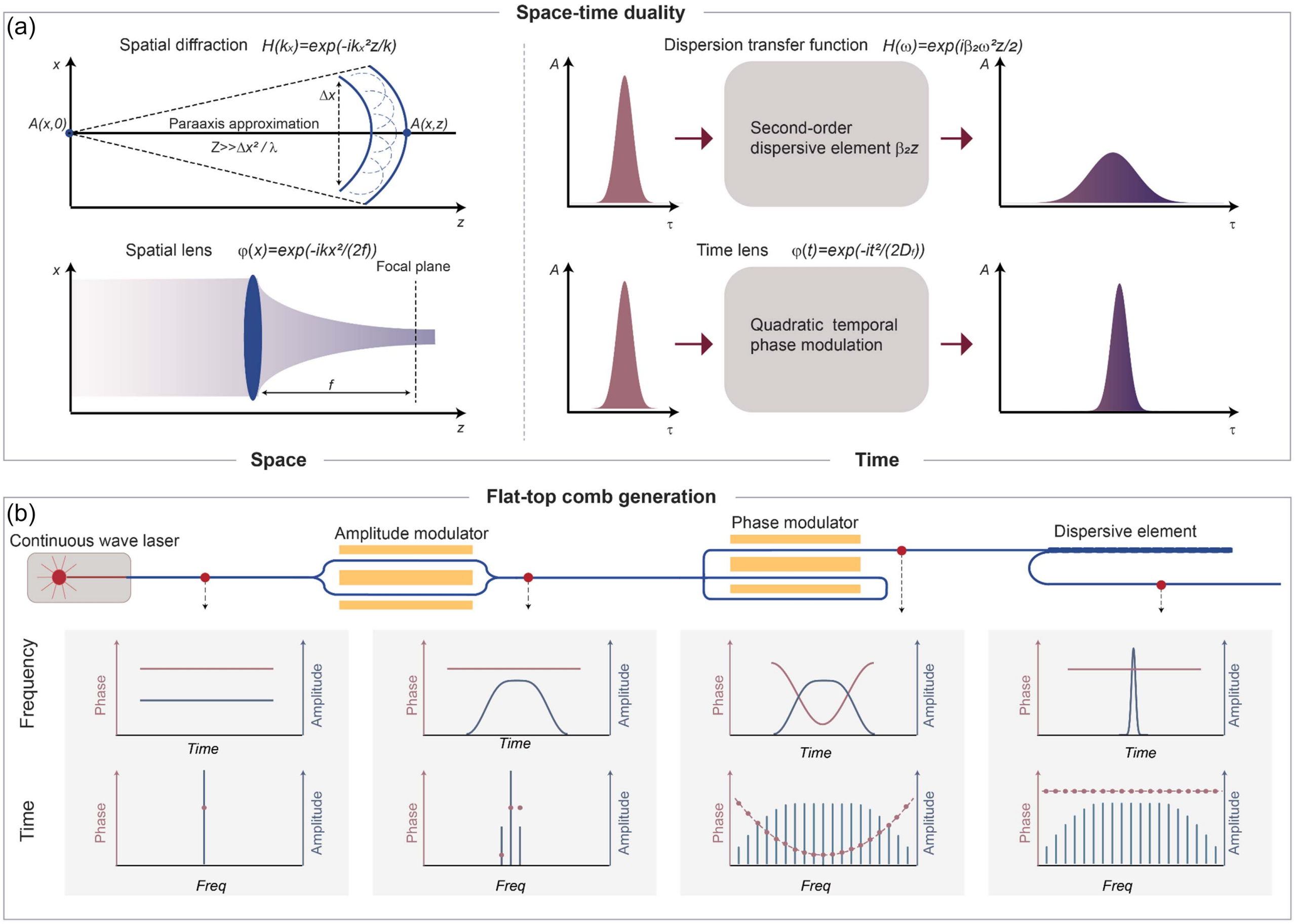

**Fig. 4** Principle of flat-top EO comb generation based on space–time duality. (a) Analogy between spatial diffraction of light in free space, as a function of spatial angular $k$, and fiber dispersion of light as a function of angular frequency $\omega$. Spatial and temporal lenses share the same mathematical form, with quadratic phase profiles across the spatial and temporal apertures, respectively. (b) Joint time–frequency view of a flat-top EO comb generation process. The comb is generated by cascading an amplitude modulator (AM) and a phase modulator (PM). After appropriate dispersion compensation, the output compresses to a Fourier-transform-limited pulse train.

Following the principle of space–time duality, a time-lens-based EO comb generation scheme addresses these limitations by producing spectrally flat-top EO combs with quadratic spectral phase, which could be further compensated for by a second-order dispersion medium, enabling Fourier-transform-limited ultrashort pulses[58–61].

The basic principle of a time-lens system for flat-top comb generation following the space–time duality[62] is shown in Fig. 4(a). In free-space optics, a finite aperture limits the transverse extent of a beam, while a spatial lens applies a quadratic phase profile across the aperture, leading to an output beam that is focused to a small spatial spot after propagation through a focal length. The temporal analogue follows the same logic. An amplitude modulator (AM) defines the useful temporal window of the time lens, i.e., a temporal aperture, whereas a phase modulator (PM) acts as a time lens that provides a quadratic phase profile across the temporal aperture. With proper design, the temporal aperture and quadratic phase jointly map the output signal into a broad and nearly flat optical spectrum. Furthermore, a dispersive element could serve the role of free-space propagation, leading to focusing in the time domain, compressing the comb into a train of ultrashort pulses [Fig. 4(b)].

Specifically, within the temporal aperture $A(t)$ defined by the AM, the time-dependent phase imposed by the phase modulator can be approximated by a nearly quadratic temporal profile near the maxima and minima points:

$$\varphi(t) = \pm\beta\cos(\omega_m t) \approx \pm\beta\left(1 - \frac{\omega_m^2 t^2}{2}\right), \tag{15}$$

where $\omega_m$ is the modulation amplitude and the frequency, respectively.

This quadratic temporal phase induces a linear instantaneous-frequency chirp:

$$\Delta\omega(t) = \frac{d\varphi(t)}{dt} \approx \mp\beta\omega_m^2 t. \tag{16}$$

This expression shows that the time lens maps the temporal profile into optical frequency. If the AM-defined temporal aperture has a full temporal width $T_a$, the frequency span across the aperture is approximately

$$-\frac{|K|T_a}{2} \le \Delta\omega \le \frac{|K|T_a}{2}, \tag{17}$$

where $K = \beta\omega_m^2$ denotes the chirp rate coefficient.

Therefore, the total angular-frequency bandwidth generated by the time lens is

$$\Delta\omega_{\text{comb}} \approx |K|T_a = \beta\omega_m^2 T_a = \frac{\pi V_m}{V_\pi}\omega_m^2 T_a. \tag{18}$$

Since the comb lines are spaced by the modulation frequency $\omega_m$, the maximum comb order on each side of the carrier is approximately

$$n_{\max} \approx \frac{\Delta\omega_{\text{comb}}}{2\omega_m} = \frac{|K|T_a}{2\omega_m} = \frac{\beta\omega_m T_a}{2}. \tag{19}$$

Hence, the total number of comb lines within the time-lens-generated bandwidth can be estimated as

$$N_{\text{comb}} = 2n_{\max} + 1 \approx \beta\omega_m T_a + 1 = \frac{\pi V_m}{V_\pi}\omega_m T_a + 1. \tag{20}$$

When the RF modulation frequency $\omega_m$ and the AM-defined temporal aperture $T_a$ are fixed, the number of generated comb lines scales approximately linearly with the phase-modulation depth $\beta = \frac{\pi V_m}{V_\pi}$. A broader flat-top EO comb therefore requires a larger effective modulation depth, which can be achieved by increasing the RF driving voltage $V_m$, reducing the half-wave voltage $V_\pi$, or cascading multiple phase-modulation stages, which we discuss in more detail in the next section.

To further compress the spectrally flat-top comb into a temporal pulse, we consider the spectral phase of the modulated signal, which follows an approximately quadratic profile:

$$\Phi_{\text{TL}}(\Delta\omega) \approx -\frac{\Delta\omega^2}{2K}. \tag{21}$$

A dispersive element with group-delay dispersion $D$ adds

$$\Phi_{\text{disp}}(\Delta\omega) = \frac{1}{2}D\Delta\omega^2. \tag{22}$$

Pulse compression is obtained when the quadratic phase is compensated for, i.e.,

$$|D| \approx \frac{1}{|K|} = \frac{1}{\beta\omega_m^2} = \frac{V_\pi}{\pi V_m \omega_m^2}. \tag{23}$$

When the above condition is satisfied, a Fourier-transform-limited pulse can be generated, whose pulse duration is primarily determined by the comb bandwidth. For an ideal flat-top spectral amplitude, the intensity full width at half maximum width of the sinc-squared pulse is approximately

$$\tau_{\text{FWHM}} \approx \frac{0.886}{\Delta f_{\text{comb}}}. \tag{24}$$

In summary, increasing the effective modulation depth $\beta$ broadens the comb, reduces the required dispersion, and shortens the transform-limited pulse duration.

### 3.1.2 Schemes for efficient non-resonant EO comb generation

As discussed above, reducing modulation $V_\pi$ could significantly increase the EO interaction strength under a given RF drive, leading to broader EO combs and shorter temporal pulses. Compared with traditional EO comb systems that rely on cascaded bulk LN modulators[67,68], TFLN-based EO combs benefit from substantially improved EO overlap and reduced $V_\pi$ on the thin-film platform[63–66,69–78]. For example, Xu *et al.* (2020) demonstrated an integrated TFLN flat-top EO comb generator with one AM and one PM, producing 13 comb lines with less than 1.2 dB power variation under less than 2 W RF drive[69,70]. Beyond the reduced $V_\pi$, the scalable fabrication and versatile photonic–electrode design enabled by TFLN provide additional freedom to accumulate stronger EO interaction in more compact and integrated architectures, offering a route toward more efficient EO comb generation. As a result, several new design strategies have been proposed and implemented in recent years to further reduce the effective $V_\pi$ of TFLN phase modulators, including recycled modulation paths[63,71,77], mode multiplexing or mode circulation[64,78], and cascaded phase modulations[66].

To increase the effective modulation path length, Yu *et al.* (2022) first proposed to recycle the optical signal through the modulation section twice, generating an ultrabroadband EO comb that could be compressed to femtosecond pulses[63,71]. As shown in Fig. 5(a), the comb spanned more than 18 nm and was compressed using an integrated chirped Bragg grating, whose group-delay dispersion was engineered by periodic waveguide-width modulation. Later, Du *et al.* (2024) proposed to further combine a folded-electrode design with the recycling phase modulation architecture that folds the RF-optical interaction path within a compact footprint[77].

Zhang *et al.* (2023) proposed to further reduce the effective $V_\pi$ by reusing the same RF modulation region through mode multiplexing. As shown in Fig. 5(b), they used $TE_0/TE_1$ mode multiplexers and waveguide crossings to route light through the modulation electrodes for multiple passes, increasing phase-modulation efficiency by a measured factor of about 3.9 and reducing the required electrical power by about 15 times[64]. This multi-pass, non-resonant scheme generated 47 comb lines at a 25 GHz spacing with 28 dBm RF power. Gao *et al.* (2026) later extended mode circulation to four spatial modes ($TE_0$, $TE_1$, $TE_2$, and $TE_3$), achieving an eightfold enhancement of the equivalent modulation index[78].

Another route to increase modulation depth is by cascading multiple phase modulators. For example, Wang *et al.* (2025) demonstrated a fully packaged TFLN EO comb generator consisting of a Mach–Zehnder modulator followed by three cascaded PMs, as shown in Fig. 5(d). The device generated 148 comb lines at a 25 GHz repetition rate with a 10 dB power variation[66]. This approach is effective for spectral broadening, but the need for multiple RF amplifiers and phase-controlled drive paths can increase external RF-system complexity and power consumption.

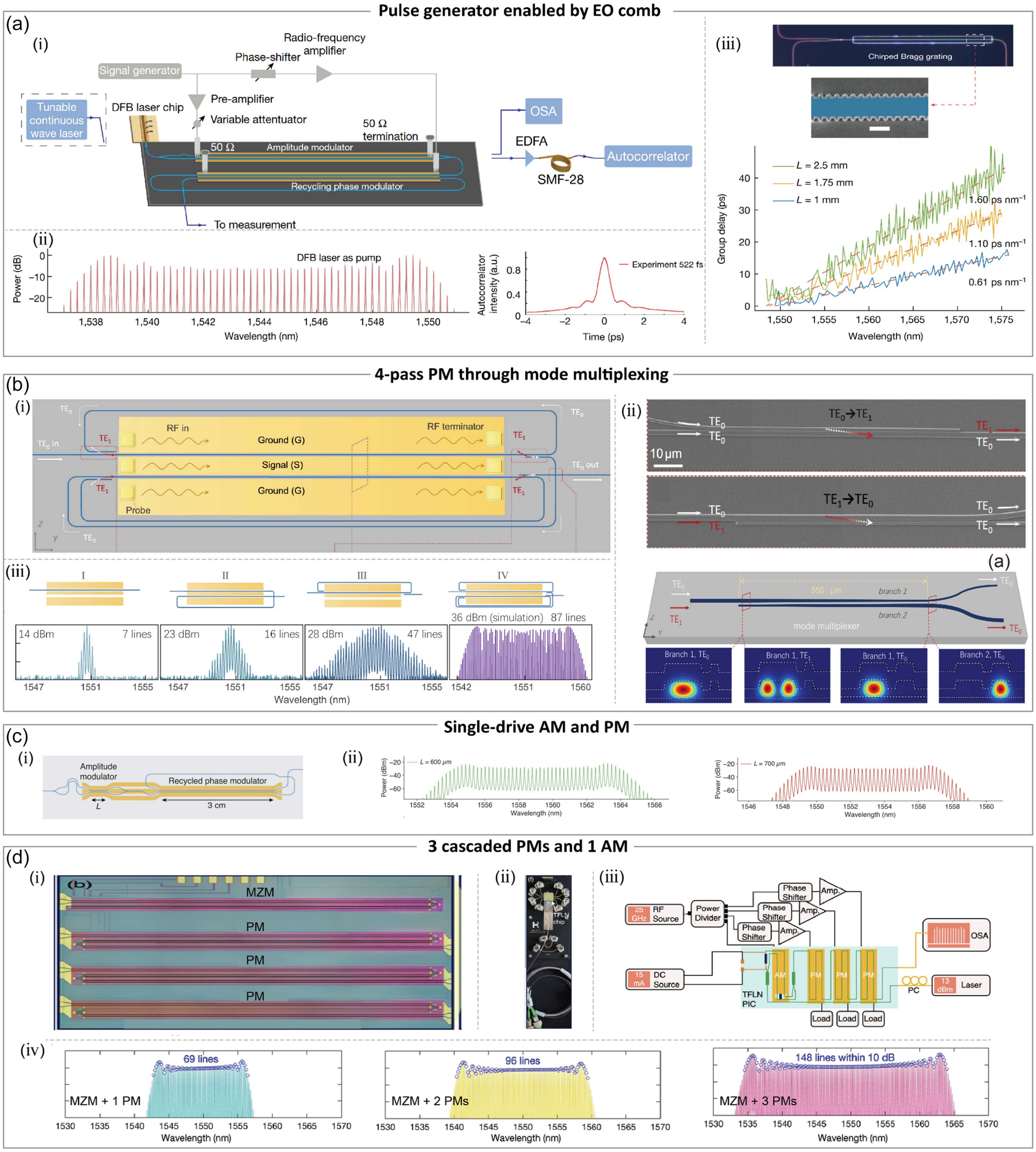

**Fig. 5** Experimental demonstration of flat-top EO comb on TFLN platform. (a) Pulse generator enabled by EO comb. i: chip layout; ii: comb and compressed pulse; iii: chirped Bragg grating. (b) 4-pass PM through-mode multiplexing. i: chip layout; ii: mode multiplexer; iii: generated comb. (c) Single-drive comb generator. i: chip layout; ii: generated comb. (d) Three cascaded PMs and one AM. i: chip layout; ii: packaged chip; iii: experimental setup; iv: generated comb using different numbers of PMs. (a) Adapted with permission from Ref. [63], © 2022 Springer Nature. (b) Adapted from Ref. [64] under CC BY 4.0. (c), (d) Adapted with permission from Refs. [65,66], © 2024/2025 Optica Publishing Group.

In addition to the above demonstrations that focus on improving the phase modulation efficiencies, significant efforts have also been made to simplify the RF driving circuit. In Fig. 5(c), Cheng *et al.* (2024) proposed and demonstrated a single-drive flat-top EO comb generation scheme, where a single RF source successively drives an AM and a PM cascaded on chip. By engineering the on-chip RF phase delay and relative modulation lengths, this architecture removes the necessity of external RF phase shifters and splitters[65]. Similarly, Chen *et al.* (2025) proposed a hybrid-mode Mach–Zehnder modulator to utilize the different EO efficiencies of the $TE_0$ and $TE_1$ modes to realize the required AM–PM response within a single device[79]. Finally, owing to the broad transparency window of LN, the same non-resonant EO-comb concept has also been extended toward the mid-infrared spectrum[74,80].

## 3.2 Optical-Resonator EO Combs

### 3.2.1 Single-optical-resonator EO combs

The flat-top non-resonant EO combs discussed above offer excellent pump-frequency and repetition-rate tunability, but the optical field usually interacts with the modulator only once or through a limited number of recycled passes. This non-resonant nature limits the achievable sideband number unless high RF power or cascaded modulators are used[63,64]. Optical-resonator EO combs address this limitation by placing the EO interaction inside an optical resonator[87]. In the canonical configuration shown in Fig. 6(a), an EO crystal is inserted in a Fabry–Pérot (FP) cavity and driven by a microwave field[81]. The intracavity optical field circulates repeatedly through the modulator. The optical field exchanges energy with microwave photons through sum-frequency generation and difference-frequency generation, creating optical sidebands around the pump[82]. When the modulation frequency matches the cavity FSR, these sidebands are resonantly enhanced and cascade into an EO comb. The resonant EO coupling is often characterized by $\Omega = \mathrm{FSR}\cdot\beta/2\pi$[81,85,88], where $\beta = \pi \cdot V/V_\pi$ is the intracavity modulation depth.

While this resonant EO comb configuration has been explored in bulk LN modulators, recent advances in the TFLN platforms have offered new opportunities through significantly enhanced EO interaction, dispersion and mode engineering capability (to be discussed in Sec. 3.2.3), and scalability toward more advanced architectures like coupled cavities (to be discussed in Sec. 3.2.4). Zhang *et al.* (2019) first proposed to combine an X-cut TFLN racetrack resonator with a high loaded optical $Q$ factor ($Q_{\mathrm{load}} \approx 1.5M$) with coplanar modulation electrodes for efficient EO comb generation. The device achieved a large modulation depth ($\beta \approx 1.2\pi$), generating combs with more than 80 nm bandwidth and about 900 comb lines [Fig. 6(c)][81]. The racetrack geometry allowed the X-cut phase-modulation electrodes to overlap efficiently with the optical mode. As such, this layout has since become a common architecture for optically resonant EO combs. In parallel, Rueda *et al.* (2019) used a mechanically polished microdisk resonator with a much higher optical quality factor ($Q \approx 1.4 \times 10^8$), combined with a microwave cavity, for broadband EO comb generation [Fig. 6(d)][82]. It should be noted that these optically resonant EO combs, while featuring much broader bandwidths than their non-resonant counterparts, have more rigid comb spacing tied to the cavity FSR. In applications that require smaller and more flexible comb spacing, the cascaded cavity–waveguide architecture in Fig. 6(e) provides one workaround, where an FSR-matched microwave drive first creates a fixed-spacing comb in the optical cavity, and a subsequent waveguide phase modulator fills the comb spectrum with lower and reconfigurable repetition rates[84].

### 3.2.2 Strong-coupling EO combs

While the use of optical resonances substantially increases EO-comb span, such resonant EO combs still face two deployment limitations. First, the output sidebands are often weak because only a small fraction of the optical pump is converted; for example, Zhang *et al.* (2019) reported an optical conversion efficiency of about 0.3% for the early TFLN resonant EO comb[81]. Low optical-to-optical conversion efficiency reduces the optical signal-to-noise ratios of individual comb lines and limits applications. This limitation was later addressed by coupled-cavity designs to be discussed in Sec. 3.2.4. Second, the comb generation dynamics is dictated by the weak-coupling model, which gives a relatively fixed synthetic-frequency lattice, leading to comb bandwidths, spectral envelopes, and temporal pulse structures that are not easily reconfigured by microwave control alone.

Strong-coupling EO combs address the second limitation by moving cavity EO modulation beyond the nearest-neighbor, weak-coupling regime. In this regime, both the EO coupling strength and the bandwidth of the microwave drive signal are comparable to or larger than the cavity FSR, so the microwave drive can programmably induce long-range coupling in the synthetic frequency dimension. Lei *et al.* (2025) established a theoretical framework for cavity EO modulation in this regime[85], whereas Song *et al.* (2026) experimentally accessed this regime using a relatively large optical resonator (with reduced FSR) and extended-length traveling-wave electrodes[86].

The starting point of the comb-generation dynamics is still the tight-binding model of cavity EO modulation [Fig. 6(f)], where the EO coupling strength $\Omega$ is smaller than FSR and only adjacent frequency modes are coupled. However, once the system enters the strong-coupling regime ($\Omega$ > FSR), qualitative changes to the cavity-modulation dynamics appear, analogous to strong coupling in cavity QED systems[89]. Specifically, the EO-mediated microwave–optical interaction must be integrated over the round-trip time $t_{\mathrm{R}}$ and evaluated at discrete intervals separated by $t_{\mathrm{R}}$. The nearest-neighbor tight-binding Hamiltonian is therefore replaced by an integrated Hamiltonian formalism:

$$H_{ig} = \sum_{n=-\infty}^{\infty}\Bigg[\omega_n a_n^\dagger a_n + i\frac{\mathrm{FSR}}{2\pi}\left(e^{i2\pi\frac{\Omega}{\mathrm{FSR}}\cos\omega t} - 1\right)\sum_{m=-\infty}^{\infty} a_{n+m}^\dagger a_n\Bigg]. \tag{25}$$

In the weak-coupling limit ($\Omega \ll \mathrm{FSR}$), this expression reduces to the conventional Hamiltonian[85,86,90]

$$H = \lim_{\Omega/\mathrm{FSR}\to 0} H_{\mathrm{ig}} = \sum_n \omega_n a_n^\dagger a_n - \Omega\cos(\omega t)\sum_n a_n^\dagger a_{n+1}. \tag{26}$$

The integrated formalism produces explicit long-range interactions along the frequency dimension under a single-tone microwave drive, making the dynamics fundamentally different from the weak-coupling case [Fig. 6(f)][85]. In the frequency domain, the comb can develop a nonconventional periodic envelope instead of a triangular spectrum; in the time domain, this corresponds to multi-pulse states. The practical regime of low FSR and long modulation electrodes also relaxes microwave-source requirements and expands the accessible parameter space [Fig. 6(g)]. With multi-tone microwave drives, the same framework enables repetition-rate-flexible comb generation, bandwidth extension beyond conventional scaling laws, and resonantly enhanced flat-top spectra across more than 100 modes[86]. Moreover, the strong-coupling EO combs also exhibit improved pump-detuning robustness that is highly appealing in practical applications. This can be understood from strong-coupling-induced band overlap [Fig. 6(h)]: separated synthetic-frequency bands in the weak-coupling regime create forbidden gaps, whereas overlapping bands allow comb generation over a wider pump-detuning range.

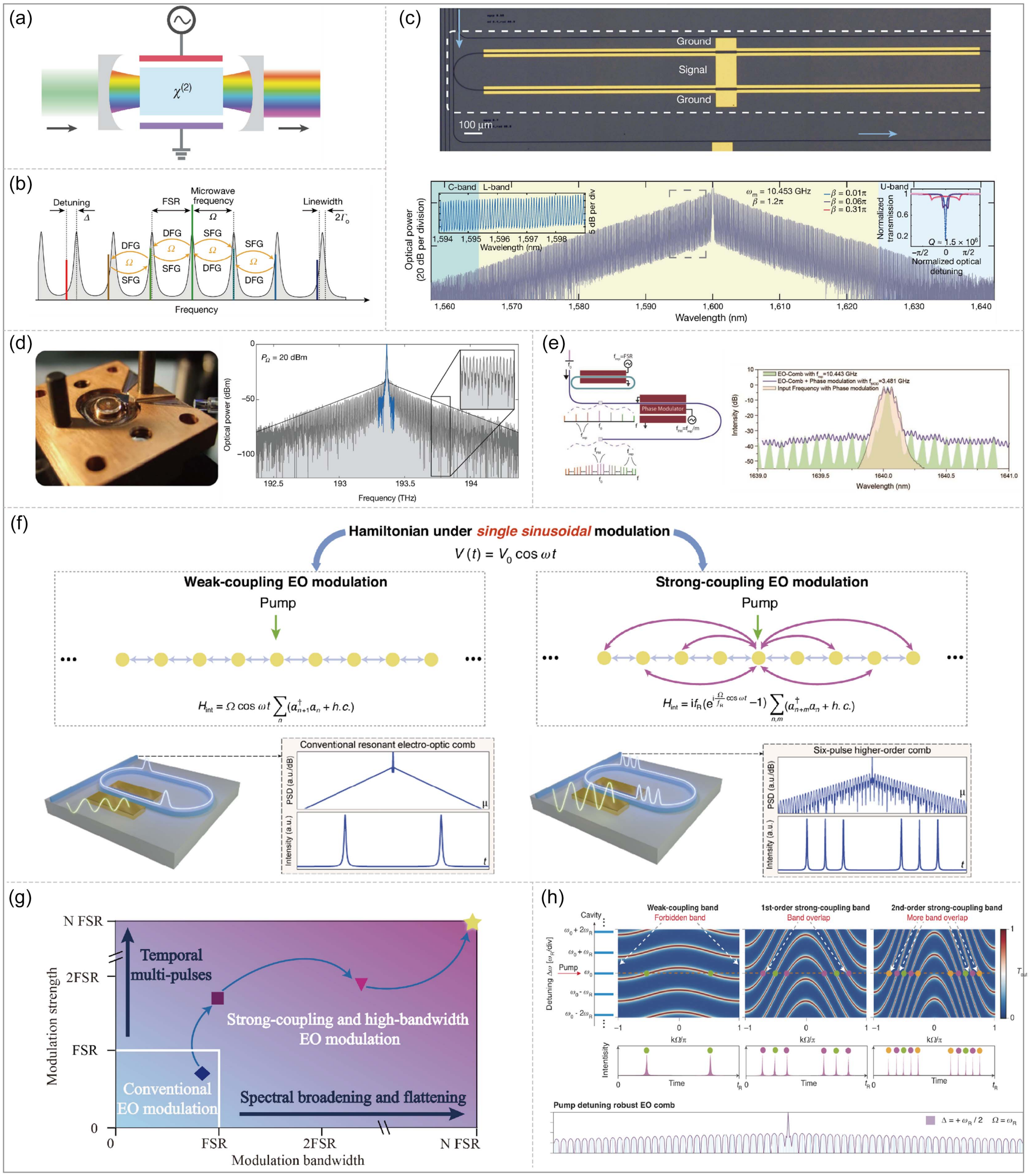

**Fig. 6** Optical-resonator EO combs. (a) Basic configuration of a single-cavity EO comb[81]. (b) Principle of cavity EO modulation. SFG: sum-frequency generation. DFG: difference frequency generation[82]. (c) On-chip microring single-cavity EO comb generator and its broadband cavity EO comb[81]. (d) Microdisk-based single-cavity EO comb generator coupled with a microwave cavity and the corresponding EO comb[82,83]. (e) Cascaded cavity–waveguide EO comb architecture and the corresponding EO comb[84]. (f) Strong-coupling EO comb and multi-pulse state[85,86]: explicit long-range interactions along the frequency dimension under only a single-tone microwave drive, yet fundamentally different from the weak-coupling case. (g) The strong-coupling regime expands the accessible parameter space of cavity EO modulation, offering greater tunability and additional degrees of freedom[85]. (h) Strong-coupling-induced energy band overlap and pump-robust EO comb[85]. (a)–(d) Adapted with permission from Refs. [81,82], © 2019 Springer Nature.

#### 3.2.3 Spectral control through mode and dispersion engineering

Compared with bulk LN, TFLN waveguides offer stronger optical confinement, multiple transverse modes, and geometry-dependent effective indices, leading to much more flexible mode and dispersion engineering capabilities for broadband and profile-controllable resonant EO combs. By changing waveguide width, etch depth, sidewall profile, and crystal orientation, designers can tailor dispersion across a broad wavelength range. This design freedom enables broader comb span by compensating for the accumulated round-trip phase deviations, as well as configurable comb spectral profile, by controlling mode degeneracies or splitting to reshape energy flow in the synthetic frequency dimension.

Figure 7(a) compares the dispersion curves of bulk LN, a TFLN waveguide, and a simulation-optimized design[81]. Material and waveguide dispersion can cause the longitudinal modes of the resonator to accumulate a mode-dependent round-trip phase, leading to non-uniform FSRs across the spectrum. As a result, detuning between optical resonances and respective comb lines could appear and exacerbate away from the optical pump frequency, weakening resonant enhancement and limiting the bandwidth of cavity EO combs. It should be noted that under active EO modulation, the system allows more tolerance to optical detuning than that in a static, passive scenario. This is because the EO modulation periodically shifts the intracavity refractive index, leading to a periodically varying cavity resonance frequency that allows the cavity to tolerate a finite accumulated phase error. This tolerance window scales with modulation depth $\beta$ [gray region in Fig. 7(a)]. As a result, a TFLN waveguide with optimized dispersion and sufficiently strong modulation strength can in principle achieve an octave-spanning EO comb [Fig. 7(a)], a regime prohibitive for bulk LN. Residual dispersion can also be compensated for by a slight microwave detuning [Fig. 7(b)][91]. In a linear dispersion case, microwave detuning leads to a narrowed comb on one side of the pump and a broadened comb on the other side.

In addition to global dispersion engineering, another strategy to locally shape the comb spectrum is through mode engineering. Hu *et al.* (2022) proposed to make use of the coupling between TE and TM modes to induce a frequency mirror in the synthetic lattice [Fig. 7(c(i))], which effectively creates a defect mode that breaks the translational symmetry of the frequency lattice and reflects the energy flow in the frequency dimension. The reflected and incident components form a standing-wave-like pattern, flattening the cavity EO comb spectrum [Fig. 7(c(ii))][92]. Zhang *et al.* (2024) further showed that a similar defect mode can be created in a photonic-crystal ring resonator (PhCR), where periodic sidewall modulation couples clockwise and counter-clockwise modes [Fig. 7(d(i))]. This coupling produces mode splitting at selected wavelengths [Fig. 7(d(ii))] and can also create boundary states that redirect the synthetic-frequency energy flow [Fig. 7(d(iii))][93].

Apart from optically defined synthetic boundaries, microwave detuning can also provide a complementary, electrically reconfigurable soft boundary. Unlike hard mode-splitting boundaries, detuning gradually increases phase mismatch for higher-order modes and suppresses boundary-induced spectral oscillations. With fast electrical tuning[85], this approach enables flat-top cavity EO combs with a tunable bandwidth, as illustrated in Fig. 7(e)[86].

#### 3.2.4 Coupled-resonator EO combs

A key limitation of the above single-resonator EO combs is the low pump-to-comb conversion efficiency, which is mainly because active EO modulation increases the effective intracavity loss $\kappa_{\mathrm{eff}}$ due to dynamic coupling with other cavity longitudinal modes. This drives the cavity into a strongly undercoupled state [inset of Fig. 8(d)] where most pump power cannot enter the cavity. Simply increasing the waveguide–cavity coupling strength is not effective since it inevitably reduces the $Q$ factors of other modes, leading to a narrower comb span.

To address this problem, Hu *et al.* (2022) proposed to achieve a generalized critical coupling (GCC) state using a coupled-cavity architecture [Fig. 8(d)], which improved the pump-to-comb conversion efficiency by about two orders of magnitude[88,98]. Instead of increasing bus–ring coupling for all wavelengths, the auxiliary ring creates wavelength-selective coupling, where the pump mode remains accessible and over-coupled (at static state) to the bus, while most other modes of the main ring become bus-inaccessible dark states [Fig. 8(a)]. Under the low-loss GCC condition $\kappa_{e1} = \kappa_{1\mathrm{eff}}$, turning on microwave modulation brings the pump transmission to critical coupling [inset of Fig. 8(d)]. The pump then couples efficiently into the system and generates a broadband EO comb with about 30% pump-to-comb conversion efficiency over a 132 nm span [Fig. 8(d)]. The higher intracavity comb power further enables 336 fs on-chip pulses, useful for optical atomic clocks, precision metrology, and related applications[88].

Based on this general concept, several coupled-cavity architectures have been proposed and demonstrated. In the auxiliary-ring scheme above [Fig. 8(a)], a small auxiliary ring isolates the bus waveguide from the main ring, while the EO modulator remains in the main ring[81,88]. This converts wavelength-insensitive bus–ring coupling into wavelength-selective effective coupling. In the dual-ring modulation scheme [Fig. 8(b)], two identical coupled rings are modulated synchronously, and the electrode layout couples the symmetric and antisymmetric supermodes formed by the split resonances[94,97]. In multi-ring architectures, such as the tri-ring example in Fig. 8(c), additional coupled rings create richer mode splitting features and multiple energy-flow pathways, expanding the design space for high efficiency and programmable spectral shaping[95,99]. Representative implementations are shown in Fig. 8(e).

### 3.3 Microwave-Resonator EO Combs

Microwave resonance provides another dimension for broadening EO combs. An optical resonator increases the effective EO

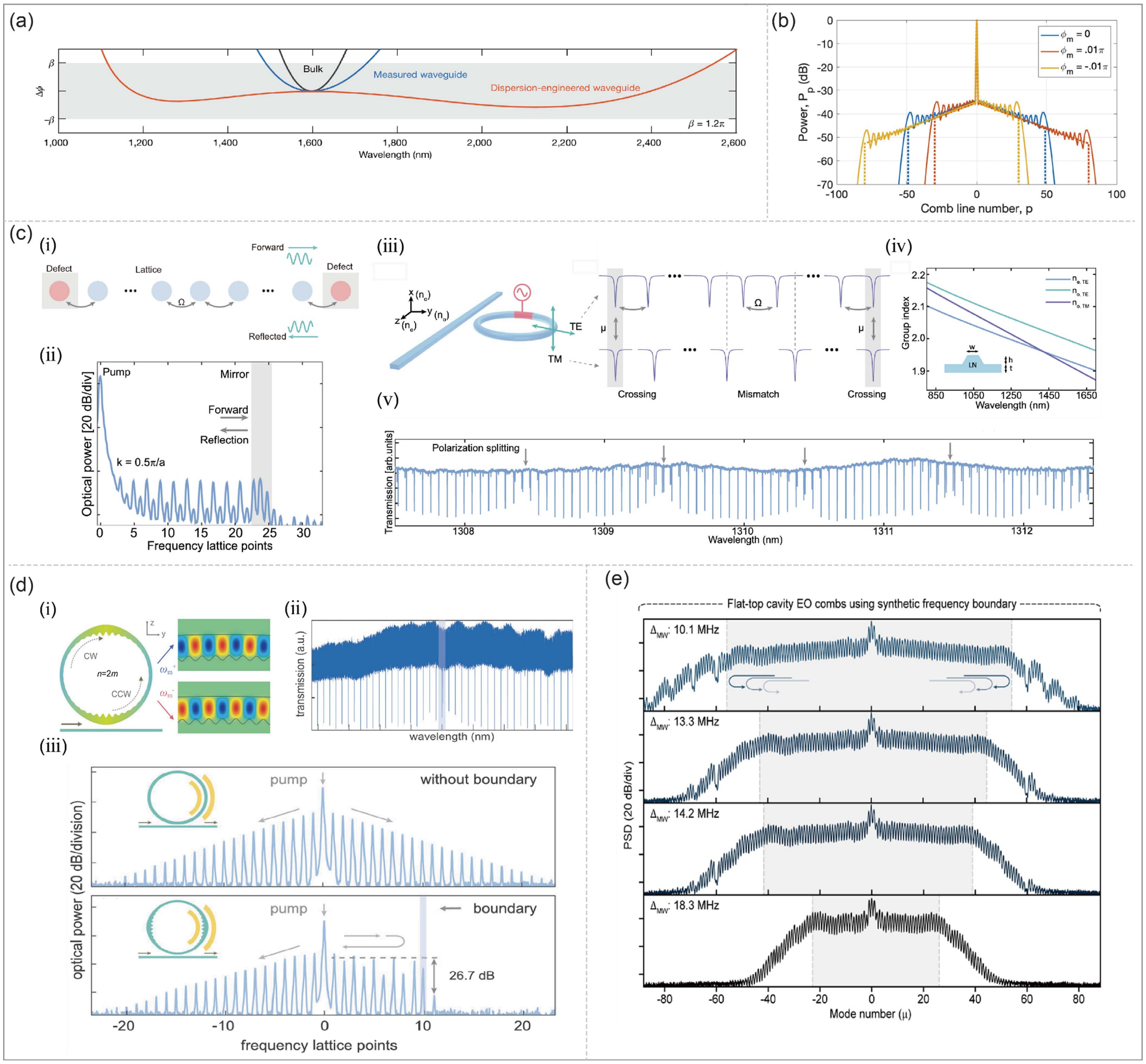

**Fig. 7** Mode and dispersion engineering for EO comb spectral control. (a) The dispersion of bulk LN, TFLN waveguide, and a simulation-optimized design[81]. (b) Asymmetric EO comb spectrum with microwave detuning under linear dispersion condition. Properly chosen microwave detuning can compensate for dispersion and increase bandwidth[91]. (c) i: Schematic of an artificial frequency mirror; ii: frequency-mirror-induced spectral flattening of the EO comb; iii–v: waveguide-geometry design yielding TE–TM coupling over a broad wavelength range[92]. (d) i: Engineering clockwise and counter-clockwise coupling in the cavity via waveguide sidewall periodic structure; ii: single-mode splitting at target wavelength created by mode engineering; iii: comb spectrum engineered by the split mode[93]. (e) Microwave detuning enables a soft frequency boundary and reconfigurable bandwidth and flat-top cavity EO combs[86].

interaction length by recirculating light, whereas a microwave resonator confines and enhances the modulating microwave field. Combining the two resonances increases the phase-modulation index and can therefore further broaden the comb span.

The Hamiltonian of a microwave cavity can be written as[100]

$$\hat{H} = \hbar\omega_b \hat{b}^\dagger \hat{b} + i\hbar\sqrt{\kappa_{\mathrm{ex},m}}(\hat{b}_{\mathrm{in}}\hat{b}^\dagger e^{-i\omega_D t} - \hat{b}_{\mathrm{in}}^\dagger \hat{b} e^{i\omega_D t}), \tag{27}$$

where $\omega_m$ is the cavity resonance angular frequency, $\omega_D$ is the drive frequency, $\kappa_{\mathrm{ex},m}$ is the external coupling rate to the microwave bus, and $\hat{b}^\dagger(\hat{b})$ is the microwave photon-number operator. Transforming to the frame rotating at $\omega_D$ and including the total loss $\kappa_{\mathrm{tot}}$, the effective Hamiltonian becomes

$$\hat{H} = \hbar\left(\Delta_m - \frac{i\kappa_{\mathrm{tot}}}{2}\right)\hat{b}^\dagger\hat{b} + i\hbar\sqrt{\kappa_{\mathrm{ex},m}}(\hat{b}_{\mathrm{in}}\hat{b}^\dagger - \hat{b}_{\mathrm{in}}^\dagger\hat{b}), \tag{28}$$

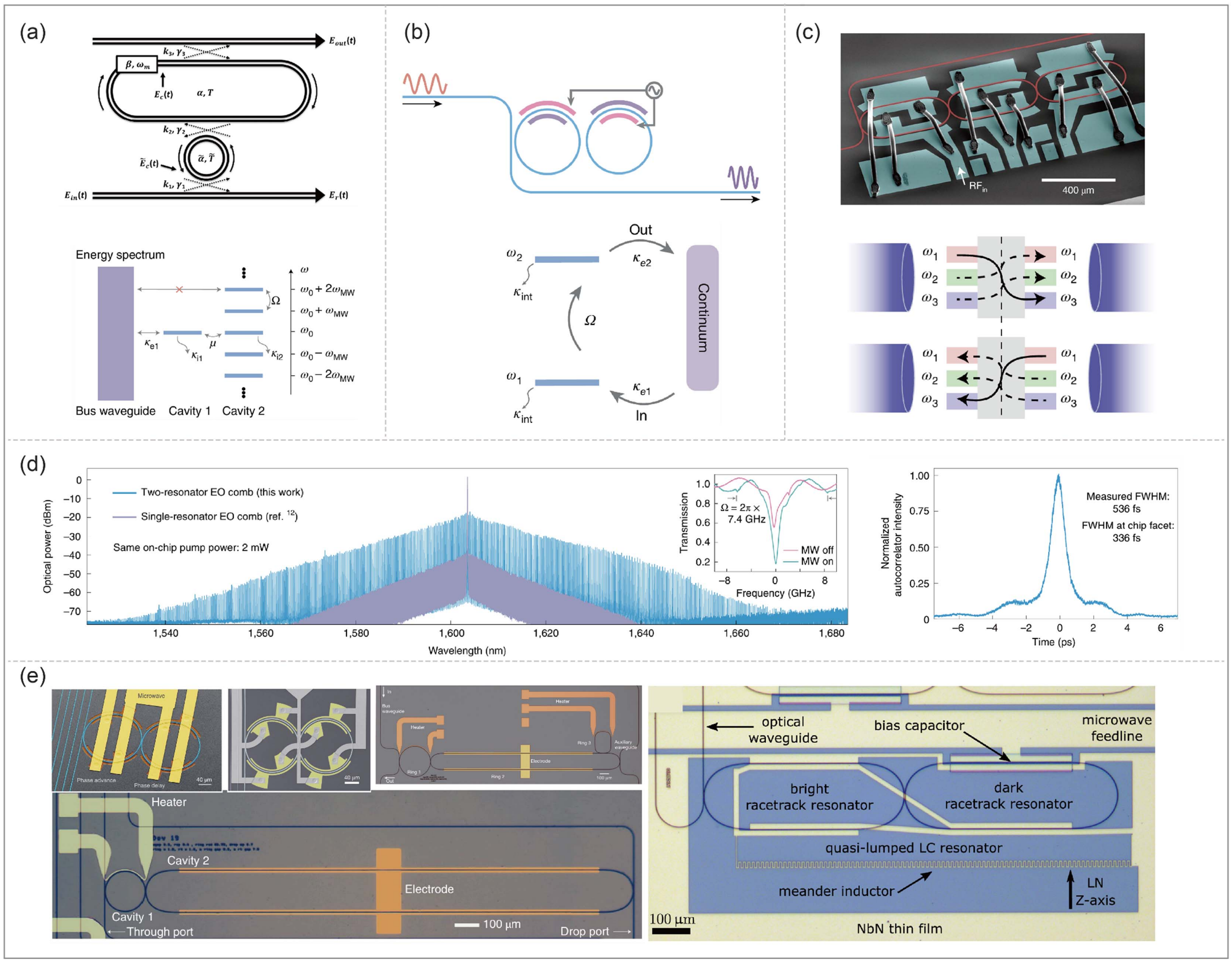

**Fig. 8** Coupled-resonator EO combs and representative coupled-cavity architectures. (a) Auxiliary-ring scheme: the auxiliary ring serves as a frequency-selective coupler between the input bus and the main resonator, effectively isolating the main ring from the bus. It provides strong coupling at the pump wavelength and weak coupling for other wavelengths[88,91]. (b) Dual-ring system: a reversed-electrode layout enforces synchronous modulation of the two ring resonances, thereby coupling the two-level system[94]. (c) Multi-ring system (tri-ring for example here) offers greater modulation degrees of freedom, enabling more complicated inter-frequency energy-flow pathways and spectral shaping[95]. (d) Generalized critical coupling (GCC) with an auxiliary ring yields an ultrabroadband high-efficiency cavity EO comb and femtosecond pulses[88]. (e) Photographs of coupled-cavity EO modulators implementing the above architectures[88,94,96,97]. (a), (e) Adapted with permission from Ref. [91], © 2020 IEEE; Ref. [96], © 2020 Optica Publishing Group.

with detuning $\Delta_m = \omega_m - \omega_D$. The Heisenberg equation yields

$$\frac{\mathrm{d}}{\mathrm{d}t}\hat{b}(t) = \left(i\Delta_m - \frac{\kappa_{\mathrm{tot}}}{2}\right)\hat{b}(t) + \sqrt{\kappa_{\mathrm{ex},m}}\hat{b}_{\mathrm{in}}(t). \tag{29}$$

Under strong coherent drive, we replace operators by their expectation values, $\langle\hat{b}_{\mathrm{in}}\rangle = s_{\mathrm{in}}, \langle\hat{b}\rangle = b$, leading to the steady-state intracavity amplitude

$$b = \frac{\sqrt{\kappa_{\mathrm{ex},m}}s_{\mathrm{in}}}{i\Delta_m - \frac{\kappa_{\mathrm{tot}}}{2}}. \tag{30}$$

The intracavity microwave power is $P_{\mathrm{cav}} = \hbar\omega_m|b|^2 = P_{\mathrm{in}}\frac{2Q_{\mathrm{total}}}{Q_{\mathrm{ex}}}\frac{F}{\pi}$, where $Q_{\mathrm{total}}$ is the loaded quality factor, $Q_{\mathrm{ex}}$ is the coupling $Q$, and $F$ is the cavity finesse. Under the critical-coupling condition, this reduces to

$$P_{\mathrm{cav}} = \frac{F}{\pi}P_{\mathrm{in}}. \tag{31}$$

Compared with direct coplanar-waveguide (CPW) traveling-wave electrodes, a microwave cavity in principle multiplies the delivered RF power by $\sqrt{F/\pi}$, thereby increasing the phase-modulation efficiency by $\sim\sqrt{F/\pi}$. Building on this idea, Chen *et al.* (2025, on LN)[101] and Zhang *et al.* (2025, on lithium tantalate)[100] experimentally verified the microwave-cavity enhancement: the former used a quarter-wave ($\lambda/4$) cavity with open circuit on one end and short circuit on the other end, while the latter used a half-wave ($\lambda/2$) cavity with short-circuited terminations at both ends. Both designs significantly broadened the optical-comb span at given RF input powers, showing the effectiveness of microwave resonators for more efficient EO comb generation. In addition to enhancing the microwave field, microwave-resonant electrodes can also improve RF power transfer by suppressing microwave reflection, making the EO comb generator more RF-circuit-friendly[100,101].

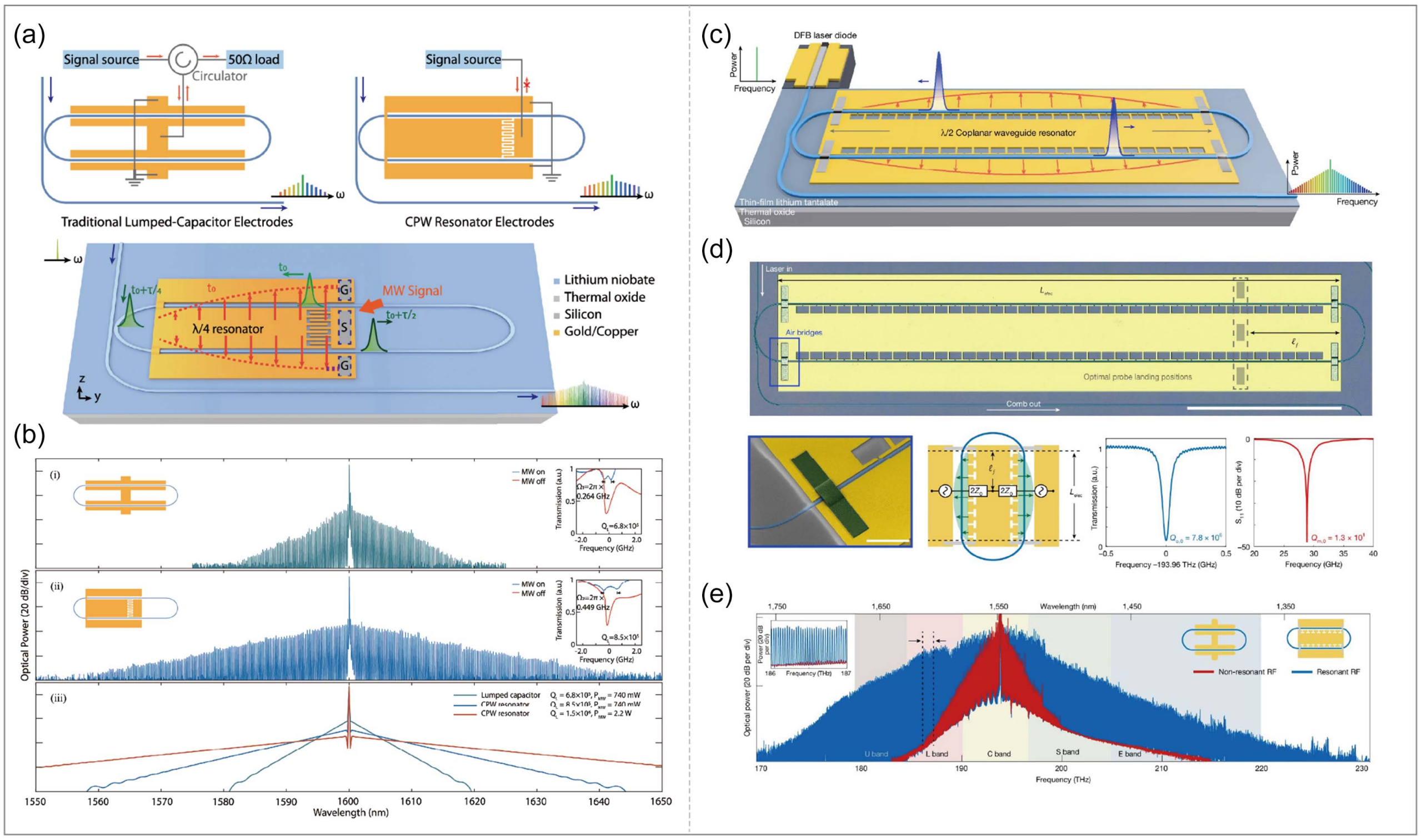

**Fig. 9** Microwave-resonator EO combs. (a) Schematic comparison of EO comb generation using a traditional lumped electrode and a quarter-wavelength cavity. (b) Measured and simulated comb spectra: i. lumped electrode, ii. quarter-wavelength cavity, and iii. simulation results for different optical $Q$ factors. (c) Schematic of half-wavelength cavity-based EO comb generation. (d) Fabricated devices, as well as microwave and optical resonances. (e) Comparison of comb span between resonant and traveling-wave electrodes. (a), (b) Adapted with permission from Ref. [101], © 2025 Optica Publishing Group. (c)–(e) Adapted with permission from Ref. [100], © 2025 Springer Nature Limited.

In the quarter-wave electrode design reported by Chen *et al.* (2025) [Fig. 9(a)], the open end contains an interdigitated-electrode capacitor, which reflects most of the circulating microwave but also provides certain capacitive coupling with the input coplanar waveguide[101]. At the shorted end, the ground and signal are directly connected, producing an impedance discontinuity and ideally providing a total reflection of the microwave. The two reflectors form a Fabry–Pérot-like microwave cavity. By varying the cavity–bus coupling strength at the interdigitated electrodes, the system can be tuned to a critically coupled state. With this design, the authors realized a microwave resonator with a loaded $Q \approx 6$ at a resonant frequency of 25.67 GHz, which is aligned with the FSR of the optical cavity ~25.6 GHz. Based on this device, the authors demonstrated an EO comb at a repetition rate of 25.612 GHz with >85 nm span and >430 lines using a moderate RF drive power of 28.7 dBm, as shown in Fig. 9(b). The comb span in this work was mainly limited by the moderate optical $Q$ factor of $8.5 \times 10^5$. Nevertheless, the achieved comb span is significantly improved by 3.6 times compared with a non-resonant-electrode device (normalized for electrode length and cavity $Q$), corresponding to ~13× reduction in RF power consumption. With further process optimization to push the optical loaded $Q$ beyond $1.5 \times 10^6$ and an RF input of 2.2 W, the approach is projected to support a comb span exceeding 150 nm.

The $\lambda/2$ electrode design reported by Zhang *et al.* (2025) is shown in Fig. 9(c)[100]. Here, impedance discontinuities at both ends are created by shorting the ground and signal, leading to microwave reflections that form a Fabry–Pérot cavity. Unlike the previous design, the coupling region lies inside the cavity by direct contact with a probe. By adjusting the distance between the coupling point and the terminations (fine-tuning $L_f$), the effective coupling to the intracavity field can be tuned to an impedance-matched state. Using this approach, the authors report a microwave cavity with intrinsic $Q \approx 13$, as shown in Fig. 9(d). Leveraging advanced fabrication on lithium tantalate, this work achieved a higher optical $Q$ factor than that reported by Chen *et al.* (2025)[101], i.e., an intrinsic $Q_0 \sim 7.8 \times 10^6$, enabling substantially broader combs. Figure 9(e) shows a span exceeding 450 nm with more than 2000 comb lines. The broadened comb span also makes the previously discussed dispersion control more important.

## 4 Kerr Microcomb Generation in TFLN

Kerr microcombs arise from cavity-enhanced third-order nonlinear interactions rather than from externally imposed microwave sideband generation. Compared with more established amorphous Kerr-comb platforms such as $Si_3N_4$, TFLN possesses not only a native third-order nonlinearity, but also effective Kerr responses induced by cascaded second-order nonlinearity and high-speed EO control on the same platform. The resulting dynamics and system functions are therefore richer than those of pure-Kerr platforms. As outlined in Sec. 2.2.3, practical Kerr-microcomb generation in TFLN requires the joint engineering of dispersion, cavity loss and coupling, mode, polarization, and Raman response.

This section therefore proceeds from theory to implementation: it first introduces the standard theoretical framework, then discusses dispersion engineering and Kerr–Raman competition, next reviews experimental progress in TFLN platforms, and finally outlines the prospects for high-efficiency, broadband, and multifunctionally integrated Kerr-microcomb systems[9,30,40–42,45,46,49,51–54,93,102–127].

### 4.1 Theoretical Framework of Kerr Microcombs

#### 4.1.1 Dispersion in microresonators

A convenient starting point is the standard resonance-frequency expansion of a single-mode family around the pumped resonance $\omega_0$:

$$\omega_\mu = \omega_0 + D_1\mu + \sum_{n\geq 2}\frac{D_n}{n!}\mu^n, \tag{32}$$

where $\mu$ is the integer mode index relative to the reference, typically pumped, mode $\omega_0$, $D_1/2\pi$ is the microresonator FSR at $\omega_0$, and $D_n \equiv \partial^n\omega/\partial\mu^n$ are the $n$-th-order dispersion coefficients that quantify deviations from an equally spaced mode grid [Figs. 10(a) and 10(b)]. In practice, the sign of the second-order dispersion $D_2$ plays a central role in determining the accessible comb state: dark-pulse (platicon) microcombs are typically supported under normal dispersion ($D_2 < 0$), whereas bright-soliton microcombs require anomalous dispersion ($D_2 > 0$) [Figs. 10(c)–10(e)].

It is also useful to define the integrated dispersion,

$$D_{\text{int}}(\mu) = \omega_\mu - (\omega_0 + D_1\mu), \tag{33}$$

which quantifies the deviation of the actual resonance frequencies from perfect equidistance. Kerr microcombs are fundamentally governed by the interplay between integrated dispersion and Kerr nonlinearity, which together determine parametric phase matching, the modulation-instability (MI) gain spectrum, and the existence and stability of coherent nonlinear attractors such as dissipative Kerr solitons (DKSs).

#### 4.1.2 Lugiato–Lefever equation

While the resonance expansion above provides a frequency-domain description of the cavity spectrum, coherent temporal states are more conveniently described by a mean-field equation for the circulating field envelope that includes detuning, loss, dispersion, nonlinearity, and pump. In the form considered here, only second-order dispersion is retained; higher-order terms can be incorporated when needed in more refined models[129]. The intracavity field in a Kerr microresonator is therefore described by the Lugiato–Lefever equation (LLE)[39,121,122]

$$\frac{\partial A(T,t)}{\partial T} = -\left(\frac{\kappa}{2} + i\delta\omega\right)A - \frac{ic\beta_2}{2n_0}\frac{\partial^2 A}{\partial t^2} + \sqrt{\kappa_{\text{ext}}P_{\text{in}}} + igA\int R(t')|A(t-t')|^2\mathrm{d}t', \tag{34}$$

where $T$ is the slow evolution time in the laboratory frame, and $t$ is the fast (retarded) time in a frame co-rotating at the FSR. Here, $A(T,t)$ is the slowly varying envelope of the intracavity field, normalized such that $|A|^2$ corresponds to the intracavity energy. The total cavity decay rate is $\kappa = \kappa_{\text{ext}} + \kappa_{\text{int}}$, where $\kappa_{\text{int}}$ accounts for intrinsic losses such as material absorption and scattering, and $\kappa_{\text{ext}}$ is the external coupling rate to the bus waveguide. The laser–cavity detuning is $\delta\omega = \omega_0 - \omega_L$, with $\omega_0$ and $\omega_L$ denoting the angular frequencies of the pumped cavity resonance and the pump laser, respectively. $\beta_2$ is the group-velocity dispersion (GVD) of the microresonator waveguide, $c$ is the speed of light, and $P_{\text{in}}$ is the injected pump power. The Kerr nonlinearity coefficient is $g = \omega_0 n_2 D_1/(2\pi n_0 A_{\text{eff}})$, where $n_0$ and $n_2$ are the linear and nonlinear refractive indices, $D_1/2\pi$ is the FSR, and $A_{\text{eff}}$ is the effective mode area.

The material third-order response enters through the nonlinear response function[39,121]

$$R(t) = (1 - f_R)\delta(t) + f_R h_R(t), \tag{35}$$

which contains an almost instantaneous electronic (Kerr) contribution and a delayed vibrational (Raman) contribution. Here $f_R$ is the fractional Raman weight, $\delta(t)$ is the Dirac delta function, and $h_R(t)$ is the Raman response function, normalized such that $\int_{-\infty}^{+\infty} h_R(t)\mathrm{d}t = 1$. This point is especially important for LN, since the same $\chi^{(3)}$ response that enables Kerr FWM also contains a strong phonon-mediated component that can drive SRS, perturb soliton dynamics, or even dominate the onset of oscillation if the Raman threshold lies well below the Kerr threshold[40,126].

The Raman response function $h_R(t)$ is causal and therefore defined only for $t \geq 0$. The imaginary part of its Fourier transform is directly related to the Raman gain spectrum[39,121], which can be measured by Raman spectroscopy. The real part follows the Kramers–Kronig relations, so the temporal response can, in principle, be reconstructed from the measured Raman spectrum. For analysis and simulation, it is often convenient to use an approximate analytical form. A widely used choice is the damped-oscillator model[39,121]

$$h_R(t) = \frac{\tau_1^2 + \tau_2^2}{\tau_1\tau_2^2}\exp\left(-\frac{t}{\tau_2}\right)\sin\left(\frac{t}{\tau_1}\right)\mathrm{H}(t), \tag{36}$$

where $\tau_1$ is the Raman vibration period, related to the frequency

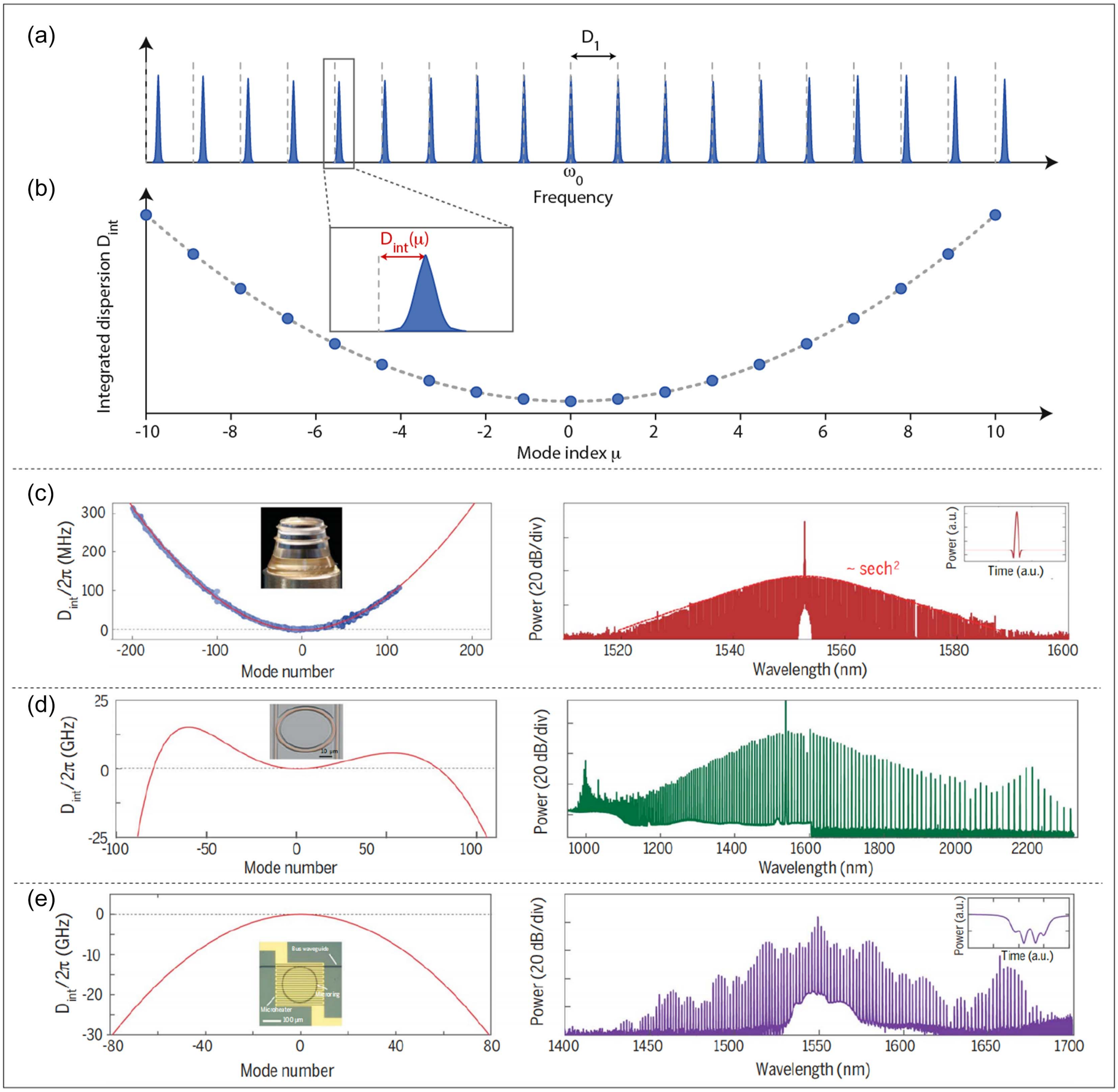

**Fig. 10** Dispersion and its role in Kerr microcomb generation. (a) Group-velocity dispersion (GVD) deviates the resonance frequencies $\omega_\mu$(blue) from the ideally equally spaced frequency grid $\omega_0 + D_1\mu$ (dashed line). (b) The deviation is quantified by the integrated dispersion, $D_{\text{int}}(\mu) = \omega_\mu - (\omega_0 + D_1\mu)$, plotted versus relative mode number $\mu$. The example shown in (a) and (b) corresponds to a predominantly second-order dispersion, for which $D_{\text{int}}(\mu)$ goes quadratically with $\mu$. (c)–(e) Representative $D_{\text{int}}(\mu)$ profiles (left) and the corresponding Kerr microcomb spectra (right)[128]: (c) an approximately quadratic $D_{\text{int}}(\mu)$ in the anomalous-GVD regime supports dissipative Kerr solitons, yielding the characteristic sech$^2$ spectral envelope; (d) an anomalous-GVD profile with two phase-matching points, where $D_{\text{int}}(\mu) \approx 0$, gives rise to resonant dispersive-wave emission on both sides of the soliton band, yielding a soliton spectrum with dual dispersive waves; and (e) an approximately quadratic $D_{\text{int}}(\mu)$ in the normal-GVD regime supports dark-pulse (platicon) states, yielding the characteristic structured spectrum of a dark-pulse microcomb. (c)–(e) Adapted with permission from Ref. [128], © 2018 American Association for the Advancement of Science.

offset of the associated Stokes line; $\tau_2$ is the Raman vibration lifetime, related to the Raman gain bandwidth; and $\mathrm{H}(t)$ is the Heaviside step function enforcing causality ($h_R(t) = 0$ for $t < 0$). If multiple Raman modes contribute, the total response can be written as a weighted superposition of the individual modes:

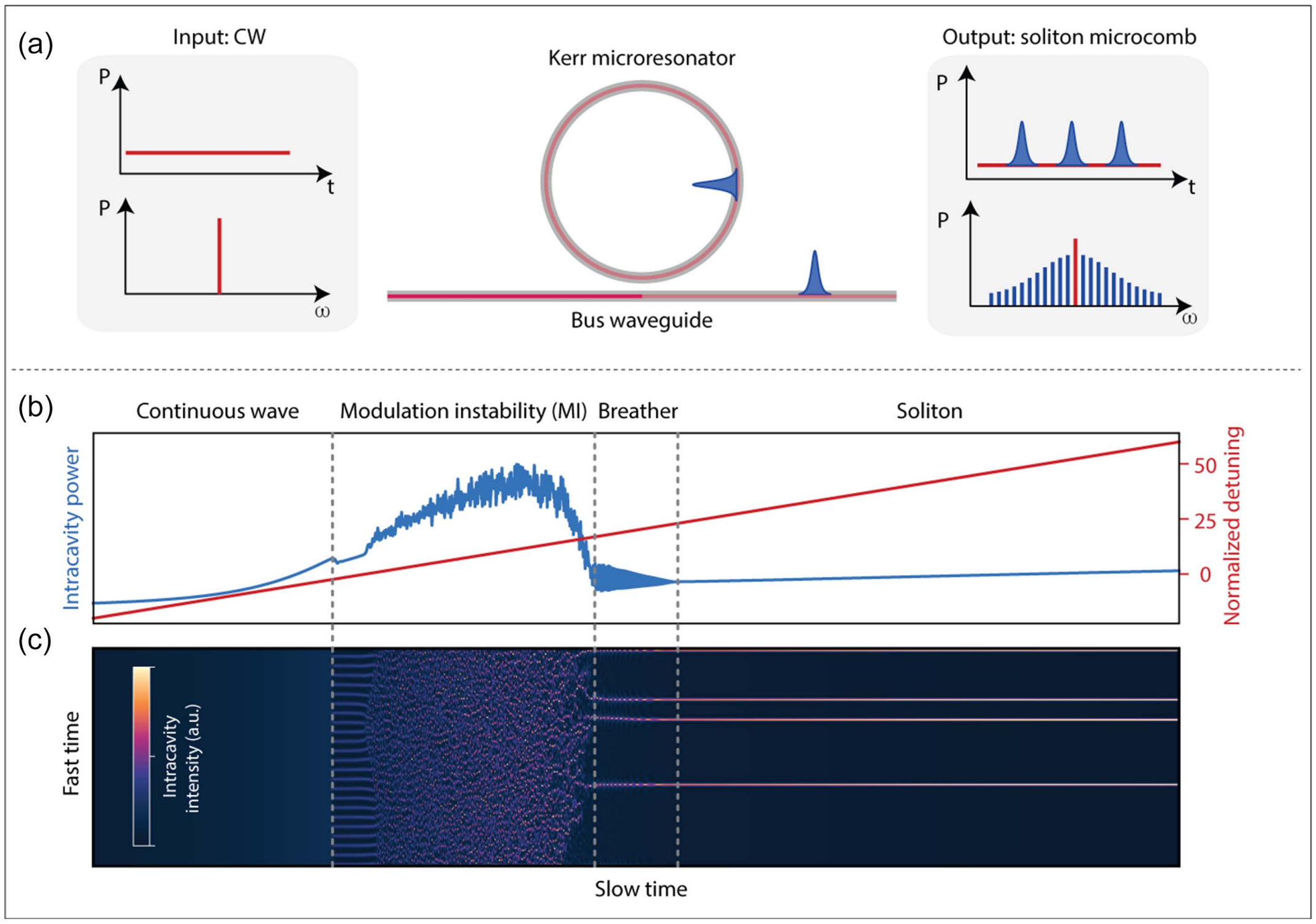

**Fig. 11** Schematic illustration of CW-driven dissipative Kerr-soliton microcomb generation. (a) A single-frequency CW pump coupled into a Kerr microresonator can generate a soliton microcomb via Kerr parametric FWM. (b), (c) Intracavity power (blue, left axis) and pump-resonance detuning (red, right axis) in (b), together with the corresponding spatiotemporal evolution of the intracavity field in (c), during tuning of the pump laser across the pumped resonance from the blue-detuned to the red-detuned side. The intracavity field initially occupies a homogeneous CW steady state, then undergoes modulation instability, evolves through a chaotic regime, and finally settles into a soliton state.

$$h_R(t) = \sum_{i=1}^{N} w_i \frac{\tau_{1,i}^2 + \tau_{2,i}^2}{\tau_{1,i}\tau_{2,i}^2} \exp\left(-\frac{t}{\tau_{2,i}}\right) \sin\left(\frac{t}{\tau_{1,i}}\right) \mathrm{H}(t), \tag{37}$$

where $w_i$ is the fractional weight of the $i$-th mode, satisfying $\sum_{i=1}^{N} w_i = 1$. The LLE, together with periodic boundary conditions in the fast-time coordinate over one cavity round trip, provides a complete mean-field description of Kerr-microcomb dynamics in microresonators[129].

In practice, numerical solutions of the LLE are commonly obtained either by split-step Fourier integration[39], which is well suited to comb build-up, transient evolution, and soliton formation, or by the Newton–Raphson solution of the discretized steady-state equation, which is efficient for obtaining stationary intracavity states, including soliton solutions[129].

### 4.1.3 Dissipative Kerr soliton

Under CW pumping, Kerr-microcomb formation typically proceeds through a characteristic sequence of dynamical states (Fig. 11). As the pump is tuned into resonance and the intracavity power increases, the CW background first becomes unstable to modulation instability (MI), which seeds the primary parametric sidebands. Further tuning drives cascaded FWM and often chaotic comb states. A stable soliton state is reached only within an appropriate, red-detuned window, where the fluctuating background collapses into one or several phase-locked localized pulses. In practice, access to solitons is tied to the bistability of the driven Kerr cavity: the red-detuned branch provides the regime in which the CW background and the localized pulse coexist as a stable dissipative state. The appearance of MI sidebands or a broadband chaotic spectrum therefore indicates oscillation or spectral broadening, but does not by itself establish a fully coherent OFC; that stronger classification requires independent evidence of stable phase relationships, a low-noise repetition-rate beat note, phase/timing-noise characterization, or reproducible formation of a stable temporal state.

Solitons in Kerr microresonators are closely analogous to those in optical fibers. Kerr nonlinearity induces self-phase modulation (SPM) and therefore a positive nonlinear phase chirp, whereas anomalous GVD ($D_2 > 0$) provides an opposing negative dispersive chirp. At the same time, Kerr parametric gain compensates for resonator dissipation. A dissipative Kerr soliton (DKS) is therefore a dissipative attractor in which nonlinearity and dispersion, as well as gain and loss, are simultaneously balanced, yielding a stable intracavity pulse with an essentially invariant temporal envelope from round trip to round trip.

In the absence of Raman nonlinearity ($f_R = 0$), an analytical soliton ansatz for the LLE is given by[103,123]

$$A(t) = B\,\mathrm{sech}\left(\frac{t}{\tau_s}\right)e^{i\phi} + A_{\mathrm{cw}},$$
$$B = \sqrt{\frac{2\delta\omega}{g}},$$
$$\tau_s = \sqrt{\frac{-c\beta_2}{2n_0\delta\omega}},$$
$$\cos\phi = \sqrt{\frac{2\kappa^2\delta\omega}{\pi^2\, g\kappa_{\mathrm{ext}}P_{\mathrm{in}}}}. \tag{38}$$

This form indicates that the soliton pulse sits on a CW background, i.e., $A_{\mathrm{cw}} \approx -if/\delta\omega$, with $f \equiv \sqrt{\kappa_{\mathrm{ext}}P_{\mathrm{in}}}$ denoting the coherent pump-driving term. Here $B$ is the soliton amplitude, $\tau_s$ is the soliton duration, and $\phi$ is the phase of the soliton envelope relative to the pump. These expressions imply that bright solitons exist only in the anomalous-dispersion regime and under red detuning ($\delta\omega > 0$). Increasing the detuning shortens the soliton duration and therefore broadens the comb bandwidth. For a fixed pump power, the maximum detuning is set by $|\cos\phi| \le 1$, yielding

$$\delta\omega \le \frac{g\pi^2\kappa_{\mathrm{ext}}P_{\mathrm{in}}}{2\kappa^2}. \tag{39}$$

Combining the above expressions shows that the soliton microcomb bandwidth scales as $\sqrt{P_{\mathrm{in}}}$, so broadband operation generally requires high pump power. More recently, Helgason *et al.* (2023) and Zhu *et al.* (2025) proposed coupled-resonator-based high-efficiency pumping schemes to enhance pump-to-comb conversion efficiency and enable broadband or octave-spanning soliton microcombs at reduced pump powers[124,125].

In the normal-dispersion regime ($D_2 < 0$), Kerr microcombs can also arise as dark-pulse states, or platicons, which may be understood as interlocked switching waves on a finite background[130,131]. Compared with bright-soliton microcombs, dark-pulse states typically offer higher pump-to-comb conversion efficiency and higher output power. Their spectral envelope, however, is less predictable than that of bright solitons. It also remains difficult for these dark-soliton states to achieve octave-spanning bandwidth required for $f$−$2f$ self-referencing.

### 4.2 Dispersion Engineering in TFLN

Bulk LN exhibits normal material dispersion in the telecom C band (∼1550 nm). Therefore, soliton-microcomb generation in LN requires engineering of the waveguide geometry to compensate for the intrinsic normal material dispersion and realize net anomalous dispersion in a guided mode. Dispersion engineering is also crucial for broadband microcombs, especially octave-spanning combs, which are a prerequisite for $f$−$2f$ self-referencing[9,104]. In practice, achieving octave span requires engineering the integrated dispersion not only to provide anomalous GVD near the pump for soliton formation, but also to tailor higher-order dispersion ($D_3, D_4$) such that phase-matched dispersive-wave (Cherenkov) emission occurs near the spectral edges[53]. These dispersive waves act as spectral "anchors" that extend the soliton spectrum toward the band limits; moreover, they can boost the spectral power density near the $f$ and $2f$ wavelengths, thereby increasing the signal-to-noise ratio of the $f$−$2f$ beat note and facilitating self-referencing.

The most widely used strategy for dispersion engineering on the TFLN platform is to tailor the waveguide cross section (Fig. 12). The refractive-index contrast between the LN core ($n \approx 2.2$) and the surrounding media, such as silica ($n \approx 1.45$) and air ($n \approx 1$), enables geometric control of modal dispersion. By adjusting the waveguide width, film thickness, etch depth, and sidewall angle, the zero-dispersion wavelength (ZDW) can be shifted to realize anomalous GVD in the wavelength band of interest. For example, in a standard 600-nm-thick Z-cut TFLN film, a waveguide width of approximately 1.2–2.0 μm, with an etch depth exceeding 350 nm typically yields anomalous dispersion at 1550 nm, facilitating soliton microcomb generation[53,105–108]. For microresonators with a protective upper cladding, achieving anomalous dispersion may require deeper etching[132] to compensate for the reduced refractive-index contrast. Through systematic optimization of the waveguide cross section and integrated dispersion, near-octave and octave-spanning soliton microcombs have been demonstrated in air-clad TFLN microresonators[53,109,110].

Beyond resonator dispersion engineering, waveguide-based studies show that dispersion and quasi-phase-matching can provide complementary control of broadband frequency generation in TFLN. Periodic, chirped, or imperfect poling can combine ultrabroadband spectral generation with harmonic translation, including visible-to-ultraviolet comb generation and simultaneous supercontinuum and harmonic generation[133–135].

Beyond direct cross-section optimization, several additional dispersion-engineering strategies based on modal coupling developed across the broader microcomb community are also relevant to TFLN (Fig. 13), which we briefly discuss below. Although many of these ideas were not first demonstrated in LN, their physical basis is general and can, in principle, be translated to TFLN once the material anisotropy and fabrication constraints are properly considered.

#### 4.2.1 Mode-interaction engineering[111,112]

When two mode families approach frequency degeneracy, linear coupling hybridizes the corresponding resonances and locally repels their eigenfrequencies [Fig. 13(a)]. This inter-mode interaction perturbs the resonance spectrum and can produce sharp features in $D_{\mathrm{int}}(\mu)$, which may degrade soliton robustness if left uncontrolled[117]. The coupling can occur either within a single resonator, between different transverse mode families [Fig. 13(b)], or between adjacent resonators, forming symmetric and antisymmetric supermodes [Fig. 13(c)]. The same mechanism, however, can also be used constructively. Deliberately engineered mode interactions can create or position dispersive-wave emission and locally improve phase matching, thereby extending the comb bandwidth [Fig. 13(d)][111]. More generally, they can reshape the local dispersion landscape and create an effectively anomalous window even when the background dispersion is normal [Fig. 13(e)], providing a route to soliton-like states in nominally normal-dispersion microresonators[112].

#### 4.2.2 Coupled-resonator (e.g., concentric/inter-ring) dispersion engineering[113,114,118–120]

Conceptually, this approach extends the same hybridization idea to modes belonging to different cavities. Coupling two or more microresonators creates hybridized supermodes whose resonance frequencies can be tailored, providing additional and tunable degrees of freedom beyond a single ring. For example, concentric [Fig. 13(f)] or evanescently coupled [Fig. 13(g)] resonators can be designed such that the hybridized supermode exhibits an effectively anomalous-dispersion window even when each individual resonator is normally dispersive, thereby enabling soliton formation in otherwise inaccessible regimes[113,114,118,119]. More generally, multi-resonator configurations allow simultaneous control of second- and higher-order dispersion and are therefore a powerful route to dispersive-wave engineering[120].

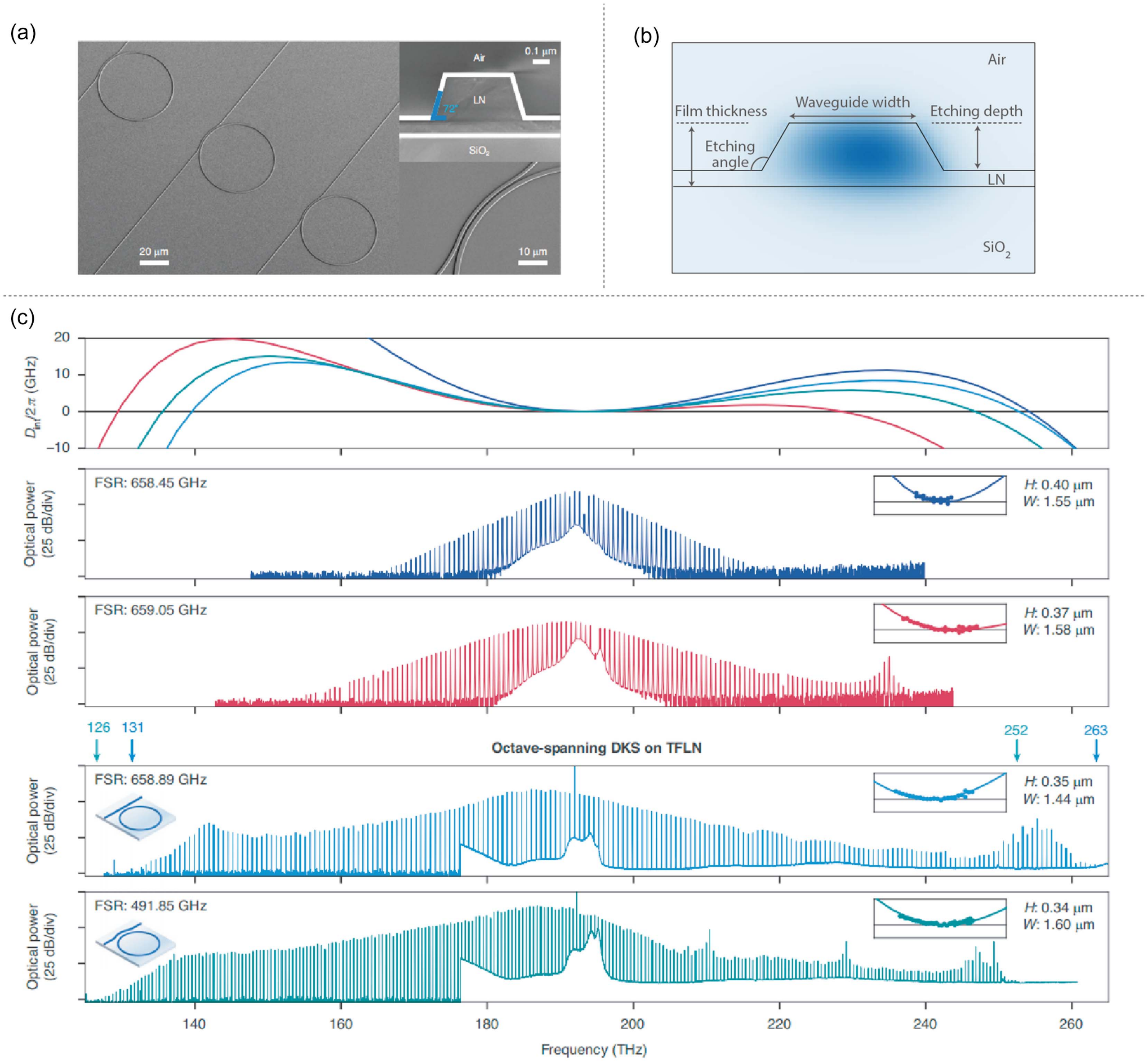

**Fig. 12** Dispersion-engineered TFLN microresonators for broadband Kerr solitons[53]. (a) Scanning electron microscope (SEM) image of fabricated TFLN microring devices. Insets show cross-sectional SEM of the waveguide (top), and a close-up view of the pulley-coupler region (bottom). (b) Cross-sectional schematic of the waveguide with the simulated optical mode profile. The geometric parameters used to engineer the dispersion, namely, the film thickness, sidewall angle, waveguide width, and etch depth, are indicated. (c) Calculated integrated dispersion $D_{\mathrm{int}}/2\pi$ for representative waveguide geometries (top) and the corresponding measured microcomb spectra (bottom). By varying the etch depth $H$ and waveguide width $W$, the integrated-dispersion profile is systematically modified, which in turn reshapes the comb bandwidth and spectral envelope. The insets in the spectral panels show the corresponding $D_{\mathrm{int}}$ profiles for each geometry; the lower examples illustrate octave-spanning dissipative Kerr-soliton spectra in TFLN. (a)–(c) Adapted from Ref. [53] under CC BY 4.0.

The practical implementation of these dispersion-engineering approaches requires consideration of fabrication tolerances. Variations in film thickness, waveguide width, and etch depth modify the waveguide dispersion, while deviations in inter-resonator spacing and resonance alignment can alter the strength and spectral location of mode hybridization. These changes can shift anomalous-dispersion windows and dispersive-wave phase-matching points[53,113]. Assessing fabrication tolerance therefore requires evaluating the resulting variations in integrated dispersion across the intended comb bandwidth, with particular attention to perturbations near the pumped mode, where mode interactions can inhibit soliton formation[117].

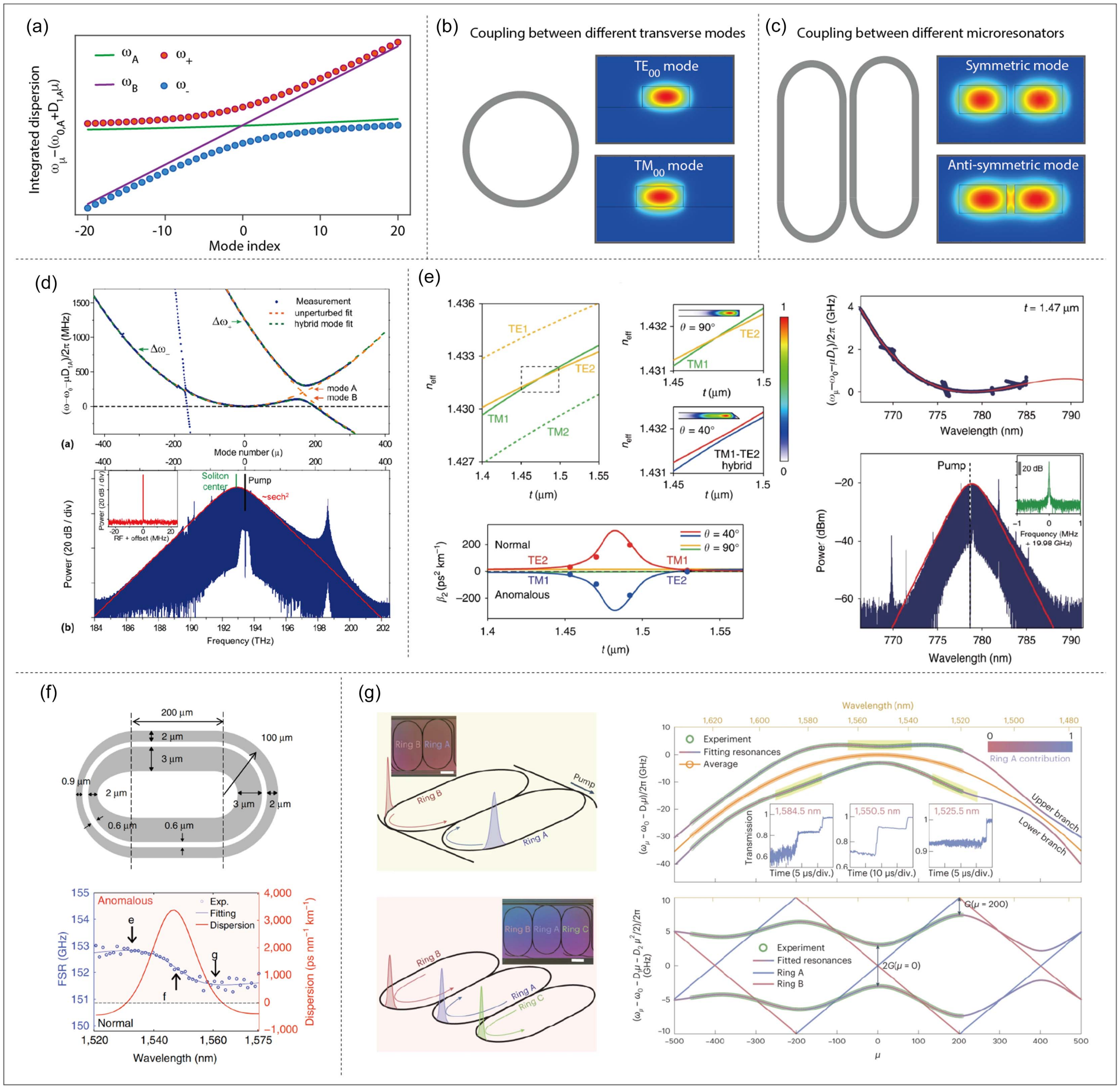

**Fig. 13** Dispersion engineering through modal coupling. (a) When two resonance families become nearly degenerate, modal coupling lifts the degeneracy and produces an avoided crossing between the hybrid supermodes. (b), (c) Such coupling can occur either within a single resonator, between different transverse mode families (b), or between adjacent resonators, forming symmetric and antisymmetric supermodes (c). (d) A localized avoided crossing between two spatial mode families perturbs the dispersion of the soliton-forming branch, creates a phase-matching point, and generates dispersive wave on the otherwise $\text{sech}^2$-like soliton envelope[111]. (e) Near 778 nm, $TM_1$–$TE_2$ hybridization converts a crossing into an avoided crossing when vertical reflection symmetry is broken; one hybrid branch then acquires strong anomalous GVD, and the measured mode-frequency dispersion supports visible soliton microcomb generation[112]. (f) In a concentric racetrack resonator, controlled coupling between the inner and outer racetrack modes produces symmetric and antisymmetric branches; although the uncoupled modes are normally dispersive, the antisymmetric branch becomes anomalously dispersive[113]. (g) In Vernier-coupled microresonators with slightly different FSRs, inter-ring coupling is mode-number non-conserving, so spectral folding creates recurring anomalous-dispersion windows. The two- and three-ring geometries shown in the left panel therefore support pulse-pair and three-pulse soliton states, while the measured hybrid-mode dispersion in the right panel exhibits multiple anomalous windows and associated soliton steps even though the individual rings are normally dispersive[114]. (a)–(d) Adapted with permission from Ref. [111], © 2016 Optical Society of America. (e) Adapted from Ref. [112] under CC BY 4.0. (f) Adapted from Ref. [113] under CC BY 4.0. (g) Adapted from Ref. [114] under CC BY 4.0.

### 4.3 Kerr and Raman Interaction

As discussed in Sec. 2.2.3, LN supports strongly Raman-active vibrational modes. Raman gain therefore opens a competing oscillation channel that can deplete the intracavity pump, destabilize coherent Kerr-comb formation, or even completely dominate the oscillation process when the SRS threshold lies well below the Kerr-comb threshold[126].

A quantitative comparison of Kerr and Raman oscillation requires a common pump normalization. In terms of the intracavity CW pump-mode energy, the ideal Kerr threshold is $E^{(0)}_{\mathrm{th},K} \simeq \frac{\kappa}{2g}$ for optimally phase-matched sidebands with loaded linewidths comparable to that of the pump. With the effective Raman-gain convention in Eq. (14), the threshold for a candidate Stokes mode $(m)$ is $E_{\mathrm{th},R,m} = \frac{\kappa_{R,m}}{2g_{R,m}}$. Here, the effective Raman gain $g_{R,m}$ has taken into account the contributions from the Raman spectral overlap, polarization dependence, and modal overlap. When the lowest Raman lasing threshold exceeds the Kerr oscillation threshold, Kerr oscillation is favored to occur first. Common engineering strategies for increasing the Raman lasing threshold include FSR and polarization engineering, which primarily modify the effective Raman gain, as well as dissipation engineering that increases the Stokes-mode loss[40,53,106]. It should be noted that this threshold comparison only characterizes the onset of oscillation but does not guarantee soliton formation; Raman competition must also be evaluated at the pump power level during soliton operation. Thus, Kerr/Raman threshold ordering identifies the first oscillation channel, but not whether the resulting spectrum is a low-noise, mutually coherent OFC; this requires separate coherence, phase-noise, or stable-state measurements.

In short, managing the Kerr–Raman interplay is a key requirement for LN microcomb design. Representative strategies for suppressing competing SRS by raising the Raman lasing threshold relative to the Kerr threshold are summarized below.

#### 4.3.1 FSR control[53,126]

The Raman gain bandwidth associated with a given vibrational mode in TFLN is typically on the order of hundreds of gigahertz. By designing resonators with an FSR exceeding that bandwidth, the discrete cavity resonances can be shifted away from the Raman-gain maximum, thereby increasing the Raman lasing threshold [Fig. 14(a), upper panel].

#### 4.3.2 Dissipation engineering[53,106,127,136–138]

Wavelength-dependent coupling schemes, such as pulley couplers [Fig. 14(c)][53,127] or self-interference couplers [Fig. 14(b)][106,136], can introduce additional dissipation near the Raman Stokes wavelengths while preserving relatively high $Q$ for the pump and Kerr-comb modes. In addition, Fabry–Pérot resonators incorporating Bragg gratings[137,138] can be designed to provide high reflectivity for the pump and Kerr-comb modes while placing Raman-active Stokes modes outside the stopband, so that they experience higher effective loss [Fig. 14(d)].

#### 4.3.3 Polarization and crystal-orientation engineering[41,42,54,102]

Since the strongest Raman modes are accessed by optical fields polarized along the Z-axis, the crystal orientation, waveguide orientation, and optical mode polarization can be designed such that the guided mode for Kerr comb formation has its electric field mainly perpendicular to the Z-axis [Fig. 14(e)]. This provides a passive route to Raman suppression without requiring elaborate wavelength-selective filtering or additional cavity-loss engineering [Fig. 14(f)].

Taken together, these strategies show that Raman management in TFLN is not a secondary optimization, but part of the primary microresonator design problem.

### 4.4 Photorefractive Effects

In addition to Kerr and Raman nonlinearities, the photorefractive effect should also be taken into careful consideration in TFLN comb designs[139,140], which could significantly influence the resonance frequencies and detuning dynamics of TFLN microresonators. Photorefraction arises from the excitation, transport, and trapping of photocarriers, which generate a space-charge field and modify the refractive index through the Pockels effect. Its magnitude and relaxation dynamics depend on the optical excitation and material properties, including defect and interface characteristics. Consequently, photorefractive resonance shifts can depend strongly on the fabrication process, as demonstrated by their sensitivity to oxide cladding and annealing in TFLN microrings[139].

Unlike the nearly instantaneous Kerr response, photorefraction involves the buildup and relaxation of space charge and therefore introduces a dependence on the optical excitation history. Its temporal response can involve multiple relaxation timescales ranging from milliseconds to seconds[140].

Interestingly, while the slow resonance shifts induced by photorefraction can complicate the wavelength fine-tuning and resonance-locking processes, in certain cases they can also facilitate soliton access in Kerr comb generation. He *et al.* demonstrated self-starting soliton generation and bidirectional switching between soliton states in Z-cut TFLN microresonators supporting quasi-TE modes. In these devices, a power-dependent resonance blueshift dominated the relatively weak ordinary thermo-optic response, stabilizing operation in the red-detuned regime required for conventional bright solitons[105]. This demonstration shows that photorefractive feedback can assist soliton initiation under suitable operating conditions. However, self-starting behavior does not by itself establish repeatable initialization into a prescribed soliton state or long-term frequency stabilization.

For TFLN Kerr microcombs, useful design considerations include both the photorefractive resonance shift relative to the cavity linewidth and its response time relative to the timescales of pump tuning and intracavity-power variations. Characterizing this response over the intended pump-power range and for the specific device and fabrication process is therefore important for developing reproducible soliton sources and selecting suitable detuning-control strategies.

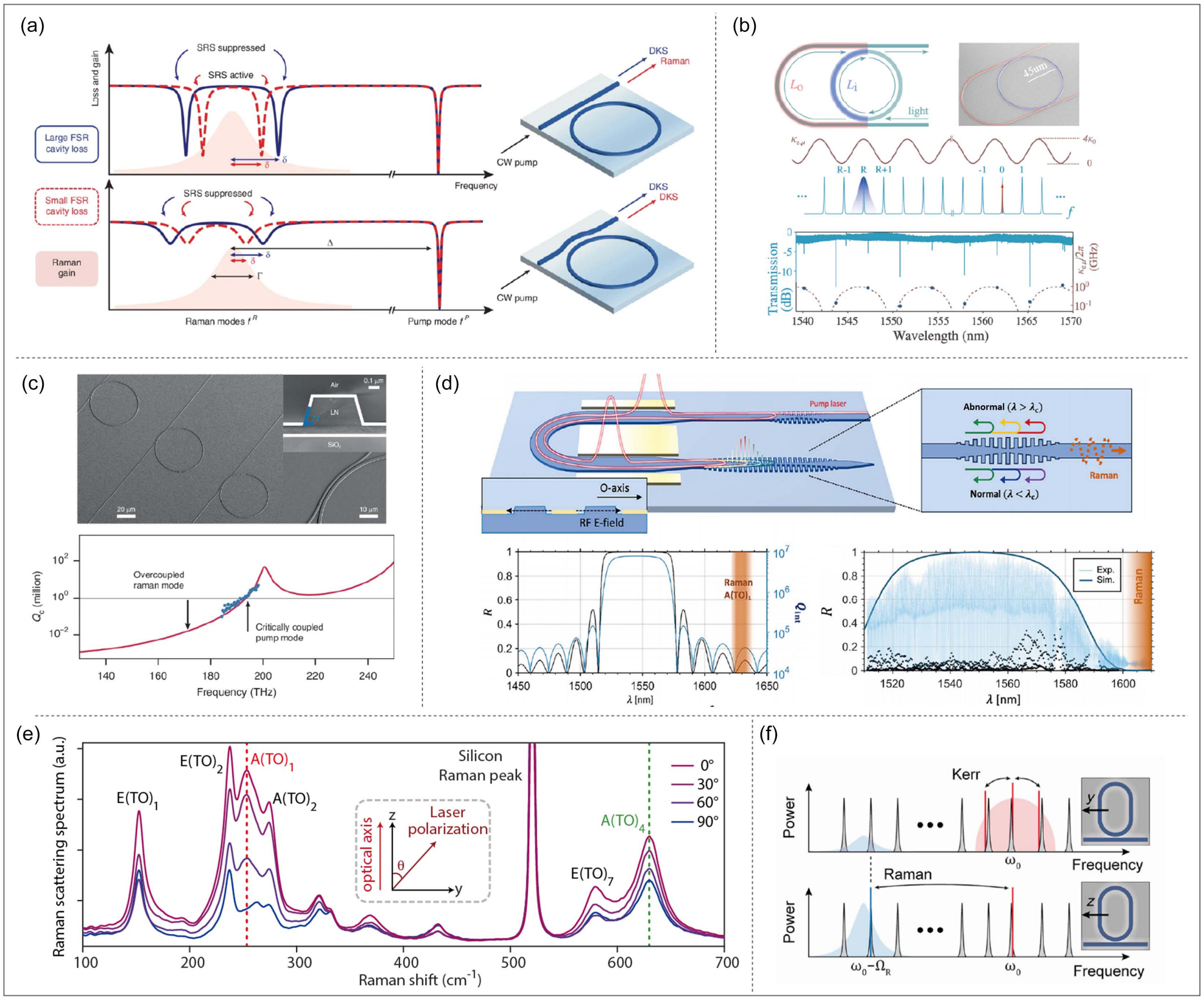

**Fig. 14** Raman-mitigation strategies for Kerr microcomb generation in TFLN. (a) Two generic routes for suppressing stimulated Raman scattering (SRS) are illustrated: FSR control, which places the Raman gain maximum between adjacent cavity modes, and dissipation engineering, which introduces excess loss for Raman-shifted modes while maintaining low loss for the pump and comb-forming modes[53]. (b) Dissipation engineering implemented with a self-interference (U-ring) coupler, where interference-modulated external coupling can selectively overcouple the Stokes mode and raise the Raman-lasing threshold[106]. (c) Dissipation engineering in pulley-coupled TFLN microrings, where wavelength-dependent phase matching in the extended coupling region creates a high-loss band overlapping with the Raman-shifted wavelengths[53]. (d) Dissipation engineering via Bragg grating[137]. (e) Polarization-dependent Raman spectra of X-cut TFLN, showing that the Raman response is substantially weaker when the optical polarization is perpendicular to the Z-axis[41]. (f) Polarization engineering in X-cut TFLN racetrack resonators: orienting the straight sections along the Z-axis reduces the round-trip-averaged Raman gain and thereby favors Kerr-soliton microcomb generation[54]. (b) Adapted with permission from Ref. [106], © 2020 American Physical Society. (a) Adapted from Ref. [53] under CC BY 4.0. (e) Adapted from Ref. [41] under CC BY 4.0. (c) Adapted with permission from Ref. [137], © 2026 The Author(s), published by Springer Nature (CC BY-NC-ND 4.0). (f) Adapted with permission from Ref. [54], © 2026 American Association for the Advancement of Science.

### 4.5 Kerr Microcombs in TFLN Platform

The early phase of research in this field established that LN can support coherent microcombs despite its normal material dispersion and strong Raman response. Early Z-cut demonstrations showed that these obstacles can be overcome through sufficiently strong geometric dispersion engineering together with careful control of nonlinear dynamics. He *et al.* (2019) observed self-starting bichromatic soliton microcombs, showing that coherent soliton states can be reached in an integrated LN platform under CW pumping[105]. Later, Gong *et al.* (2020) reported near-octave soliton microcombs[107], followed by octave-spanning operation in dispersion- and dissipation-engineered LN microresonators[53,109,110]. Importantly, coherent LN microcombs in Z-cut devices are not limited to the conventional bright-soliton branch: Lv *et al.* (2025) showed that normal-dispersion LN resonators can support dark-pulse microcombs with substantially higher conversion efficiency[127]. Recently, turnkey operation has been demonstrated in TFLN, bringing electrically pumped integrated microcombs within reach on this platform[141]. Collectively, these results established Z-cut TFLN as the baseline platform on which the core Kerr-microcomb physics commonly achieved in other platforms was verified. Future TFLN turnkey operation may also benefit from nonlinear self-injection-locking strategies developed on other microcomb platforms, provided that TFLN-specific Raman, photorefractive, and thermal dynamics are properly managed[142–144].

A second line of progress emphasized tunability and state control rather than spectral span alone. Because TFLN combines Kerr nonlinearity with strong EO tunability, it offers the dynamic control capability unavailable in most passive Kerr platforms. He *et al.* (2023) exploited this capability in high-speed tunable soliton microcombs, where rapid electrical control over cavity resonance and comb dynamics was demonstrated directly on the LN microresonator[108]. These results are important not only at the device level, but also conceptually: they show that Kerr-microcomb physics in TFLN is not merely repeating those demonstrated by other materials, but can be intrinsically intertwined with the platform's EO functionality and enable novel dynamics and phenomena.

More recently, attention has shifted toward X-cut TFLN. This transition matters for future fully integrated comb-based photonic circuits, as X-cut geometries are naturally compatible with high-performance EO devices and a broader class of $\chi^{(2)}$-based components, such as frequency converters. For a long time, however, Kerr microcombs remained absent on the X-cut platform because the stronger and more complex Raman interaction severely constrained access to coherent Kerr states[50]. Recent work has shown that this obstacle can be overcome when polarization and waveguide design are used to suppress Raman gain while retaining adequate Kerr nonlinearity[41,54]. Nie *et al.* (2025) demonstrated soliton microcombs in X-cut LN microresonators[41], followed by Song *et al.* (2025) on stable gigahertz- and mmWave-repetition-rate X-cut soliton microcombs[42], and Song *et al.* (2026) on high-efficiency broadband coherent comb generation in integrated X-cut LN resonators[54]. These experiments establish that X-cut TFLN can support coherent Kerr-comb dynamics while preserving compatibility with strong EO control and $\chi^{(2)}$-based frequency-conversion functionality—an alignment that is likely to be essential for future LN comb systems where coherent combs can be generated, tuned, stabilized, and spectrally converted on the same chip.

Taken together, these demonstrations outline the development of TFLN Kerr microcombs: Z-cut devices established bright-soliton operation with octave-spanning spectra; normal-dispersion devices demonstrated high-efficiency dark-pulse states; electrically tunable devices enabled active control of coherent LN comb states; and X-cut devices opened opportunities to combine coherent Kerr-comb generation with the distinctive EO and $\chi^{(2)}$ capabilities of TFLN. Translating these advances into practical TFLN soliton sources requires reproducible initiation and sustained coherent operation, supported by coordinated resonator design, management of material responses, and control of operating conditions. Reproducibility should be evaluated through repeated initialization under a defined control procedure and across nominally identical devices. Device-to-device variations in the required pump power, accessible detuning range, and initialization success rate, together with the duration and stability of sustained operation, provide practical measures for connecting fabrication tolerances to source performance.

Looking forward, developing suitable upper-cladding processes remains an important challenge for the practical integration of TFLN Kerr microcombs. An appropriately designed upper cladding can facilitate the integration of microheaters and EO electrodes, support reliable packaging, and protect the device surface from moisture, dust, and other contaminants, thereby supporting long-term operation. Depending on the cladding material and fabrication process, however, it may reduce the resonator $Q$, enhance photorefractive effects, and complicate anomalous-dispersion engineering. Joint optimization of low-loss deep-etching[132] and upper-cladding processes[139] therefore offers a route toward robust, packaging-compatible TFLN Kerr microcombs. Packaging compatibility further requires verifying that optical coupling, electrical access, and thermal management preserve soliton access and stability after assembly. At the system level, the required pump power and tuning feedback procedures should be evaluated together with the stability of the repetition rate and comb frequencies over the intended operating period.

More broadly, the next stage of the field is likely to be defined by three converging directions. The first is the realization of high-efficiency, broadband microcombs using coupled-resonator configurations. Helgason *et al.* (2023) demonstrated this high-efficiency coupled-resonator strategy on the SiN platform, and the same design logic is directly applicable to TFLN[124,125]. Combined with appropriate dispersion engineering, such efficient pumping schemes could enable octave-spanning microcombs at moderate power budgets compatible with integrated-laser pumping, bringing self-referenced operation closer to practical realization. The second is robust, high-speed state control built around the native EO tunability of TFLN. Fast EO actuation can greatly expand the bandwidth of feedback and locking loops for detuning control, soliton capture, and repetition-rate stabilization, which is directly relevant to optical atomic clocks and ultralow-noise microwave synthesis. The third is tighter functional integration between Kerr-microcomb generation and the broader LN capabilities already available on the same platform, including EO modulation, nonlinear frequency conversion, and microwave–photonic interfacing. In this sense, TFLN Kerr microcombs are evolving from proof-of-principle soliton devices into multifunctional integrated nonlinear systems. It is precisely this shift—from isolated comb generation to co-designed generation, control, and transduction—that makes Kerr-microcomb physics on TFLN both scientifically distinctive and technologically important.

## 5 Hybrid Nonlinearities for Comb Generation

Apart from the two most popular comb generation schemes discussed above, namely, EO combs and Kerr combs, recent years have also seen comb-generation mechanisms involving the interplay between more than one physical process, including EO modulation, second- and third-order nonlinearities, and laser dynamics, which we review in this section. The rich nonlinearities of LN make it particularly suitable for such hybrid schemes, enabling novel comb physics and more versatile device functionalities than single-mechanism approaches.

### 5.1 Active Mode-Locked Lasers

A mode-locked laser (MLL) produces an OFC by phase-locking the longitudinal modes of a laser cavity. In an MLL, a gain medium amplifies a set of dissipative cavity modes over a finite bandwidth, while nonlinear effects or actively applied EO modulation enforce a fixed phase relation among them. The coherent superposition of these modes forms a periodic train of ultrashort pulses and, equivalently, an evenly spaced and phase-locked OFC [Fig. 15(a)].

Although free-space and fiber-based systems have defined the state of the art of MLLs, their size, complexity, and power consumption hinder widespread deployment. Integrated approaches promise compact, energy-efficient, and scalable ultrafast sources that can be co-packaged with photonic and electronic circuits. TFLN is emerging as a strong integrated platform for MLLs, especially those locked by active EO modulation. The EO response in the TFLN platform in principle supports modulation bandwidths beyond 100 GHz, ideally suited to enforcing phase locking at gigahertz-to-terahertz repetition rates.

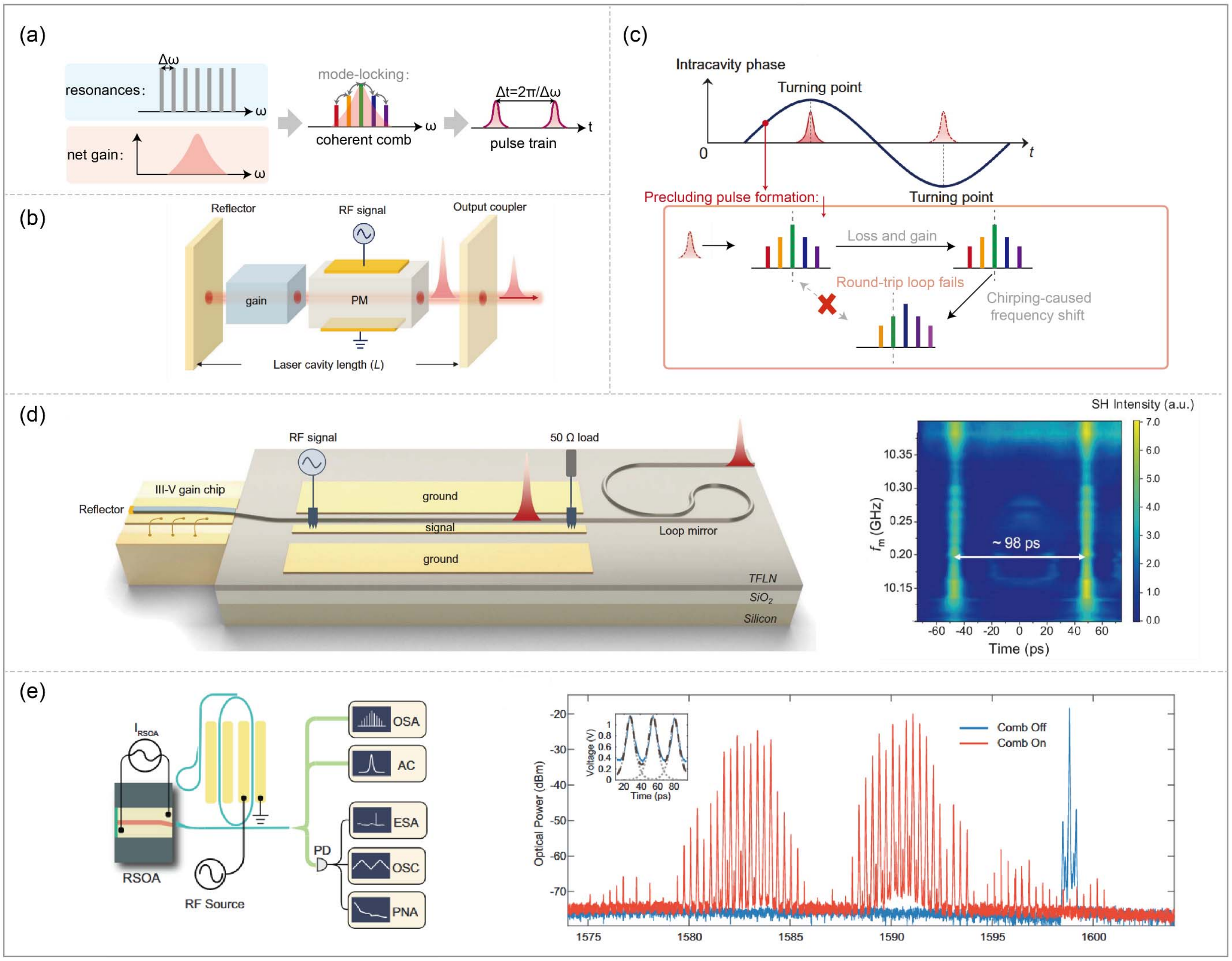

**Fig. 15** Active mode-locked lasers (MLLs). (a) Basic working principle of MLLs. (b) Typical configuration of an MLL with active EO modulation. RF: radio frequency; PM: phase modulator. (c) In this configuration, a pulse will only emerge at one of the turning points of the sinusoidal modulation. (d) The first TFLN-based MLL using a hybrid integration scheme that combined III–V gain with a TFLN cavity containing an EO phase modulator[145]. (e) Electrically powered integrated comb laser[146]. (d) Adapted with permission from Ref. [145], © 2023 American Association for the Advancement of Science. (e) Adapted from Ref. [146] under CC BY 4.0.

Figure 15(b) provides a canonical picture of an MLL with phase-modulation-based mode-locking. A sinusoidal drive applied to the PM imposes a periodic phase modulation on the intracavity field, forcing stable pulses to form only at one of the turning points of the modulation—the maxima or minima of the sinusoidal waveform—where the instantaneous phase variation vanishes. A simple intuitive picture is as follows: if a pulse were to appear at a nonzero modulation slope, it would experience a chirp that produces a frequency shift for all comb lines, and this asymmetric effect precludes a stationary comb solution [Fig. 15(c)]. In addition, a finite cavity dispersion is required to compensate for the quadratic chirp at one of the turning points, so that only one single stable pulse is sustained within the cavity.

MLLs on TFLN typically incorporate a hybrid integration architecture [Fig. 15(d)], where a III–V reflective semiconductor optical amplifier (RSOA) chip provides optical gain and optical reflection on one end of the hybrid cavity. The RSOA chip is butt-coupled to a TFLN chip integrated with an on-chip reflector to close the loop of the cavity, along with other functional devices like EO modulators and filters. Based on this configuration, the first TFLN-based MLL was demonstrated by Guo *et al.* (2023) using a TFLN chip containing an EO phase modulator[145]. They achieved stable active mode-locking, generating ~4.8 ps pulses with energies above 2.6 pJ and peak powers exceeding 0.5 W. This milestone proved that TFLN's ultrafast EO response can be directly harnessed for compact, electrically tunable ultrafast lasers [Fig. 15(d)].

Building on this foundation, Ling *et al.* (2024) presented an electrically powered integrated comb–laser that combines semiconductor gain and Kerr dynamics with TFLN microresonators[146]. They demonstrated turnkey operation and spectral flexibility in a cavity-enhanced MLL comb. Within the same architecture, the comb beat note can be fed back to drive the system, enabling self-sustained operation without an external driving source. This design delivers flexibility and self-sustainability that are difficult to achieve in conventional on-chip MLLs, underscoring TFLN's potential to support electrically reconfigurable pathways toward integrated ultrafast sources [Fig. 15(e)].

The main challenges for these hybrid integration schemes include relatively large coupling loss, thermal mismatch, and fabrication complexity. To address these issues, researchers have explored monolithic rare-earth-doped TFLN, where gain is incorporated directly into the host material. This approach reduces interfaces, improves stability, and simplifies fabrication. For example, Huang *et al.* (2025) reported a high-power Yb-doped Fabry–Pérot microcavity laser with >30 dB side-mode suppression[147].

### 5.2 $\chi^{(2)}$ Optical Parametric Oscillators

Optical parametric oscillation (OPO) based on second-order nonlinearity provides another comb-generation route. In an OPO cavity, a strong pump drives parametric down-conversion; coherent signal and idler fields begin to self-oscillate once the parametric gain exceeds the round-trip cavity loss. When multiple signal–idler mode pairs oscillate and become mutually phase-locked, the resonator produces an OPO frequency comb. Compared with Kerr or EO combs, OPO combs use $\chi^{(2)}$ gain rather than $\chi^{(3)}$ FWM or direct phase modulation, and often provide access to broader spectral ranges, including the mid-infrared.

The TFLN platform, with its large $\chi^{(2)}$ nonlinearity and wide transparency window, provides a highly suitable platform for implementing such combs. Most notably, LN is a ferroelectric crystal whose spontaneous electric polarization can be reversed by applying an external electric field stronger than its coercive field. This enables periodically poled lithium niobate (PPLN) [Fig. 16(b)], in which the sign of the second-order nonlinear coefficient $d_{33}$ is modulated with a controllable period $\Lambda$. A $\chi^{(2)}$ nonlinear process involves three optical fields whose frequencies satisfy $\omega_1 = \omega_2 + \omega_3$. Maximum nonlinear output occurs under the phase-matching condition, where the wavevectors $\beta$ (or momenta) of the three waves should satisfy $\beta_1 = \beta_2 + \beta_3$. Due to material and geometric dispersions, the wavevector mismatch $\Delta\beta = \beta_1 - \beta_2 - \beta_3$ is usually non-zero in a TFLN waveguide. The periodic domain inversion in a PPLN waveguide induces an effective $\chi^{(2)}$ grating that provides an additional nonlinear wavevector, $\frac{2n\pi}{\Lambda}$, which compensates for the intrinsic phase mismatch, leading to quasi-phase matching (QPM) when $\Delta\beta + \frac{2n\pi}{\Lambda} = 0$ [Fig. 16(a)]. Such PPLN techniques have been widely used, historically in ion-diffused LN waveguides and more recently in TFLN waveguides, to enhance $\chi^{(2)}$ nonlinear processes, including second-harmonic generation (SHG), sum-frequency generation (SFG), and difference-frequency generation (DFG).

To generate OPO-based frequency combs, PPLN is typically embedded in an optical cavity to provide efficient $\chi^{(2)}$ parametric gain through a parametric down-conversion process. In this process, one pump photon is down-converted into a signal and an idler photon [Fig. 16(a)], subject to energy conservation $\omega_p = \omega_s + \omega_i$ and momentum conservation $\beta_p = \beta_s + \beta_i$; $\omega_{p,s,i}, \beta_{p,s,i}$ denote the frequencies and wavevectors of the pump, signal, and idler fields. The PPLN is designed such that the intrinsic wavevector mismatch $\Delta\beta = \beta_p - \beta_s - \beta_i$ can be compensated for by $\Delta\beta + \frac{2n\pi}{\Lambda} = 0$. By engineering the poling period $\Lambda$, one can selectively enhance parametric gain and continuously tune the signal and idler wavelengths over a broad spectral range. This capability makes PPLN-based OPOs highly versatile and efficient sources for frequency-comb generation. To evolve from OPO operation to OPO-based comb formation, the narrowband gain at the signal and idler must be extended into broadband comb generation. This process relies on cascaded $\chi^{(2)}$ interactions[152], in which multiple $\omega_s, \omega_i$ photons undergo cascaded SHG, SFG, and DFG processes to produce new frequency components.

Broadband comb formation usually requires pump powers well above the basic oscillation threshold. This can be achieved either by raising the CW pump power or by synchronously pumping the cavity with a high-peak-power pulse train, where the comb spacing is determined by the pump repetition rate[152]. Dispersion engineering, including control of GVD and group-velocity mismatch (GVM), is also essential for OPO combs[148], because it governs phase matching and enables phase-synchronized coupling among multiple frequency components.

Roy *et al.* (2023) reported one of the early demonstrations of OPO combs in TFLN[148] [Fig. 16(c)]. Using synchronous pumping, they generated signal and idler combs tunable across 1.5–3.3 μm, while on-chip nonlinear conversion extended the spectrum into the visible near 620 nm. The system operated at a ~19 GHz repetition rate, with a femtojoule-level per-pulse threshold. Coherence was also directly verified. This work emphasized the ability of a single TFLN device to host multiple nonlinear processes—OPO, SHG, and SFG—simultaneously, creating a versatile platform for broadband comb generation.

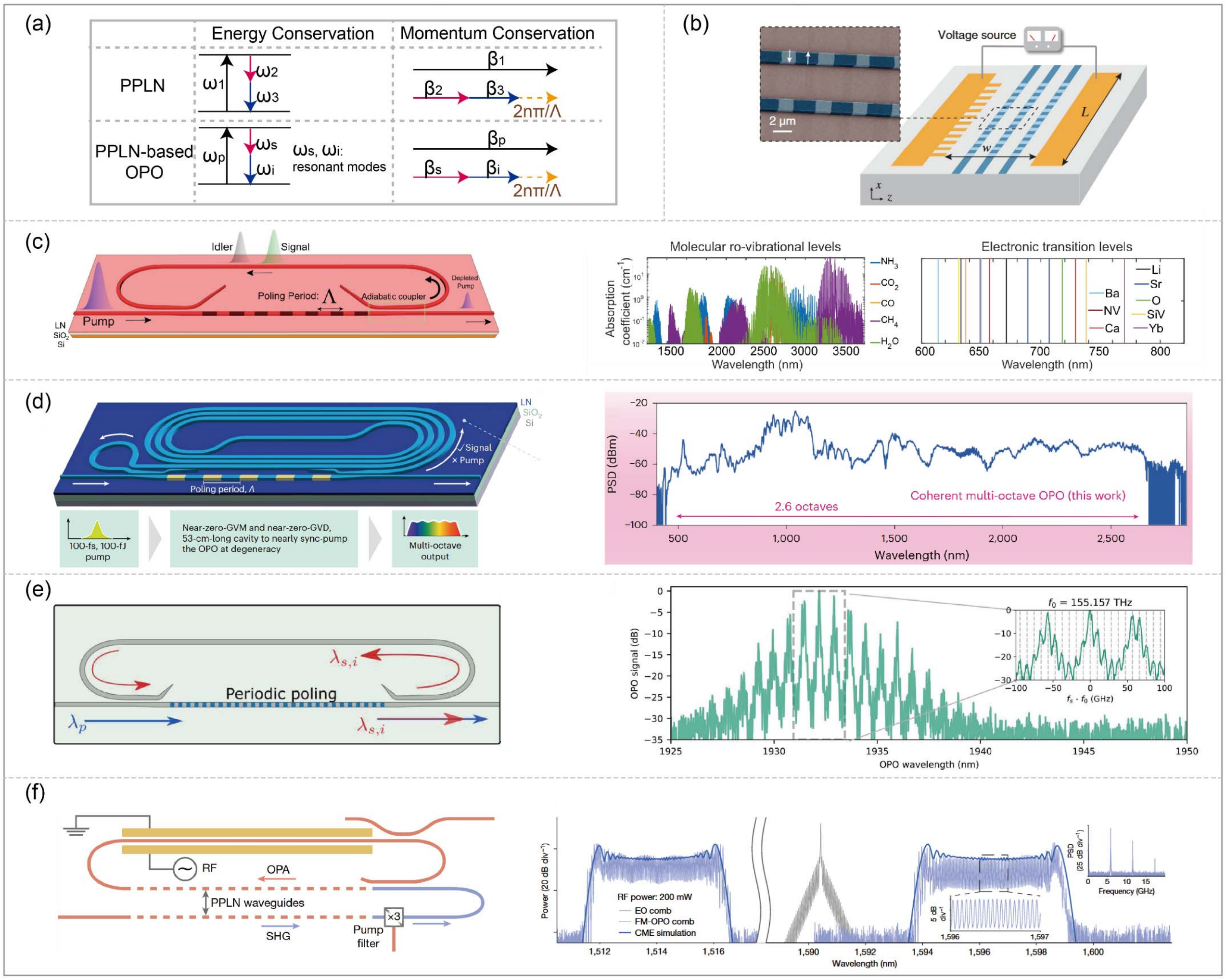

**Fig. 16** Optical parametric oscillation (OPO) combs. (a) Energy- and momentum-conservation conditions for PPLN-based $\chi^{(2)}$ process and PPLN-based OPOs. PPLN: periodically poled lithium niobate. (b) Schematic of the periodic poling process of a TFLN waveguide. (c) The first demonstrations of an OPO comb in TFLN using pure PPLN-based $\chi^{(2)}$ process[148]. (d) PPLN-based combs exceeding 2.6 octaves[149]. (e) OPO combs under quasi-CW pumping[150]. (f) Frequency-modulated OPO (FM-OPO) for flat-top comb generation[151]. (c) Adapted from Ref. [148] under CC BY 4.0. (d), (f) Adapted with permission from Refs. [149,151], © 2024/2025 Springer Nature Limited.

Interestingly, at higher powers, synchronously pumped TFLN OPOs do not simply broaden the spectrum. Sekine *et al.* (2025) showed that the coherence state changes with pump energy[149] [Fig. 16(d)]: near threshold the device enters a coherent OPO regime; at higher pump energy the oscillation becomes incoherent; and with further pumping it recovers into a coherent state. In this re-coherent regime, the source produced multi-octave spectra exceeding 2.6 octaves, establishing a low-energy route to ultrabroadband $\chi^{(2)}$ comb formation. The reported energy scales were very low, with oscillation beginning at ~18 fJ and re-coherence appearing at ~121 fJ, enabled by a near-zero group-velocity dispersion (GVD) and a 53-cm-long near-synchronously pumped cavity.

Ledezma *et al.* (2023) also studied OPO combs under quasi-CW pumping with ~100 ns pulses[150] [Fig. 16(e)]. Their device achieved 1.53–3.25 μm tunability, with ~15% efficiency and ~75% pump depletion. Above the OPO threshold, the output evolved from single-mode to multimode operation and finally to clustered states arising from dispersion-induced FSR mismatches. These clusters were identified as precursors for stabilized coherent combs through dispersion engineering.

Stokowski *et al.* (2024) further demonstrated a frequency-modulated OPO (FM-OPO) based on an EO-$\chi^{(2)}$ hybrid architecture. As illustrated in Fig. 16(f), EO modulation first generates microwave-spaced sidebands around the circulating field, while the PPLN section supplies quasi-phase-matched $\chi^{(2)}$ wavelength conversion and hence the intracavity parametric gain[151]. The authors obtained a nearly flat-top spectrum spanning approximately 1000 modes (~6 THz) and an internal pump-to-comb conversion efficiency exceeding 93%. Because the microwave frequency is intentionally detuned from the cavity FSR, the system avoids conventional mode locking; it should therefore be described conservatively as a highly efficient frequency-modulated OPO comb rather than as a pulsed, phase-locked mode-locked comb.

### 5.3 Hybrid Combs based on EO and Third-Order Nonlinearity

The fast Pockels modulation and strong Kerr/Raman effect in the TFLN platform also support comb generation and manipulation based on hybrid dynamics of EO and third-order optical nonlinearities. The $\chi^{(2)}$-based EO processes provide microwave-rate control over comb spacing, phase, and electronic stabilization, whereas $\chi^{(3)}$-based Kerr and Raman processes can be exploited to generate broad optical spans. Combining these effects addresses a central trade-off: Kerr combs can be very broad but often have large native spacings, while EO combs offer electronically accessible spacings but usually require additional broadening. In TFLN, co-locating EO modulation with third-order nonlinear broadening enables combs that are simultaneously broadband, electronically accessible, and reconfigurable.

Wang *et al.* (2019) reported the first monolithic TFLN circuit that simultaneously supports Kerr and EO responses, demonstrating modulation-instability comb formation together with on-chip EO filtering and modulation of individual comb lines using a single chip[50] [Fig. 17(a)]. Building on a similar monolithic integration architecture, Gong *et al.* (2022) introduced a dual-pump hybrid Kerr–EO architecture[153], in which a high-repetition-rate Kerr soliton (334 GHz spacing) was spectrally interleaved with a low-repetition-rate EO comb (~16 GHz). This architecture enabled direct electronic detection and stabilization of the soliton repetition rate [Fig. 17(b)].

More recently, Song *et al.* (2025) further explored the performance boundaries of EO–Kerr hybrid combs using two TFLN chips that are separately optimized for each function: a Z-cut, anomalous-dispersion resonator generating a broadband dissipative Kerr soliton, and an X-cut non-resonant EO phase modulator for densifying the comb spacing[154] [Fig. 17(c)]. The Kerr source had a native spacing of 410 GHz, while EO modulation produced a 29.308 GHz microwave-rate grid with 2589 lines across a 588 nm bandwidth. This architecture makes the division of labor clear: Kerr dynamics supply the optical span and high pump-to-comb conversion efficiency, whereas EO modulation supplies electronic access and stabilization. A complementary demonstration of high-efficiency broadband Kerr comb generation in normal-dispersion X-cut LN microresonators[54] suggests a path toward monolithically integrating efficient Kerr-comb generation and EO comb manipulation on the same X-cut TFLN platform.

Another hybrid route is synchronous pumping, which uses a flat-top non-resonant EO comb introduced in Sec. 3.1 to form high-peak-power femtosecond pulses before pumping the Kerr resonator. Cheng *et al.* (2024) used on-chip amplitude and phase modulation to synchronously pump a $\chi^{(3)}$ TFLN resonator[155] [Fig. 17(d)]. The pulsed pump increases the intracavity peak power and therefore lowers the average pump threshold for Kerr-comb generation. The device produced a 30.14 GHz comb spanning about 400 nm over ~1400 lines with >10% pump-to-comb conversion efficiency; the pulsed-pumping scheme also mitigated SRS without additional dispersion or coupling engineering.

SRS, rather than being suppressed, can also be exploited as a constructive nonlinear mechanism. Wan *et al.* (2025) introduced the concept of a self-locked Raman–EO microcomb[156] [Fig. 17(e)]. In REO combs, EO modulation of the pump establishes a primary frequency grid, while Raman lasing and FWM introduce Stokes and anti-Stokes channels that generate additional EO combs. With self-locking mechanisms, the distinct combs self-synchronize into a single broadband, phase-locked spectrum, spanning over 300 nm with ~1400 lines at a 26.03 GHz spacing, without the need for external feedback.

## 6 Applications and Perspectives

This section turns to the application landscape of TFLN combs, including optical communications, microwave photonics, optical computing, and optical metrology. The discussion emphasizes how the distinctive properties of different TFLN comb schemes can be matched to different system-level demands, highlighting the broader potential of TFLN for highly integrated photonic systems.

### 6.1 Optical Communications

Demand for high-capacity optical communication continues to surge, driven by artificial intelligence, 5G/6G, and cloud computing[157], with the most intense growth in data-center interconnects (DCIs)[158]. This demand has outpaced traditional scaling—adding fibers or increasing lane rates—prompting a shift toward multiplexing, especially wavelength-division multiplexing (WDM). Conventional WDM sources based on discrete laser arrays face significant footprint, cost, and thermal management issues in a DCI use case. Consequently, compact, low-cost, and power-efficient multi-wavelength sources are highly desirable. Integrated OFCs—with broad bandwidth, low noise, and precise line spacing—are promising WDM light sources. While ultrahigh-capacity WDM data transmissions have been demonstrated with Kerr combs[4,159] and MLLs[160,161], key components of the full systems, such as multiplexers and modulators, have remained discrete and bulky, owing to their limited compatibility with the comb platforms used.

The TFLN platform offers a distinct opportunity to monolithically integrate on-chip comb sources, especially EO combs, with wavelength-multiplexing components and ultrabroad-bandwidth EO modulators. This section reviews recent progress in these subsystems and outlines a roadmap toward fully integrated TFLN transceivers, highlighting their potential to improve bandwidth density and energy efficiency in optical networks.

#### 6.1.1 EO combs for WDM communications

A particularly suitable comb source for WDM optical communications is the flat-top non-resonant EO comb discussed in Sec. 3.1. This source is attractive because its repetition rate can be electronically tuned to align with ITU channel grids, its flat spectral envelope gives relatively balanced comb-line powers and more uniform channel SNRs, and all wavelengths remain mutually coherent because they originate from a single optical pump. These features distinguish it from arrays of independent lasers, where wavelength locking and mutual coherence are more difficult to maintain. Xu *et al.* (2024) demonstrated an 18-channel WDM transmission over a 53 km field-deployed link using 18 TFLN EO-comb lines with 50 GHz spacing and a 45 Gbaud polarization-multiplexed QAM modulation format, achieving a total bit rate of 6.48 Tb/s[162] [Fig. 18(a)]. Zhang *et al.* (2024) later demonstrated ultra-high-capacity transmission by spectrally broadening a TFLN EO comb with a highly nonlinear fiber[163] [Fig. 18(b)]. Using 132 comb lines replicated across six spatial copies in a 2 km multicore fiber, a total data rate of 0.879 Pb/s was achieved. An additional, unloaded comb copy was reserved for homodyne detection. This design reduces phase noise, simplifies DSP algorithms, and can lower DSP power consumption in practical systems. Zhang *et al.* (2023) also showed that the homodyne reference comb can be reconstructed at the receiver from three seed EO-comb lines[164]. This "comb clone" technique can be more difficult to implement with Kerr combs or MLLs because of their cavity-dependent resonance properties.

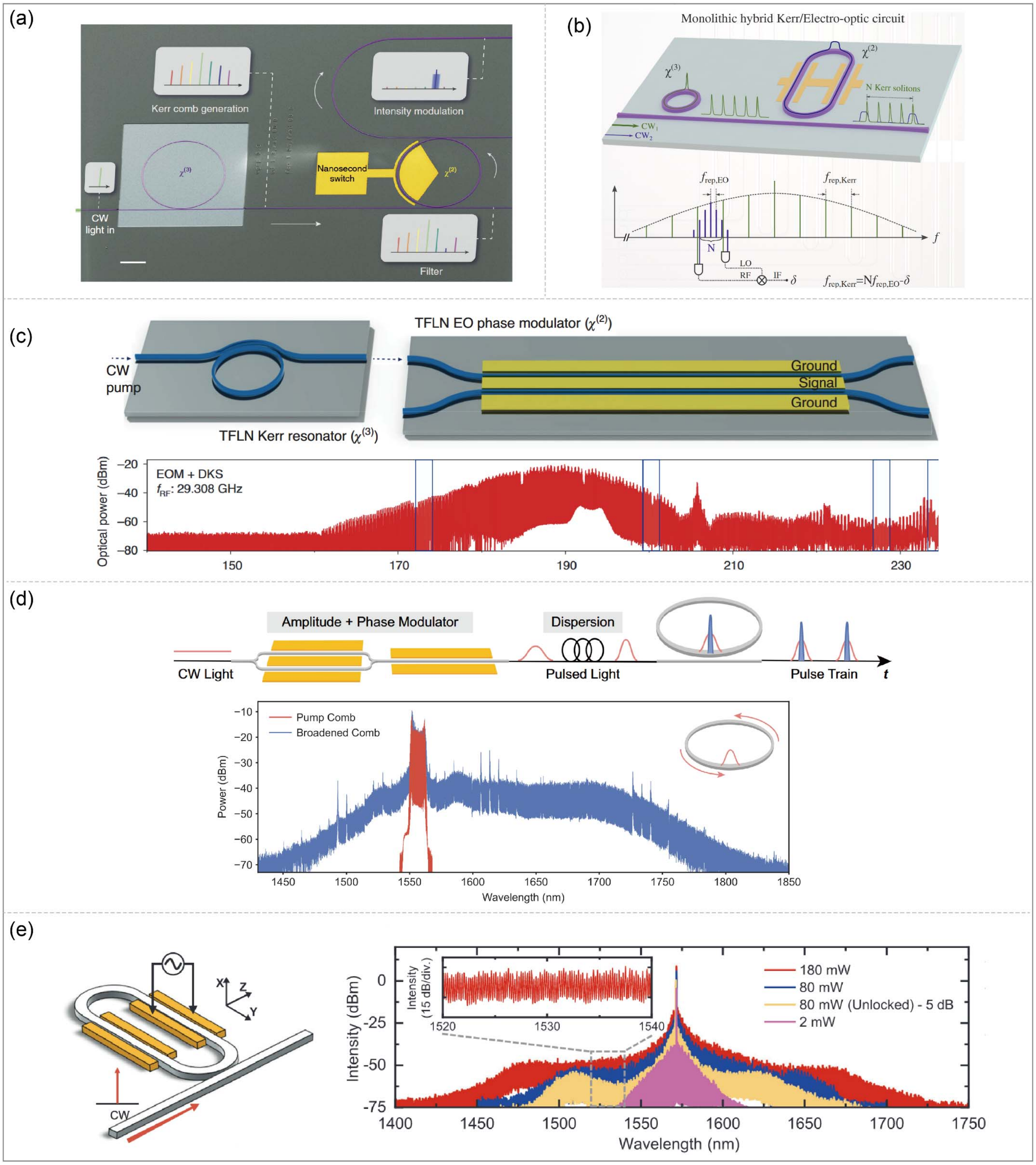

**Fig. 17** Hybrid combs based on EO and third-order nonlinearity. (a) Monolithic TFLN circuit for Kerr comb generation and integrated EO filtering/modulation[50]. (b) Hybrid Kerr–EO architecture enabling electronic access and stabilization of soliton repetition rates[153]. (c) Kerr–EO cascaded combs combining octave-level span with electronically accessible spacing for stabilization[154]. (d) Integrated synchronously pumped Kerr combs[155]. (e) Self-locked Raman–EO comb combining Raman gain, EO modulation, and Kerr effect[156]. (b) Adapted with permission from Ref. [153], © 2022 Optica Publishing Group. (a), (c), (d) Adapted from Refs. [50,154,155] under CC BY 4.0. (e) Adapted with permission from Ref. [156], © 2025 the Author(s), published by Springer Nature (CC BY-NC-ND 4.0).

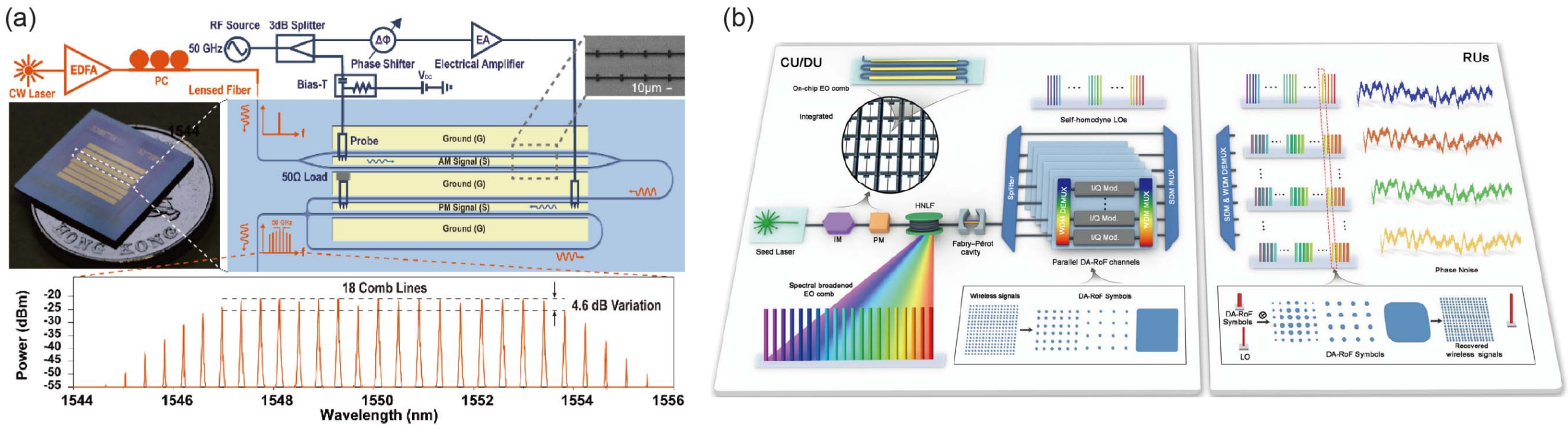

**Fig. 18** EO combs on TFLN for data transmission. (a) EO comb for high-capacity long-haul data transmission[162]. (b) Highly nonlinear fiber-broadened EO comb for high-capacity self-homodyne front-haul data transmission[163]. (a) Adapted with permission from Ref. [162], © 2024 Optica Publishing Group. (b) Adapted with permission from Ref. [163], © 2024 the Author(s), published by Springer Nature (CC BY-NC-ND 4.0).

### 6.1.2 TFLN device building blocks for WDM

Beyond the comb source, key components such as multiplexers based on different design principles and high-speed modulators for both intensity-modulation direct-detection (IMDD) and coherent formats have been demonstrated on TFLN, laying the groundwork for fully integrated WDM transmitters. Arrayed waveguide gratings (AWGs), the workhorse multiplexers in commercial systems, demultiplex and multiplex wavelengths by precisely controlling the optical path lengths of a waveguide array. As an anisotropic material platform, TFLN poses greater challenges for precise phase control than isotropic platforms such as silicon, InP, and SiN. In an X-cut TFLN platform commonly deployed for EO modulators and combs, the waveguide effective index for a TE mode varies with propagation angle relative to the Z-axis. As a result, Wang *et al.* (2024) demonstrated the first TFLN AWG on a Z-cut wafer[165] [Fig. 19(a)], where the Z-axis is perpendicular to the routing plane and the refractive index is invariant with routing angle.

Yi *et al.* (2024) later addressed the anisotropy challenge on X-cut TFLN by aligning the AWG symmetry axis 45° to the LN Z-axis, enabling designs compatible with conventional isotropic platforms [Fig. 19(b)][166]. Using this approach, four-channel AWGs with 400 GHz and 800 GHz spacing were demonstrated, achieving adjacent-channel crosstalk of −24.1 and −22.8 dB, respectively. Beyond AWGs, Zhu *et al.* (2025) also implemented cascaded Bragg gratings as TFLN multiplexers, with crosstalk below −25 dB[167] [Fig. 19(c)].

Wang *et al.* (2018), Xu *et al.* (2020), and Xu *et al.* (2022) demonstrated representative high-speed TFLN modulators, including amplitude modulators (AMs) for IM/DD[168] [Fig. 19(d)], in-phase/quadrature (IQ) modulators[169] for coherent transmission [Fig. 19(e)], and dual-polarization (DP) IQ modulators[170] [Fig. 19(f)], which double the data capacity via polarization multiplexing. As shown in Figs. 19(g) and 19(h), Feng *et al.* (2024) and Chen *et al.* (2025) further demonstrated SSB modulators based on an RAMZI interleaver[171] or optical delay lines[172], respectively. They offer higher spectral efficiency and better resistance to frequency fading in long-haul links than standard AMs.

These key components have collectively formed a high-performance and versatile TFLN device toolbox that can in principle be integrated onto a single chip. Liu *et al.* (2024) showed the integration of multiplexers and modulators using Bragg gratings as multiplexers, enabling 4 × 100 Gbps PAM4 transmission[173] [Fig. 19(i)]. Looking ahead, higher-level integration of these components, including flat-top EO combs, is expected to enable fully integrated comb-based transmitters and deliver compact, efficient, and low-cost optical communication solutions.

## 6.2 Microwave Photonics

### 6.2.1 Comb-based RF signal processing and measurement

Integrated microwave photonics is an emerging technology for RF signal processing, distribution, and generation based on photonic integrated circuits, leveraging their advantages such as very wide instantaneous bandwidth, immunity to electromagnetic interference, and the potential for reduced size, weight, and power[174–176]. The TFLN platform offers efficient, broadband EO conversion together with low-loss optical routing and signal processing, positioning it as a promising technology for integrated microwave photonic systems[177–179]. TFLN has been widely explored for microwave photonic systems, including low-noise signal generation[180,181], measurement[182–184], and RF signal processing[185]. These emerging systems also show strong potential for radar[178] and 5G-and-beyond wireless communications[186], offering compact, low-power, and high-bandwidth solutions for beamforming[187], frequency conversion[188], and signal processing[189].

The ability to monolithically integrate frequency combs, modulators, and signal processors on TFLN chips provides a particularly powerful platform for advanced RF manipulation with precise frequency-comb rulers. For example, by selectively filtering out one EO-comb line as a frequency-shifted local oscillator and beating it with an EO-modulated input signal, Xie *et al.* (2025) demonstrated RF frequency up- and down-conversion (or frequency mixing) with record-wide intermediate frequency and RF bandwidths and operating frequencies extending into the mmWave band on the TFLN platform[188]. As shown in Fig. 20(a), the mixer achieved fully reconfigurable frequency down-conversion over the 20–67 GHz frequency range with a 20 GHz IF bandwidth, and up-conversion to frequencies up to 110 GHz. In another recent demonstration, Tao *et al.* (2026) demonstrated an integrated TFLN real-time spectrum-sensing system in which a broadband EO modulator first transfers the unknown RF signal onto optical sidebands, while rapidly EO-tuned microring filters perform frequency-to-time mapping through the linear and fast Pockels response [Fig. 20(b)]. An on-chip EO comb serves as an absolute frequency ruler, enabling multiple microring-filter channels to be calibrated and stitched together for broadband spectrum reconstruction. The system achieves a 57.5 GHz effective spectral analysis bandwidth, a measurable RF frequency range up to 120 GHz, sub-110 ns latency, 100 ns temporal resolution, and 350 MHz frequency resolution[182].

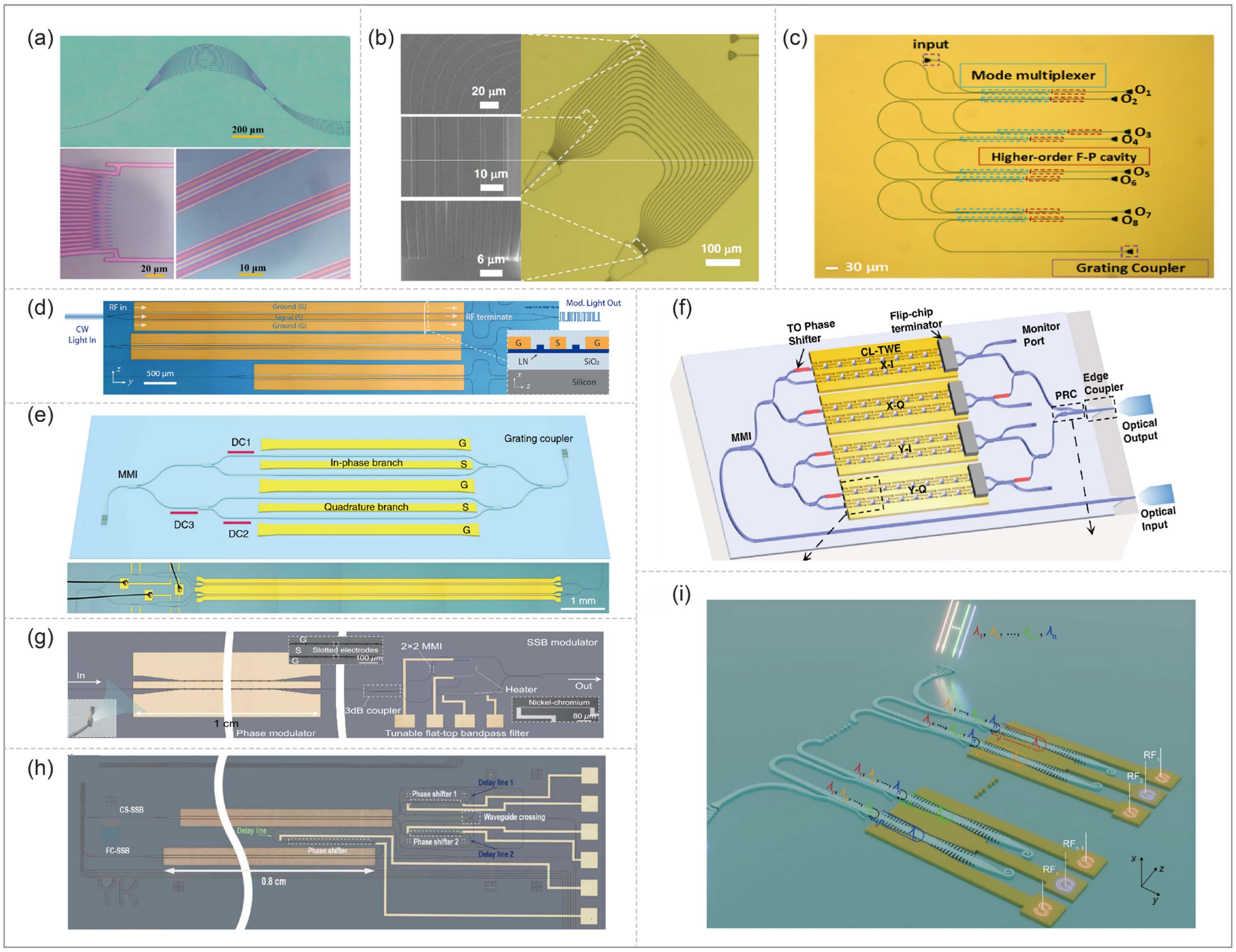

**Fig. 19** Key components for monolithically integrated EO-comb-based WDM system on TFLN. Arrayed waveguide gratings (AWGs) on (a) Z-cut[165] and (b) X-cut[166] TFLN. (c) Bragg-grating-based multiplexer[167]. (d) High-speed amplitude modulator[168]. (e) In-phase/quadrature (IQ) modulator[169]. (f) Dual-polarization (DP) IQ modulator[170]. (g) Single-sideband (SSB) modulator based on ring-assisted Mach–Zehnder interferometer (RAMZI) interleaver[171]. (h) Mono-drive SSB modulator based on optical delay lines[172]. (i) Dense wavelength-division multiplexing (DWDM) transmitter[173]. (b) Adapted with permission from Ref. [166], © 2024 Springer Nature. (c) Adapted with permission from Ref. [167], © 2025 American Chemical Society. (f), (h) Adapted with permission from Refs. [170,172], © 2022/2025 Optica Publishing Group. (g), (i) Adapted from Refs. [171,173] under CC BY 4.0. (e) Adapted from Ref. [169] under CC BY 4.0. (d) Adapted with permission from Ref. [168], © 2018 Springer Nature Limited. (a) Adapted with permission from Ref. [165], © 2024 the Author(s), published by Wiley (CC BY-NC 4.0).

Looking ahead, TFLN EO combs, with their tunable repetition rates, flat spectral profile, and compatibility with other functional elements, could enable a range of compact and high-performance microwave photonic signal processing and measurement systems with high resolution, broad bandwidth, and excellent reconfigurability. In the following, we outline two comb-based microwave photonic architectures that have been investigated and verified using bulk LN EO combs, to illustrate the working principles and discuss the potential advantages of on-chip implementations: RF channelization, where comb lines provide many parallel optical carriers for broadband reception[190–192], and bandwidth-scaling signal processing, where comb coherence, Vernier filtering, and multiheterodyne detection expand the effective resolution and processing bandwidth through microwave filtering[196], spectral analysis[197], temporal imaging[198], temporal reversal[199], and fractional Fourier transformation[200].

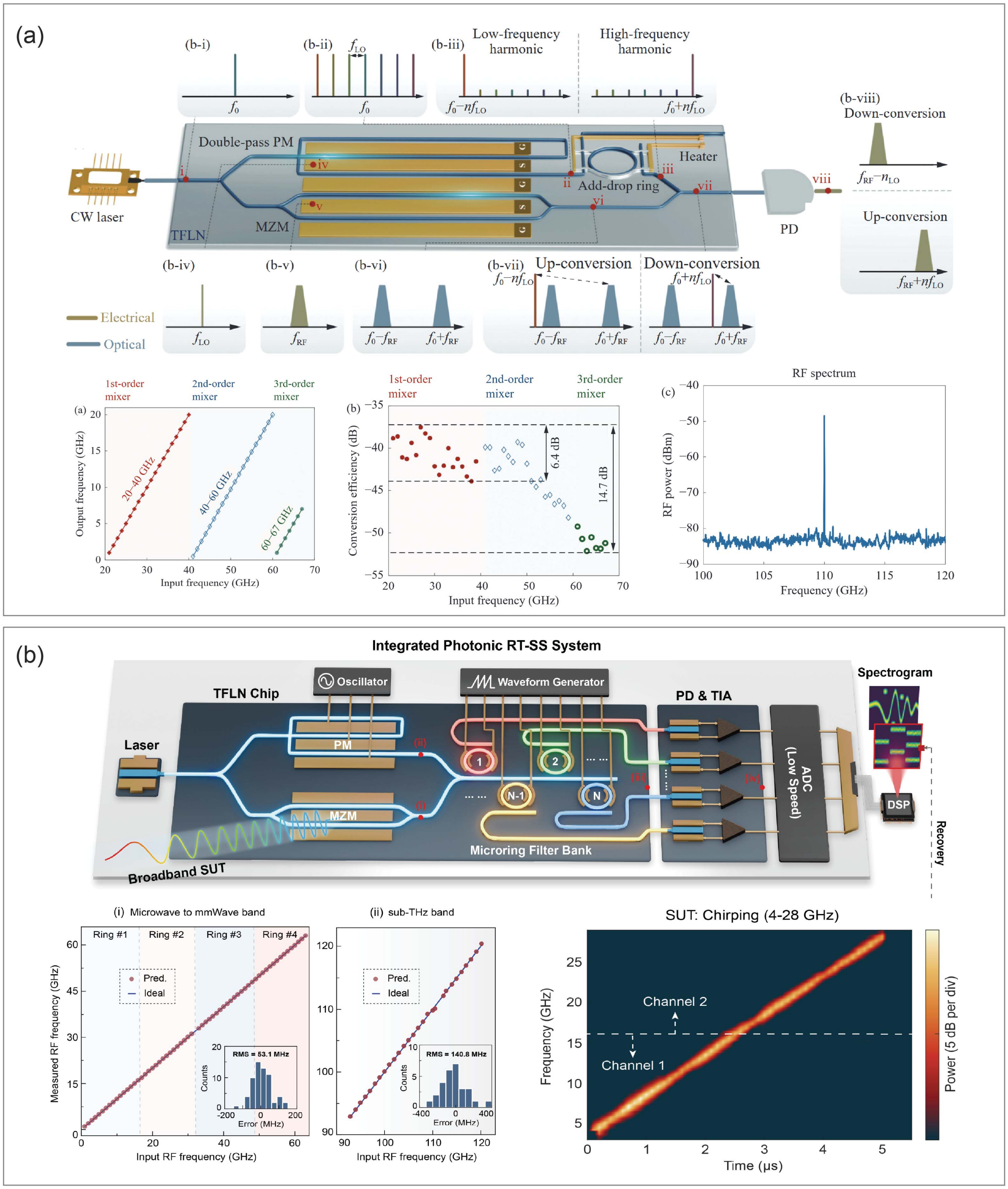

**Fig. 20** TFLN comb-based photonic mmWave frequency mixers (a)[188] and spectral sensing system (b)[182]. (a) Adapted with permission from Ref. [188], © 2025 IEEE. (b) Adapted with permission from Ref. [182], © 2026 the Author(s), published by Springer Nature (CC BY-NC-ND 4.0).

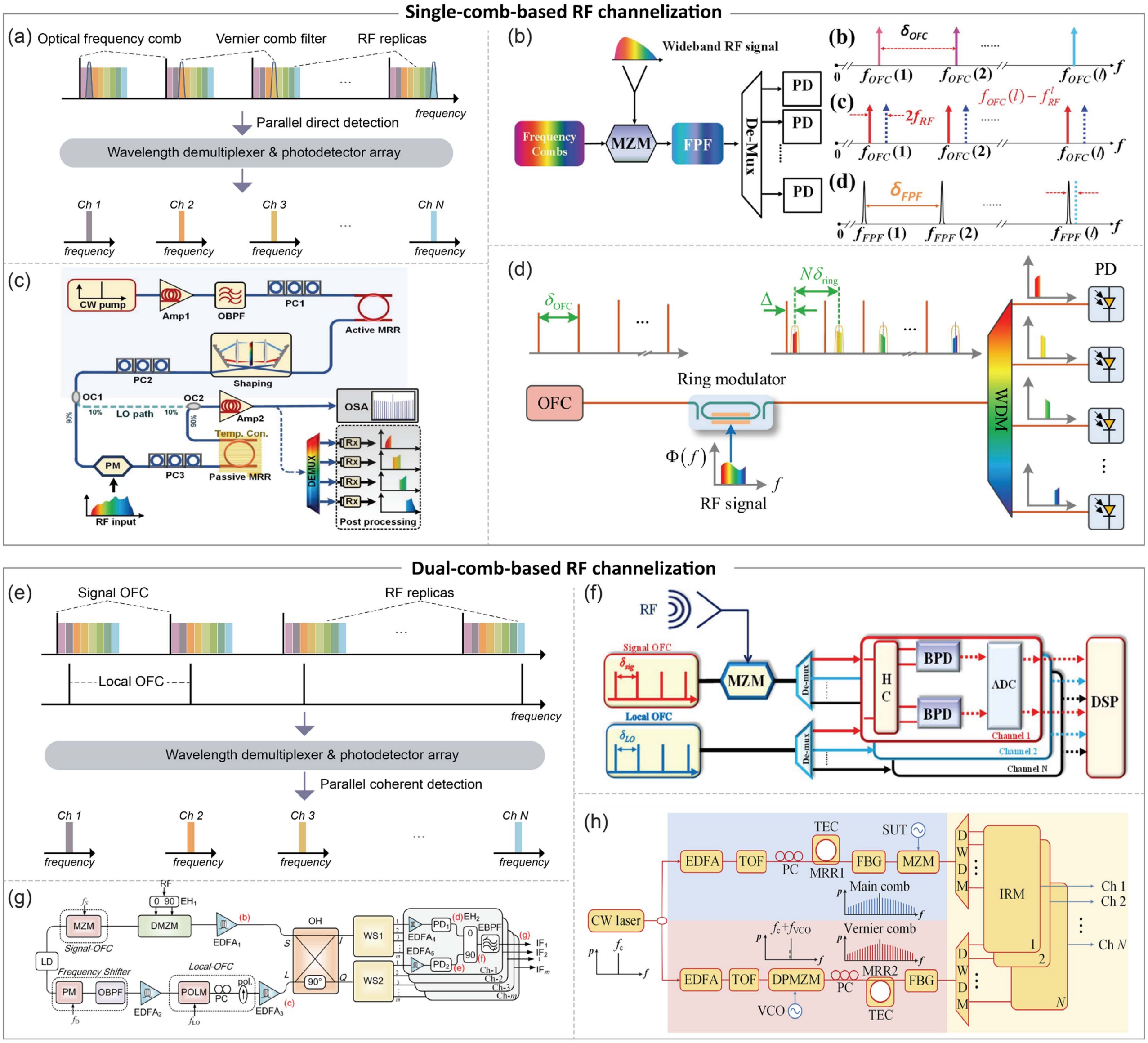

**Fig. 21** Comb-based RF channelization. (a) Basic principle of a single-comb channelizer. (b) RF channelization based on bulky commercial optical devices[190]. (c) Implementation using a Kerr comb source[191]. (d) Integrated implementation featuring a TFLN ring modulator[192]. (e) Principle of a dual-comb coherent channelizer. (f) Coherent channelizer based on dual bulk EO combs[193]. (g) Analog IQ demodulation module for coherent channelization[194]. (h) Coherent channelizer utilizing dual Kerr soliton combs[195]. (a), (b) Adapted with permission from Ref. [190], © 2012 IEEE. (c) Adapted with permission from Ref. [191], © 2018 IEEE. (d) Adapted with permission from Ref. [192], © 2024 IEEE. (e), (f) Adapted with permission from Ref. [193], © 2012 IEEE. (g) Adapted with permission from Ref. [194], © 2019 Optical Society of America. (h) Adapted from Ref. [195] under CC BY 4.0.

In photonic-assisted RF channelization, an OFC acts as a multiwavelength carrier, whereas a single EO modulator imprints the RF signal onto every comb line, creating parallel optical replicas [Fig. 21(a)]. A Vernier optical filter (a periodic filter with a slightly detuned FSR from the comb) picks up different spectral slices from each replica, mapping different spectral sections of the input RF signal to spectrally separated optical wavelengths. These channels are wavelength-demultiplexed and detected by a modest-bandwidth photodetector (PD) array, enabling broadband coverage with many low-speed receivers[190–192].

In practice, OFCs for channelization have been generated either with cascaded bulk EO modulators [Fig. 21(b)][190], which offer fine FSR tunability and simple control, or with integrated

Kerr microresonators [Fig. 21(c)][191], which provide compact footprint and larger FSR. More recently, Wan *et al.* (2024) showed that an on-chip ring-modulator architecture can merge the modulation and Vernier filtering functions in a single device, greatly reducing system complexity [Fig. 21(d)][192].

Simple channelizers based on a single comb can effectively separate frequency bands but lose the phase information under direct detection. For full-field recovery of both amplitude and phase, dual-comb architectures can be used: a signal comb carries the RF components, while a second comb with a slightly detuned FSR provides a bank of optical local oscillators for parallel coherent down-conversion [Fig. 21(e)][193–195,201]. Early demonstrations employed dual EO combs, IQ modulation, and IQ coherent receivers, enabling each channel to be down-converted to zero IF and allowing mirror-image suppression in the digital domain [Fig. 21(f)][193]. IQ modulation can also be implemented in the analog domain, where the receiver combines I/Q outputs from balanced PDs to suppress in-band interference and improve channel isolation [Fig. 21(g)][194,201]. Dual-soliton microcombs provide a compact alternative: their large soliton FSRs increase per-channel bandwidth and reduce footprint [Fig. 21(h)][195].

These architectures involve a fundamental trade-off among per-channel bandwidth, channel count, and receiver complexity. For a fixed optical span, increasing the comb FSR increases the RF bandwidth assigned to each channel and can simplify channel separation, but reduces the number of available carriers and may increase the bandwidth requirement of each receiver. A smaller FSR provides more parallel channels within the same optical span, but narrows the bandwidth per channel and places greater demands on spectral filtering, channel isolation, and calibration. Single-comb channelizers use direct detection and therefore offer simpler receivers, but discard phase information; dual-comb coherent architectures recover both amplitude and phase, at the cost of an additional comb, IQ components, and more demanding phase control.

On the other hand, bandwidth-scaling techniques have been widely explored in discrete optical systems for arbitrary microwave-photonic signal processing because they provide higher resolution and simultaneous control of amplitude and phase. The principle is shown in Fig. 22(a). Similar to channelized reception introduced above, an RF signal is first modulated onto a comb signal, creating many optical replicas. A Vernier filter maps initially closely spaced RF tones to widely separated wavelengths, enabling a bandwidth-magnified EO conversion. The magnified optical spectrum is then processed in the optical domain. Finally, a local-oscillator comb beats with the processed signal through multiheterodyne detection, which maps the processor response back to the RF domain with a compressed frequency axis.

Over the years, this bandwidth-scaling technology has been applied for a variety of microwave photonic processing functions. For example, reconfigurable microwave shapers enable high-resolution RF filters with fine amplitude control [Fig. 22(b)][196]. RF spectral analyzers use precise phase control and effective dispersion magnification to realize frequency-to-time mapping with higher spectral resolution [Fig. 22(c)][197]. Temporal imaging achieves coherent magnification of ultrafast waveforms by spectral compression while preserving phase information [Fig. 22(d)][198]. Temporal reversal can also be achieved by channelizing the RF signal into multiple sidebands with two slightly detuned combs and using multiheterodyne detection for parallel frequency down-conversion [Fig. 22(e)][199]. Together, these bandwidth-scaling methods, built on comb coherence, Vernier filtering, and multiheterodyne detection, significantly extend the resolution and capabilities of microwave photonic systems.

The comb-based channelization and bandwidth scaling schemes discussed above generally rely on a common set of building blocks—coherent multiwavelength sources, broadband EO modulators, programmable spectral filters, optical amplifiers, and high-speed PDs—which have so far mostly been implemented using discrete optical components. The key device functions are increasingly available on TFLN, including broadband and high-linearity modulators[168,202], efficient EO comb sources[63], fast tunable microring filters[182], integrated microwave photonic processors[177], heterogeneous uni-traveling-carrier PDs for high-power and efficient optical-to-electrical conversion[203,204], Er-doped waveguide amplifiers[205], and hybrid-integrated lasers[206]. Bringing these elements onto a common TFLN-based platform would reduce optical/RF interconnects, insertion loss, latency, and phase drift, while also improving RF bandwidth, modulation efficiency, tuning speed, RF output power, and scalable channel parallelism.

### 6.2.2 Optical frequency division

Optical frequency division (OFD) is an established method for synthesizing microwave signals with ultralow phase noise. Briefly, the frequency of an ultra-stable optical reference, typically $\sim 10^{14}$ Hz, is coherently divided down to the RF-frequency domain ($\sim 10^{10}$ Hz), through referencing to a frequency-stabilized optical frequency comb[209]. In the ideal limit, frequency division by $N$ reduces the phase-noise power by $N^2$ (equivalently, $20\log_{10} N$ in dB), allowing the synthesized microwave to inherit and significantly transcend the noise limit and stability of the optical references. These optical references most commonly consist of a cavity-stabilized laser locked to a high-finesse Fabry–Pérot resonator, and in the ultimate limit an optical-clock laser referenced to a narrow atomic transition (e.g., Sr, Yb, $Al^+$) [Fig. 23(a)]. Recent progress has pushed these references toward more compact implementations, including chip-scale high-$Q$ resonators that can serve as optical flywheels with remarkable spectral purity[208,210–214]. It should be noted that, although self-referenced (one-point) OFD has delivered the lowest microwave phase noise to date using discrete optical systems[215], extending the scheme to microresonator combs is hindered by the simultaneous requirements of octave comb bandwidth, nonlinear self-referencing interferometry, and microwave-level repetition rate, constraints that remain challenging under chip-scale power budgets.

As a result, a widely adopted alternative is EO frequency division, which circumvents the octave-bandwidth requirement and replaces self-referencing with a direct division of an optical frequency difference to the microwave domain. In this scheme, two optical references at $\nu_1$ and $\nu_2$ are each phase-modulated by the same microwave oscillator at frequency $f_m$, generating EO combs (phase-modulation sidebands) around both carriers. The division condition arises when a sideband of one reference appears in the vicinity of a sideband of the other reference; in its most common form, this yields an error signal:

$$\delta f = |\nu_2 - \nu_1 - N f_m|, \tag{40}$$

where the integer $N$ is the effective sideband order that participates in the overlap beat[207]. Feeding back to the microwave

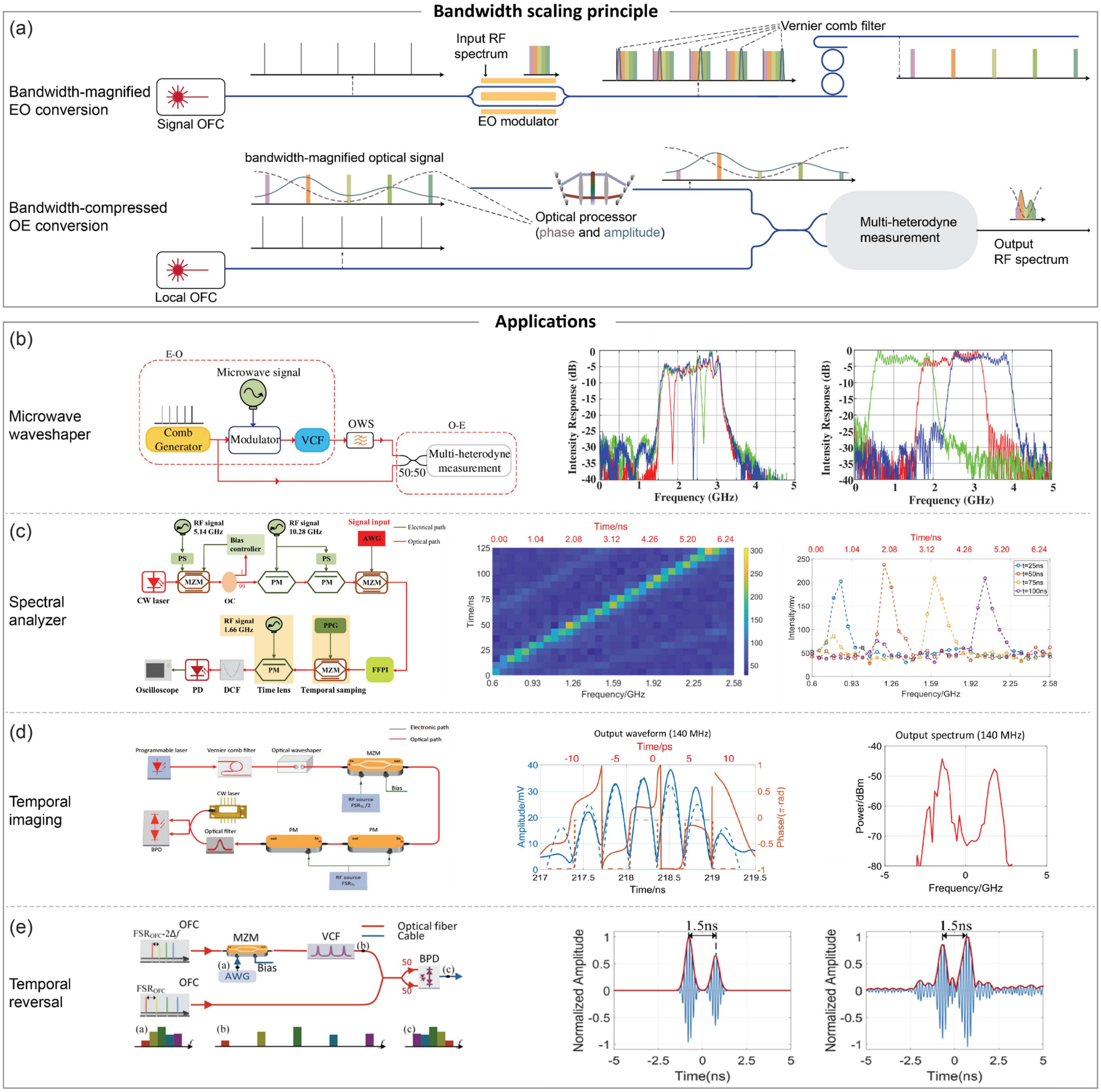

**Fig. 22** Bandwidth scaling and its applications. (a) Operating principle of the bandwidth scaling technique. (b) MHz-resolution microwave waveshaper with arbitrary frequency response[196]. (c) MHz-resolution RF spectral analyzer realized through bandwidth magnification[197]. (d) Full-field picosecond pulse measurement via temporal imaging[198]. (e) Temporal reversal achieved through multi-subband frequency conversion[199]. (c) Adapted with permission from Ref. [197], © 2021 IEEE. (a), (b), (e) Adapted with permission from Refs. [196,199], © 2018/2023 Optica Publishing Group. (d) Adapted with permission from Ref. [198], © 2021 Optica Publishing Group.

source to drive $\delta f \rightarrow 0$ enforces

$$f_m = \frac{\nu_2 - \nu_1}{N}, \tag{41}$$

so that the microwave output inherits the optical reference stability through coherent division by $N$.

This approach is particularly compatible with the TFLN platform, where efficient and high-speed EO phase modulation enables broadband comb generation and low-noise actuation in a compact form factor. The practical ceiling on $N$ is set by the highest usable sideband order, which depends on the modulation index and therefore on the modulator half-wave voltage $V_\pi$ and available microwave drive power, limitations that historically motivated cascaded bulk modulators or resonant enhancement. TFLN materially relaxes this constraint by providing a low RF $V_\pi$ in the ~1–2 V range at tens of gigahertz frequencies, enabling larger $N$ and reduced system complexity[213,214]. Recent chip-scale demonstrations have realized division factors of $N \sim 60$–90 with microwave output near 37 GHz, achieving phase noise of approximately $-129$ to $-133$ dBc/Hz at a 10 kHz offset[213,214]. These results also bring up a core system consideration for LN-based OFD: while optical-to-microwave coherence can be transferred with high fidelity, the overall system performance is often dictated by the microwave drive requirement and optical insertion loss, which together determine the end-to-end electrical and optical power budgets [Figs. 23(b)–23(e)].

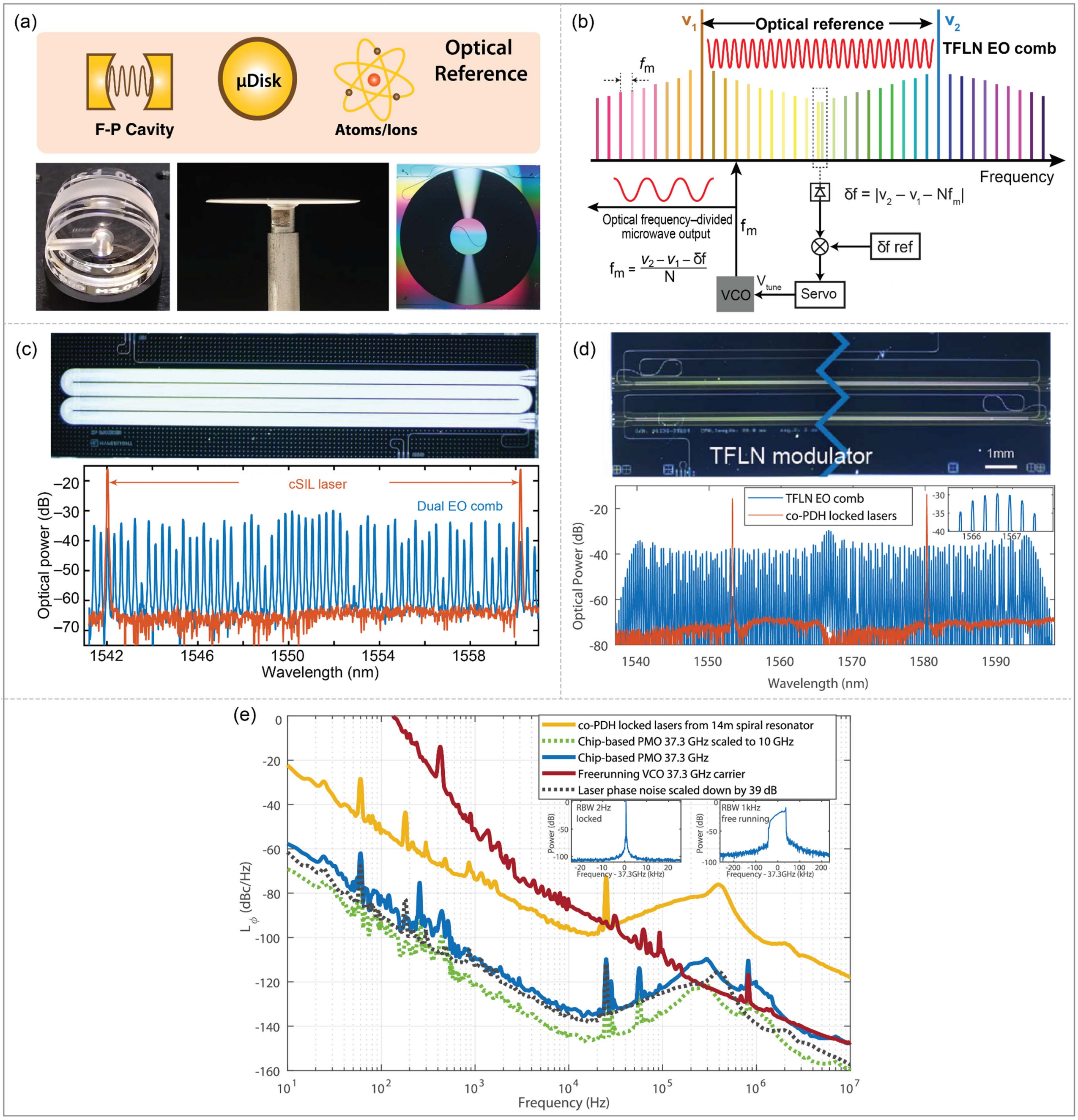

**Fig. 23** Optical frequency division. (a) Example optical references[207,208]. The bottom images show (from left to right) mini-FP cavity, $MgF_2$ whispering-gallery-mode resonator, and $Si_3N_4$ spiral micro-resonators. (b) Schematic of electro-optic frequency division. (c) Top: image of a single-pass TFLN phase modulator. Bottom: optical spectra of reference lasers and the dual EO comb. (d) Top: image of a cascaded dual-pass TFLN phase modulator. Bottom: optical spectra of reference lasers and the dual EO comb. (e) Microwave phase noise. (a) Adapted with permission from Ref. [207], © 2014 American Association for the Advancement of Science; Ref. [208] under CC BY 4.0.

Kerr microcombs provide a complementary route to OFD because their multi-terahertz bandwidth can in principle support large division ratios. However, OFD based on Kerr microcombs has not yet been experimentally demonstrated on TFLN, even though LN offers a valuable control primitive—fast electro-optic actuation—that could, in principle, enable high-speed repetition-rate feedback on chip[108]. A related LN-native concept is hybrid Kerr–EO operation (as discussed in Sec. 5.3), in which a Kerr microcomb supplies optical bandwidth while an EO phase modulator provides electronic accessibility and stabilization; recent demonstrations of such hybrid comb operation on TFLN therefore point toward a plausible pathway, but they have not yet closed the full OFD loop[154].

### 6.3 Optical Computing

Integrated frequency combs are also promising light sources for optical computing applications leveraging their massive frequency-multiplexing parallelism. A single on-chip broadband frequency comb can yield hundreds to thousands of mutually coherent lines[64,81,88,100], each functioning as an intrinsically parallel optical carrier. Such spectral multiplicity enables optical processors to increase computational throughput and density by two to three orders of magnitude within essentially the same physical footprint. Consequently, TFLN photonic computing has rapidly developed into a vibrant research direction[220–222], motivating a broad range of device concepts and system architectures that aim for full on-chip integration and scalable deployment.

Generally, comb-driven TFLN computing processors typically comprise two functional units: a comb source that drives wavelength-multiplexed processing, and a core processing network that can be partially or fully tunable via EO or thermal-optic elements. All these elements have nowadays reached high levels of maturity that support sophisticated system functionalities. In addition to frequency comb sources that we have detailed in previous sections, a variety of passive components with low loss have been demonstrated, including multimode interferences[223,224], high-$Q$ microresonators[87,225], low-crosstalk waveguide crossings[226,227], and multichannel AWGs[165,166,228]. These elements form the low-noise, low-crosstalk backbone required for large-scale comb routing and interferometric optical networks. On the active side, high-performance EO modulators—combining low half-wave voltage and ultra-broad bandwidth—are central to high-speed data encoding and dynamic weight loading across dense spectral channels. Finally, at the platform level, efforts toward higher integration density and foundry-scale chip fabrication are accelerating. Coupled with CMOS processors, FPGAs, on-chip lasers, and advanced packaging, these trends are pushing TFLN toward scalable and manufacturable optical-computing hardware.

Building on these advances, several TFLN frequency-comb-based optical computing architectures have been proposed. One representative architecture [Fig. 24(b)] employs wavelength-time interleaving to realize highly parallel convolution operations. Here, the convolution kernel is mapped onto comb lines through spectral shaping, while the input data—encoded as a temporal optical intensity sequence—modulates the comb using an MZM. Wavelength-dependent delays introduced by SMF or on-chip dispersive waveguides align different spectral components in time; summing their optical power yields the target convolution at essentially the speed of light. Xu *et al.* (2021) reported computational throughputs on the order of 11 TOP/s[7], highlighting the potency of wavelength-time interleaving for convolutional workloads. More recently, on-chip spectral shaping on TFLN has laid the groundwork for monolithically integrated implementations of this architecture.

A complementary route [Fig. 24(c)] employs WDM to construct optical tensor-computing cores and neural-network accelerators. In this architecture, broadband TFLN frequency combs are wavelength-demultiplexed on chip to provide multiple parallel inputs for the subsequent optical neural networks. Such demultiplexing devices can be realized by Fabry–Pérot-based filters (e.g., four channels with ~1.3 dB insertion loss per channel in Ref. [173]), or AWGs (e.g., 16 channels with ~2.3 dB loss and $\sim -24$ dB crosstalk in Ref. [166]). Vector-matrix-multiplication operations can then be performed on these spatially fanned-out signals using on-chip interferometric networks with partial or full EO programmability. Recent demonstration of a single-channel TFLN photonic tensor core has reached a computation speed of ~43.8 GOPS with a power consumption of only ~0.0576 pJ/OP[218], underscoring the platform's efficiency. Results based on wafer-scale modulator arrays further indicate that multi-channel MZM arrays can be realized with nearly linear channel-count scaling, highlighting the compatibility of WDM-ONN architectures with TFLN frequency combs and their suitability for large-scale tensor operations.

Taken together, these advances point toward a modular design paradigm in which a broadband comb source is co-integrated with a compact, programmable EO processing core on a single TFLN chip. Such modularity allows functional blocks to be flexibly selected and composed to meet diverse application requirements, while leveraging large-scale frequency parallelism to balance efficiency, footprint, and interconnect complexity. As device fabrication and packaging technologies continue to mature and standardize[30,99], TFLN comb-driven optical computing is positioned to play an important role in next-generation on-chip AI acceleration[221,222], massively parallel communication signal processing[177,229], and emerging quantum-information interfaces[230].

### 6.4 Optical Metrology

Since its invention, OFC has been the workhorse for precision metrology owing to its precise frequency and temporal features. A prominent example of OFC-based metrology is dual-comb spectroscopy, where two combs with slightly different repetition rates generate a multiheterodyne RF comb, allowing rapid measurement of both phase and amplitude responses, e.g., absorption spectra of gases, over a broad optical bandwidth using low-cost electronics [Fig. 25(a)][3,231–235]. Ideal comb sources for such dual-comb spectroscopy systems should feature either broad comb span or center-wavelength tunability to address more gas absorption bands, repetition-rate tunability for more flexible optimization of the optical-to-RF mapping relationship, mutual coherence between the probe and local-oscillator combs, as well as low cost and small footprint. Conventionally, EO combs with electronic repetition-rate control have been applied to spectral measurements, but most demonstrations have relied on off-the-shelf EO modulators [Fig. 25(b)][231], making the systems bulky and limited in spectral coverage. Kerr microcombs offer compact broadband comb sources and can support a wide spectral measurement range, e.g., ~20 nm in a representative dual-comb spectroscopy demonstration [Fig. 25(c)][232]. However, repetition rates of Kerr combs are typically fixed after fabrication, making it difficult to accurately control and fine-tune the repetition-rate difference between the two combs. TFLN flat-top non-resonant EO combs combine the strengths of these approaches: a relatively broad comb span can be generated around any center wavelength set by the optical pump, repetition rates are defined by and can be flexibly adjusted by the RF drive, and small form factors[63]. These features make TFLN EO combs promising for integrated dual-comb spectroscopy with broad instantaneous coverage and fine repetition-rate tunability. Beyond optical span and repetition-rate tunability, practical dual-comb spectroscopy also depends on mutual coherence time, carrier–envelope-offset control, absolute frequency calibration, and the mapping between comb-spacing difference and spectral resolution. The comb-spacing difference determines the spectral resolution, whereas the RF detection bandwidth constrains the usable alias-free optical bandwidth[3,231–235].

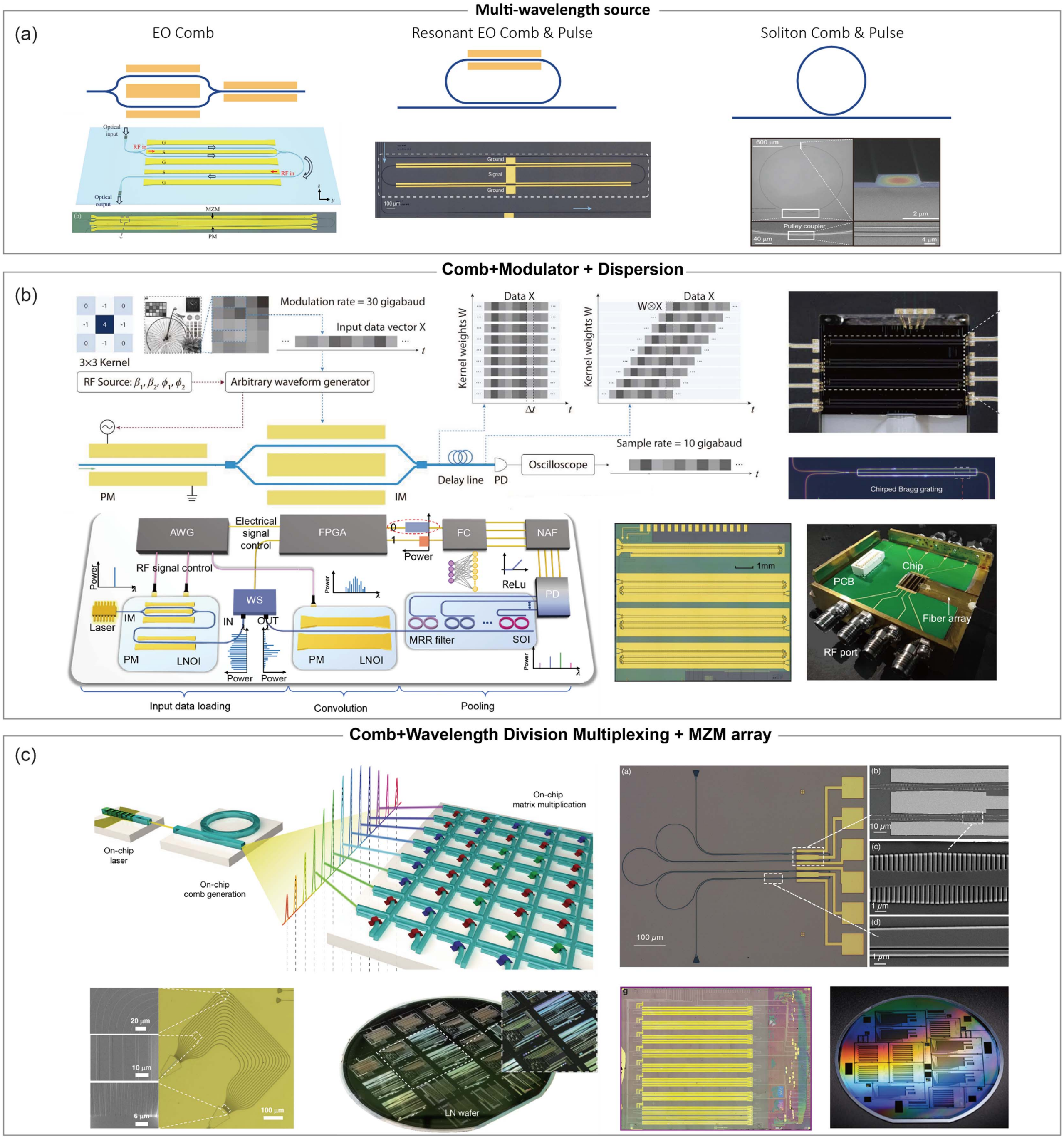

**Fig. 24** TFLN comb-driven optical computing. (a) Representative comb types and their physical realizations[70,81,127]. (b) High-speed optical convolution core based on a modulator and a dispersive component[216,217]. (c) Photonic tensor core enabled by WDM and Mach–Zehnder modulator (MZM) array[166,173,177,218,219]. (a), (b) Adapted with permission from Refs. [70,217], © 2022/2025 IEEE. (c) Adapted from Ref. [173] under CC BY 4.0; Ref. [166] under CC BY 4.0; Refs. [127,218], © 2025 the Author(s), published by Springer Nature (CC BY-NC-ND 4.0); Refs. [81,177,219], © 2019/2021/2024 Springer Nature Limited.

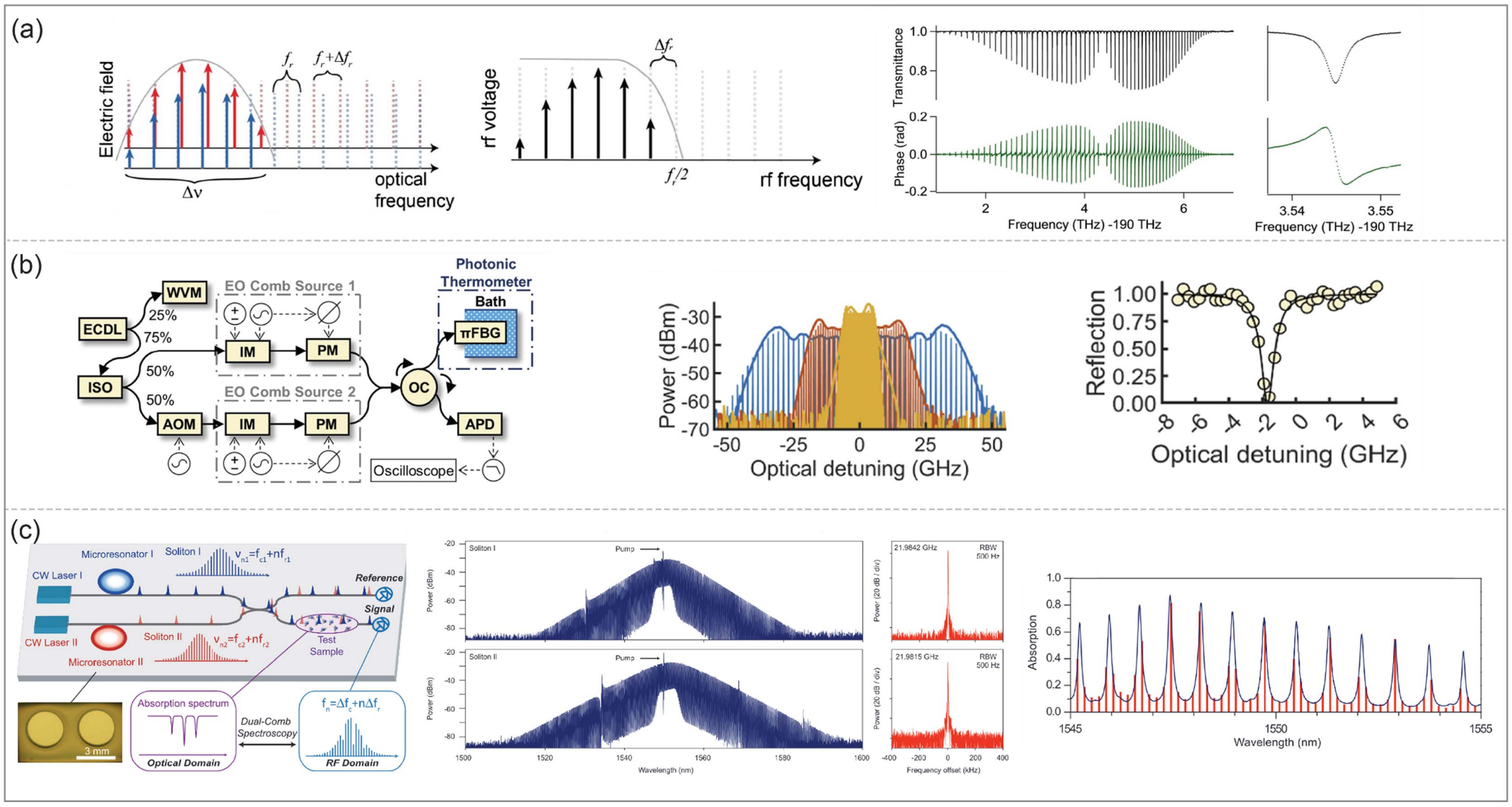

**Fig. 25** Dual-comb spectroscopy. (a) Working principle of dual-comb spectroscopy with two frequency combs with a detuned FSR[3]. (b) An experimental setup based on cascaded EO modulators for photonic thermometry[231]. (c) A demonstration of gas sensing using dual Kerr combs, each pumped by an independent laser[232]. (c) Adapted with permission from Ref. [232], © 2016 American Association for the Advancement of Science. (a), (b) Adapted with permission from Refs. [3,231], © 2016/2023 Optica Publishing Group.

Another metrology application of TFLN frequency combs is light detection and ranging (LiDAR), which, in analogy to radar technology, transmits structured light toward a target and analyzes the returned field to recover the range, velocity, and, in imaging systems, angular or spatial information of the target. Compared with microwave radar, the optical carrier enables finer spatial resolution, while coherent LiDAR can additionally retrieve Doppler shift and phase. Comb-based LiDAR offers multiple mutually coherent wavelength channels, allowing parallel ranging, ambiguity reduction, and reconfigurable field-of-view control.

Qi *et al.* (2025) demonstrated a LiDAR system with a 1.79 GHz acquisition rate by exploiting the excellent repetition-rate tunability of EO combs [Fig. 26(a)], enabling absolute distance measurements with an ultrafast refresh time[236]. In this scheme, a linearly chirped microwave signal drives a PM, so that the comb spacing is rapidly swept in time. The delayed echo signal interferes coherently with the reference signal, and the beat notes generated by multiple comb lines add coherently to form sharp peaks in the time-domain interferogram. The target distance can therefore be directly extracted from the temporal spacing between adjacent peaks, avoiding the relatively slow Fourier-domain frequency estimation used in conventional frequency-modulated continuous-wave (FMCW) LiDARs. EO combs are particularly suitable for this approach because of their fast and broadband repetition-rate tunability.

In another recent report, Chen *et al.* (2026) further leveraged the comb-spacing tunability of EO combs to realize a bionic LiDAR that can adaptively focus on selected regions of interest with higher angular resolution, mimicking the gazing feature of human visual systems [Fig. 26(b)]. In this LiDAR system, a chirped laser is first replicated into multiple mutually coherent wavelength channels by an EO comb, and these channels are spatially dispersed to different viewing angles for parallel FMCW detection. The key advantage of the EO comb lies in its reconfigurability: by reducing the comb spacing, the angular interval between adjacent probing beams can be compressed, enabling a local zoom-in operation in selected regions of interest. This dynamic gazing capability improves the angular resolution to 0.012°, without globally increasing the number of physical detection channels[237]. In addition, because EO combs can be generated from pump lasers over a broad wavelength range, the center wavelength and channel spacing can be independently adjusted, supporting adaptive multichannel signal generation and reception across a wide field of view.

Apart from directly serving as LiDAR sources, EO combs can also act as high-precision frequency rulers for FMCW LiDAR systems. Practical FMCW LiDAR based on frequency-modulated lasers (FMLs) can suffer from mode hops and chirp nonlinearity, which distort the ranging spectrum and hinder unambiguous distance retrieval. Yang *et al.* (2026) recently demonstrated a dual-EO-comb-based frequency-tracking method, in which two TFLN EO combs with slightly different repetition rates are used to measure the instantaneous frequency of an FML under test through multiheterodyne detection. The measured laser-frequency trajectory is then used to calibrate and reconstruct FMCW ranging signals, enabling mode-hop-FML-based FMCW ranging and LiDAR [Fig. 26(c)][238]. The integrated EO combs provide a 29.45 nm optical bandwidth and arbitrarily tunable repetition rates, allowing real-time tracking of FMLs with chirp rates up to $2 \times 10^{18}$ Hz/s. Using this approach, the authors demonstrated FMCW ranging using lasers with severe mode hops and nonlinearity, 3D imaging at a 20 m stand-off distance, and high-update-rate FMCW ranging with a 100 MHz update rate and 7.5 mm ranging resolution.

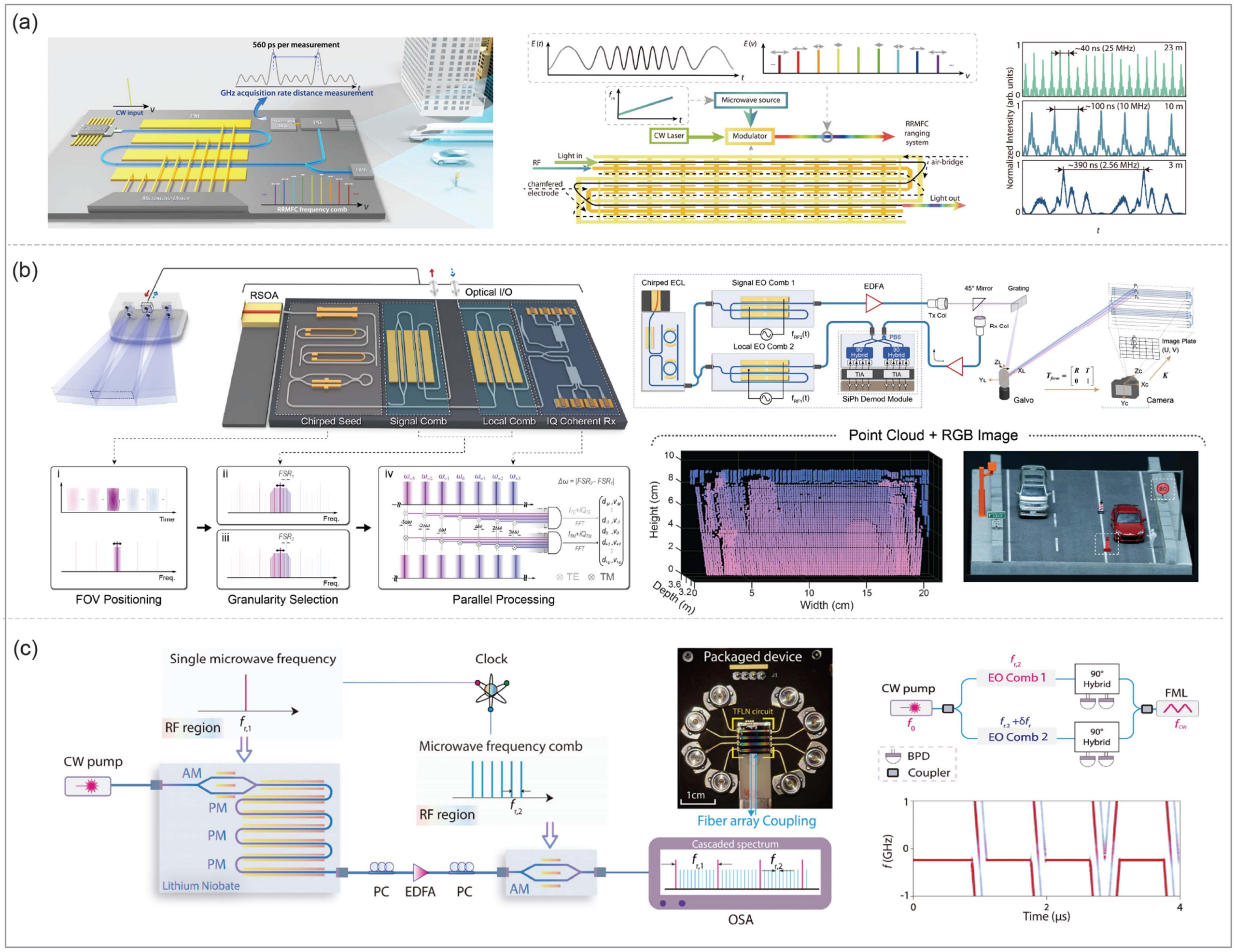

**Fig. 26** Comb-based LiDAR systems. (a) Distance measurement with 1.79 GHz acquisition rate[236]. (b) Adaptive LiDAR with 4D machine vision[237]. (c) Mode-hop-free laser precision measurement[238]. (a), (b) Adapted with permission from Refs. [236,237], © 2025 the Author(s), published by Springer Nature (CC BY-NC-ND 4.0). (c) Adapted with permission from Ref. [238], © 2026 The Author(s), under exclusive licence to Springer Nature Limited.

Microcomb-based dual-comb ranging offers a complementary pathway to EO-comb FMCW and frequency-ruler architectures for chip-scale LiDAR and absolute metrology. Early soliton-microcomb ranging established the feasibility of this approach for absolute distance measurement with chip-scale coherence[239]. Subsequent work has pushed the performance envelope in complementary directions: photon-level dual-comb detection has reached nanometer ranging precision under extremely low returned optical power[240]; dual-chirped microcombs have enabled massively parallel ranging at megapixel-line update rates[241]; and timing-engineered soliton dual-combs have further refined the temporal structure of the ranging waveform for dual-microcomb metrology[242]. Relative to EO-comb systems—which excel in electronically programmable line spacing, multi-beam "gazing," and real-time calibration of frequency-modulated lasers—microcomb dual-comb platforms trade RF agility for a broad, phase-locked optical spectrum and a shared optical clock between the two combs.

## 7 Discussion and Outlook

In this review, we started by tracing the fundamental material properties of TFLN that underpin on-chip comb generation, namely, its strong second-order $\chi^{(2)}$ and third-order $\chi^{(3)}$ nonlinearities and their intricate interplay. We then systematically examined recent advances in the distinct yet complementary paradigms of EO, Kerr, and hybrid comb generation on this platform. Importantly, each of these comb generation technologies features unique combinations of pros and cons, and therefore should be carefully chosen and tailored depending on the specific application scenarios.

This section therefore organizes the remaining discussion in three steps. First, we compare representative integrated platforms using a material-by-mechanism matrix for frequency-comb generation. Second, we classify application architectures by whether they have already been demonstrated with TFLN combs, whether they have been shown on related platforms and are transferable, or whether key active and electronic building blocks still need to be integrated. Third, we map non-resonant EO, resonant EO, Kerr, and MLL combs onto the application priorities reviewed in Sec. 6.

### 7.1 Comparison of OFC Performance across Integrated Platforms

Table 1 provides a detailed comparison of the distinctive properties of different on-chip OFC generation mechanisms across representative integrated platforms: silicon-on-insulator (SOI), silicon nitride (SiN), indium phosphide (InP) and related III–V stacks, TFLN, and TFLT[14,23,99,102]. When considering the performance of the generated combs by themselves, TFLN EO combs clearly stand out in terms of tunability and pump-to-comb conversion efficiency due to its excellent EO response and low optical loss, whereas the achieved performances of TFLN-based OPOs, Kerr combs, and mode-locked lasers are comparable to those of other popular platforms thanks to the all-around material properties of TFLN. The true merit of TFLN-based combs, however, is the possibility of co-integrating different comb generation schemes, as well as other passive, EO, and nonlinear optical functionalities on a single platform, which could inspire a generation of comb-based photonic systems with advanced architectures and novel functions.

### 7.2 Mechanism-Specific Opportunities and Application Mapping

As summarized in Table 1, each comb-generation technology carries a distinct set of trade-offs, so the appropriate scheme should be chosen based on the target application. TFLN is particularly well suited to this need because multiple comb mechanisms and EO functions can be monolithically co-integrated on one chip to better serve diverse application scenarios. We visualize the key performance metrics, namely, repetition rate, spectral span, flatness, and tunability, of the four representative TFLN comb families, i.e., flat-top non-resonant EO combs, resonantly enhanced EO combs, Kerr microcombs, and MLL combs, in the central radar plot of Fig. 27.

As shown in Fig. 27, no single comb technology can provide a one-size-fits-all solution for all applications. Instead, each technology spans a unique design space. For example, flat-top non-resonant EO combs offer the best tunability of repetition rate and center wavelength, alongside good spectral flatness, but they usually require high RF drive power and/or subsequent broadening techniques to achieve a large comb span. Resonantly enhanced EO combs improve modulation efficiency and span by recycling optical and/or microwave energy, yet the adoption of resonators constrains tunability and makes spectral flattening more difficult to achieve. Kerr combs provide much broader, often octave-spanning spectra through cavity-enhanced FWM without the need for RF drive, but their native spacings are usually set by the microresonator FSR and cannot be easily tuned. MLL combs supply pulse trains directly and can offer strong per-pulse peak power, while they require more complicated hybrid integration techniques to introduce a gain medium to the system.

In the corners of Fig. 27, we show four representative applications of TFLN combs and label the corresponding performance metrics (in Roman numerals) that are most critical for each application. Specifically, optical communications favor flat, broadband combs with line spacings matched to WDM grids, making flat-top EO combs and dark-soliton Kerr combs particularly relevant. Microwave photonics often benefits from electronically tunable comb spacing and coherent multiwavelength carriers, so non-resonant EO combs and hybrid EO/Kerr schemes are natural candidates. Optical computing requires stable, low-noise comb lines that can be routed and processed using modulators and filters; here, it is important that comb generation and high-speed EO processing can be co-integrated on a single platform (e.g., X-cut TFLN). Optical metrology and LiDAR place stronger emphasis on mutual coherence, tunable center wavelength, repetition-rate control, and broad instantaneous bandwidth, suggesting that such systems in the future will likely combine the tunability of EO combs with nonlinear broadening or resonant enhancement.

**Table 1** Representative Performance Domains of Mechanism-Resolved Integrated Frequency-Comb Platforms

| Integrated platform | Comb mechanism | Repetition rate (GHz) | Span (nm) | Flatness | Tunability | Pump-to-comb efficiency |
|---|---|---|---|---|---|---|
| SOI | Kerr bright soliton | 100–130 | 1000–2000 | Low | Low | Up to 40% |
| SiN | Kerr bright soliton | 10–1000 | 10–1000 | Low | Low | 0.1%–50% |
| | Kerr dark soliton | 20–200 | 50–250 | Medium | Low–Medium | 20%–50% |
| InP | Mode-locked laser | 1–100 | 0.1–37 | Medium | Medium | N/A (electrically pumped) |
| TFLN/TFLT | Non-resonant EO | 5–50 | 2–30 | High | High | 25%–50% |
| | Resonant EO | 10–30 | 60–140 | Low–Medium | Low–Medium | 0.3%–30% |
| | Kerr bright soliton | 30–650 | 180–1900 | Low | Low–Medium | 0.5%–10% |
| | Kerr dark-pulse | 25 | 200 | Medium–High | Low–Medium | Up to 4% |
| | OPO | 0.25–19 | ~2000 | Medium | Medium–High | 1%–5% |
| | Mode-locked laser | 10–20 | 0.35–45 | Medium | Medium | N/A (electrically pumped) |

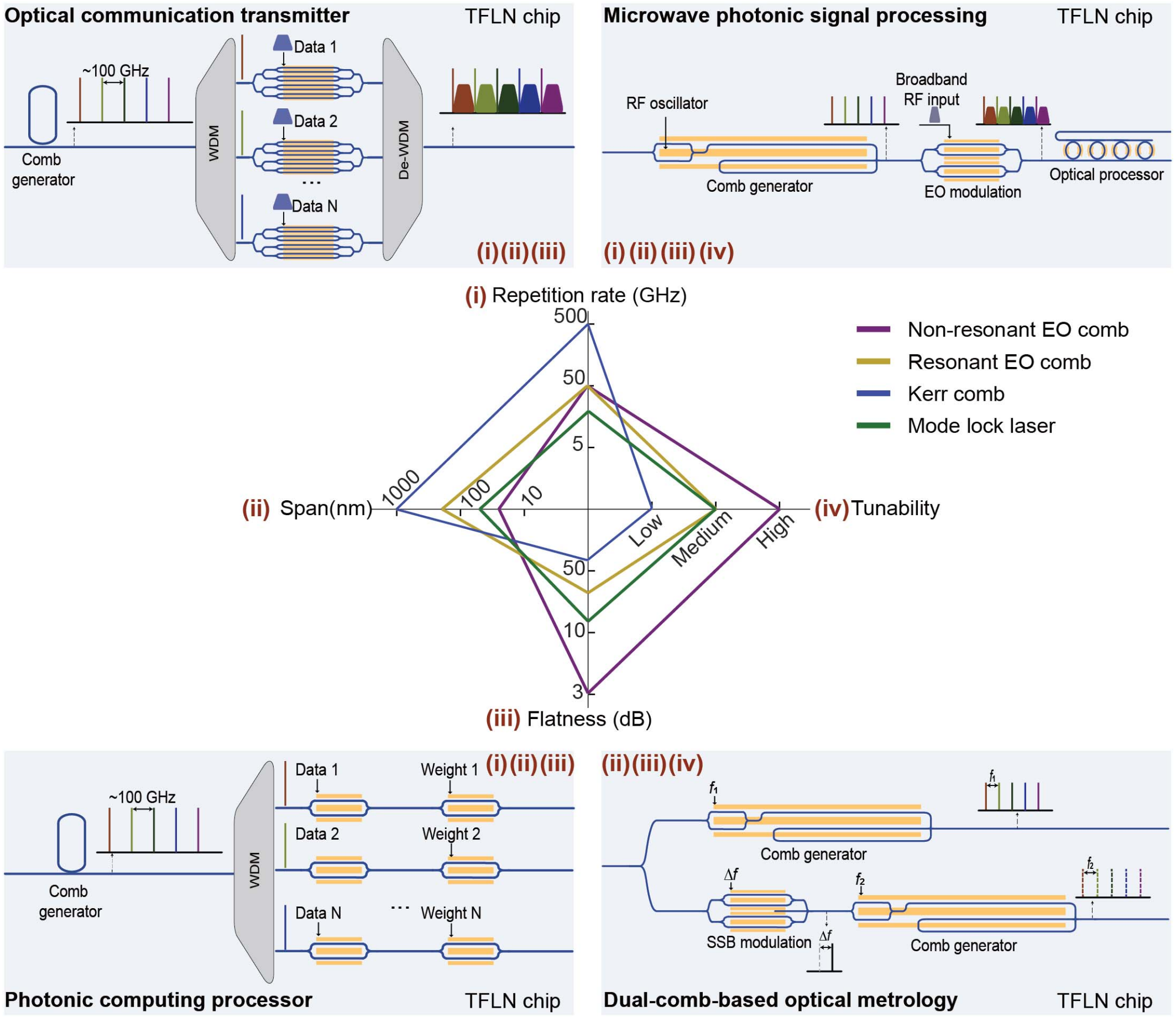

**Fig. 27** Performance comparison of TFLN comb generation schemes along with their application perspectives. The central radar plot qualitatively compares the four representative TFLN comb families, including flat-top non-resonant EO combs, resonantly enhanced EO combs, Kerr microcombs, and mode-locked laser combs, in terms of (i) repetition rate, (ii) optical span, (iii) spectral flatness, and (iv) tunability. The surrounding schematics map these comb characteristics onto major system-level applications: optical communications, microwave photonic signal processing, photonic computing, and dual-comb optical metrology. The Roman-number labels on the corners correspond to those comb performance metrics that are important in each application perspective.

### *7.3 Roadmap toward Advanced System Functionality and Higher Level of Integration*

Looking ahead, we envision that the unique material properties and excellent scalability of the TFLN platform could ultimately enable innovations and breakthroughs in three distinct aspects. First, at the level of comb generation itself, TFLN is distinguished from other integrated platforms by the monolithic combination of the EO effect, the $\chi^{(2)}$ all-optical nonlinearity, and the Kerr effect. Interplay between these properties could lead to novel comb phenomena and physics that are otherwise not available in other material platforms. Second, the ability to monolithically co-integrate additional photonic functions, including EO amplitude/phase modulation, low-loss photonic signal processing, and wavelength (de)-multiplexing, with TFLN comb sources, could inspire a new class of advanced system-level functionalities and application scenarios. Finally, fully deployable comb-based photonic systems demand heterogeneous integration or hybrid co-packaging of other necessary optoelectronic components, including laser, optical amplifier, photodetector, RF source, and RF amplifier, and associated control electronics. Toward these ambitious goals, we outline a three-stage roadmap for the research and development of TFLN combs and comb-based systems.

In the first stage, which most current demonstrations in this review fall into, comb generation mechanisms and phenomena that are well established on bulk LN or other integrated platforms are reproduced on TFLN through platform-specific engineering. For example, flat-top non-resonant EO combs based on cascaded AM and PM, a well-established technique in the bulk LN platform, have been naturally transferred to the TFLN platform, achieving a broader comb span thanks to the lowered $V_\pi$ recycled propagation, mode circulation, cascaded modulators, or simplified RF driving[63–66,69–80]. Resonant variants instead use cavity recirculation[81–84], strong coupling[85,86], dispersion and mode engineering[91–93], coupled cavities or generalized critical coupling[88,91,94,95,97,98], and microwave resonance[101]. High-$Q$ TFLN resonators with engineered dispersion have proven to support Kerr comb generation with bright- and dark-soliton states, similar to those achieved in other material platforms[41,42,53,54,105,107–110,127,141].

The second stage moves beyond simple replication of concepts first proposed in other platforms by demonstrating unique comb properties or system-level functions that are only possible on the TFLN platform. Early prominent examples include EO combs that leverage advanced on-chip devices and architectures to substantially broaden the comb span and reduce the RF power consumption compared with their bulk LN counterparts[63–66,69–80]. The more efficient high-order sideband generation process has enabled larger-ratio optical-frequency division and low-noise microwave synthesis[213,214], while rapid spacing tuning, uniform chirp, and broad optical coverage led to fast ranging[236], adaptive 4D LiDAR[237], and dynamic laser-frequency tracking[238]. Furthermore, monolithically integrated TFLN photonic subsystems that combine EO-comb generation with tunable filtering and EO modulation[50] have enabled microwave-reconfigurable frequency down-conversion of signals up to 67 GHz and up-conversion to 110 GHz[188], while potentially supporting channelized frequency measurement over a 100 GHz range[182]. The rich nonlinearities in TFLN are also inspiring a variety of novel hybrid comb operations, including Pockels control of Kerr combs[54,108,153], EO–$\chi^{(2)}$ oscillation[151], EO-driven Kerr comb generation[155], and self-locked EO–Kerr–Raman dynamics[156]. Hybrid III–V/TFLN cavities provide actively mode-locked comb lasers[145,146], while periodically poled resonators support $\chi^{(2)}$ OPO combs[148–151]. With versatile material properties and a maturing device toolbox, new comb dynamics, system architectures, and applications will continue to emerge in the coming decade.

Finally, the TFLN comb community shall start to close the remaining system integration gaps by combining the TFLN photonic core with active and electronic components at the chip or package level. TFLN chips are well suited to provide the comb generation, modulation, routing, filtering, and nonlinear processing functions, whereas lasers, optical gain, photodetectors, RF sources and drivers, and control electronics shall be incorporated through heterogeneous integration or co-packaging. Key elements already exist, including III–V/TFLN electrically pumped comb lasers[145,146], high-power and broadband photodetectors integrated with TFLN[203,204], Er-doped TFLN waveguide amplifiers[205], and electrically pumped laser transmitters[206]. Photonic wire bonding[243] and multichannel high-speed flip-chip packaging[244] provide complementary routes to laser integration and dense electrical interfacing, consistent with the broader integration and co-packaged optics roadmaps of the TFLN platform[157]. The goal, however, is not integration alone. Co-designing optical coupling, microwave delivery, thermal management, packaging, and electronic stabilization could shorten optical and RF interconnects. The resulting systems could feature reduced loss, latency, and phase drift, while improving energy efficiency, long-term stability, and channel scalability. The transition could take time, but when completed, may lead to communication, microwave–photonic, ranging, and frequency-synthesis systems built around TFLN cores with functionality and aggregate performance well beyond present solutions.

## Acknowledgements

This work was supported by the National Natural Science Foundation of China (No. 62525507) and the Research Grants Council, University Grants Committee (No. STG3/E-104/25-N).